\documentclass[structabstract]{aa}  
\usepackage{natbib}
\usepackage{array}
\usepackage{subfigure} %Sous-figures
\usepackage{graphicx}
\usepackage{txfonts}
\usepackage{aas_macros}

\begin{document}

    \title {Impact of orbital motion on the structure and stability of adiabatic shocks in colliding wind binaries}

    \author{A. Lamberts 
        \inst{1}
        \and  G. Dubus 
        \inst{1}
        \and G. Lesur
        \inst{1}
       \and  S. Fromang 
        \inst{2}
       }

     \institute{UJF-Grenoble 1 / CNRS-INSU, Institut de Plan\'{e}tologie et d\textquoteright Astrophysique de Grenoble (IPAG) UMR 5274, Grenoble, F-38041, France 
       \and
Laboratoire AIM, CEA/DSM - CNRS - Universit\'e Paris 7, Irfu/Service d'Astrophysique, CEA-Saclay, 91191 Gif-sur-Yvette, France}       

    \date{\today}% Accepted **. Received **; in original form **}

    \offprints{A. Lamberts: astrid.lamberts@obs.ujf-grenoble.fr}

%\and

\titlerunning{KHI in colliding wind binaries}
\authorrunning{A. Lamberts et al.}

\abstract
% context heading (optional)
{The collision of winds from massive stars in binaries results in the formation of a double-shock structure with observed signatures from radio to X-rays.}
% aims heading (mandatory)
{We study the structure and stability of the colliding wind region as it turns into a spiral due to orbital motion. We focus on adiabatic winds, where mixing between the two winds is expected to be restricted to the Kelvin-Helmholtz instability. Mixing of the Wolf-Rayet wind with hydrogen-rich material is important for dust formation in pinwheel nebulae such as WR 104, where the spiral structure has been resolved in infrared.}
% methods heading (mandatory)
{We use the hydrodynamical code RAMSES to solve the equations of hydrodynamics on an adaptive grid. A wide range of binary systems with different wind velocities and mass loss rates are studied with 2D simulations. A specific 3D simulation is performed to model \object{WR 104}. }
% results heading (mandatory)
{Orbital motion leads to the formation of two distinct spiral arms where the Kelvin-Helmholtz instability develops differently. We find that the spiral structure is destroyed when there is a large velocity gradient between the winds, unless the collimated wind is much faster. We argue that the Kelvin-Helmholtz instability plays a major role in maintaining or not the structure. We discuss the consequences for various colliding wind binaries. When stable, there is no straightforward relationship between the spatial step of the spiral, the wind velocities, and the orbital period. Our 3D simulation of WR 104 indicates that the colder, well-mixed trailing arm has more favourable conditions for dust formation than the leading arm. The single-arm infrared spiral follows more closely the mixing map than the density map, suggesting the dust-to-gas ratio may vary between the leading and trailing density spirals. However, the density is much lower than what dust formation models require. Including radiative cooling would lead to higher densities, and also to thin shell instabilities whose impact on the large structure remains unknown.}
% conclusions heading (optional), leave it empty if necessary
%{The colliding wind region on scales where orbital motion becomes important is strongly affected by instabilities arising on much smaller scales.}
{}
\keywords {hydrodynamics - instabilities - binaries :general  - stars:  individual : WR 104 - stars: winds, outflows}
\maketitle

\section{Introduction}

Massive stars possess highly supersonic winds due to radiation pressure on atomic lines. Wind mass-loss rates range from $\dot{M}\simeq 10^{-8} M_{\odot}$~yr$^{-1}$ for O or B type stars to $\dot{M}\simeq 10^{-4} M_{\odot}$~yr$^{-1}$ for Wolf-Rayet stars (WR) \citep {2008A&ARv..16..209P}. Most massive stars lie in binary systems, where the interaction of the two supersonic stellar winds creates two strong shocks separated by a contact discontinuity. The geometry of the colliding wind region depends on the momentum flux ratio of the winds \citep{1990FlDy...25..629L}
\begin{equation}\label{eq:eta}
\eta\equiv \frac{\dot{M}_2v_2}{\dot{M}_1v_1}
\end{equation}
where $v$ is the wind velocity. The subscript 1 usually stands for the stronger wind, the subscript 2 for the weaker one so that $\eta\leq 1$. There are several important observational signatures of the colliding wind region. The shock-heated gas generates observable thermal X-ray emission \citep[e.g.][]{1976PAZh....2..356C,Luo:1990mp,Usov:1992re,Stevens:1992on}. The presence of intra-binary structures causes variations of the emission line profiles with orbital phase \citep[e.g.][]{1988ApJ...334.1021S, 1993ApJ...407..252W}.  The high densities reached in the colliding wind region are thought to enable the formation of dust, explaining the large infrared emission from binary systems with WR stars \citep{1987A&A...182...91W,1991MNRAS.252...49U}, and the formation of spiral structures extending to distances up to 300 times the binary separation (``pinwheel nebulae", \citealt{1999Natur.398..487T}). In some systems, diffusive shock acceleration of particles leads to non-thermal radio emission \citep{2007A&ARv..14..171D}. The radio emission has been resolved by long baseline interferometry and shown to have a morphology changing with orbital phase  \citep[e.g.][]{2005ApJ...623..447D}. A new exotic class of colliding wind binaries is gamma-ray binaries, where the non-thermal emission is thought to arise from the interaction of a pulsar relativistic wind with the wind of its massive stellar companion \citep{2006A&A...456..801D}. Interpreting all this observational data requires increasingly detailed knowledge of the physics of colliding winds, hence numerical simulations, notably of the large scale regions that can be resolved in radio or infrared.

On large scales, orbital motion is expected to turn the shock structure into a spiral, although we will show that this is not always true. Orbital motion breaks the symmetry with respect to the binary axis and no analytic solution predicts the detailed structure of the colliding wind region. Material in the spiral is generally thought to behave ballistically, so that the step of the spiral is the wind velocity $v$ times the orbital period $P_{\rm orb}$. The wind velocity to use is unclear. \citet{2008ApJ...675..698T} took the speed of the dominant wind $v_1$ (dominant in the sense that $\dot{M}_1 v_1 \geq \dot{M}_2 v_2$), whereas \citet{2008MNRAS.388.1047P} assume that it is the slower wind that determines the step of the spiral but focus their study on binaries with equal wind velocities. Simple dynamical models of the shocked layer have been developed for use with radiative transfer codes, assuming that the double shock structure is infinitely thin (thin shell hypothesis) and that the material is ballistic \citep{Harries2004,2008MNRAS.388.1047P}. The spiral structure is then reproduced at small computational cost but this neglects the impact of the pressure, which creates a distinction between both arms of the spiral \citep{Lemaster:2007sl,2009MNRAS.396.1743P,2011A&A...527A...3V}, the influence of the reconfinement of the weaker wind for small $\eta$ \citep{PaperI} and the large-scale evolution of instabilities in the colliding wind region (see below). Up to now no 2D or 3D hydrodynamical simulation has modelled a complete step of the spiral. We achieve this by using the hydrodynamical code RAMSES \citep{2002A&A...385..337T} with Adaptive Mesh Refinement (AMR). AMR allows large scale simulations to be performed while keeping a high enough resolution close to the binary in order to form the shocks properly (\S2).

Small scale simulations without orbital motion (see \citealt{Stevens:1992on} and references in \citealt{PaperI}, hereafter Paper I) have shown that several instabilities are at work in colliding wind binaries. Thin shell instabilities occur when cooling is important so that the shocked zone narrows to a thin layer, which is easily perturbed \citep{1994ApJ...428..186V}. They provoke strong distortions of the whole colliding region \citep{2009MNRAS.396.1743P,2011A&A...527A...3V}. However, these instabilities are unlikely to be dominant in wide binary systems where cooling is inefficient, the shocks adiabatic, and the colliding wind region wider \citep{Stevens:1992on}. In this case, the velocity  difference between both winds triggers the Kelvin-Helmholtz instability (KHI) at the contact discontinuity (Paper I). Including orbital motion has led to contradictory results. \citet{Lemaster:2007sl} found that eddies develop even when the winds are completely identical, because orbital motion introduces a velocity difference. \citet{2009MNRAS.396.1743P} found no eddies in a simulation with a similar setup. \citet{2011A&A...527A...3V} also found no eddies, although their simulation has an initial non zero velocity difference $\beta=v_1/v_2=3/4$, and argued that orbital motion stabilises the KHI. Larger velocity differences between the winds have not been investigated. We performed a set of 2D simulations to study how orbital motion changes the shock structure and the development of the KHI, focusing exclusively on binaries with adiabatic shocks (\S3). The size of the spiral step and the stability of the spiral on large scales are addressed in \S4. 

\object{WR 104} is an example of a long orbital period binary system where the collision between the winds of the WR and its early-type companion is expected to be close to adiabatic (see \S5). The  infrared emission is very well matched by an Archimedean spiral although its brightest point is shifted by 13 milli-arcseconds from its centre, possibly because dust formation is inhibited closer in (\citealt{2008ApJ...675..698T}, hereafter T2008). The WR wind is hostile to dust formation due to its high temperature, low density, and absence of hydrogen ~\citep{2000A&A...357..572C}. The wind collision region is more favourable, providing high densities, shielding from the UV radiation of the WR star, and the possibility of mixing with hydrogen from the companion star \citep{2007ASPC..367..213M}. We carried out 2D and 3D hydrodynamical simulations using the parameters of WR 104 to investigate these questions (\S5). We then relate all our results to observations (\S6).

\section{Numerical Simulations}\label{numerics}
\subsection{Equations}
We use the hydrodynamical code RAMSES for our simulations \citep{2002A&A...385..337T}. This code uses a second order Godunov method to solve the equations of hydrodynamics
\begin{eqnarray}
	\frac{\partial\rho}{\partial t}+\nabla \cdot(\rho \mathbf{v})    &=&  0\\  
	\frac{\partial(\rho \mathbf{v})}{\partial t}+\nabla \cdot (\rho \mathbf{v}\mathbf{v})+\nabla P 	&=& 0	\\
	\frac{\partial E }{\partial t}+\nabla \cdot[\mathbf{v}(E+P)]	&=& 0
\end{eqnarray}
where $\rho$ is the density, $\mathbf{v}$ the velocity, and $P$ the pressure of the gas. The total energy density $E$ is given by
\begin{equation}\nonumber
E= \frac{1}{2}\rho v^2+\frac{P}{(\gamma -1)}
\end{equation}
$\gamma$ is the adiabatic index, set to 5/3 to model adiabatic flows. 
\subsection{Numerical parameters}
We use the MinMod slope limiter together with the exact Riemann solver (\S 3-4) or the HLLC (\S 5) Riemann solver to avoid numerical quenching of instabilities. We perform 2D and 3D simulations on a Cartesian grid with outflow boundary conditions. We use AMR which enables to locally increase the spatial resolution according to the properties of the flow. We base the refinement criterion on velocity gradients. In section \S3 we perform small scale simulations where the size of the computational domain is $l_{box}=40a$, with $a$ the binary separation. We have $n_x=64$ for the resolution of the coarse (unrefined) grid and use 7 levels of refinement. This gives an equivalent resolution which is at least two times better than in former studies \citep{Lemaster:2007sl,2009MNRAS.396.1743P,2011A&A...527A...3V}. In section \S 4 we perform larger scale simulations where $l_{box}=400a$ , the coarse grid also has $n_x=64$ but we use up to 9 levels of refinement. In some cases we adapt the size of our grid to larger or smaller values to model a complete step of the spiral. 

\subsection{Generation of the winds}
To simulate the winds, we keep the same method as used in Paper I, which was largely inspired by \citet{Lemaster:2007sl}. Around each star, we create a wind by imposing a given density, pressure, and velocity profile in a spherical zone called mask. The masks are reset to their initial values at each time step to create steady winds. We add two passive scalars $s_1$ and $s_2$ to distinguish both winds and to quantify mixing.  We initialise the passive scalars in the masks; their evolution is determined by
\begin{equation}
  \label{eq:passive_scalar}
  \frac{\partial{\rho s_i}}{\partial t}+\nabla\cdot(\rho s_i\mathbf{v})=0 \qquad i=1,2
\end{equation}
In the free wind of the first star $s_1=1$ and $s_2=0$, in the second wind it is the other way round. In the shocked zone both scalars have an intermediate value which accounts for the mixing of the winds. The rotation of the stars is clockwise in the figures, their positions are updated using a leapfrog method. 
 For each simulation, the input parameters are the mass $M$, mass loss rate $\dot{M}$, wind velocity $v$ (which we suppose to be constant), and Mach number $\mathcal{M}$ at $r=a$ for each star. The exact value of the Mach number does not matter for the colliding wind region, as long as it is high enough that pressure terms can be neglected (Paper I), which is the case for massive star winds. Here, the Mach numbers of both winds are set to 30. In all our simulations, the star with the highest momentum flux is considered as the first star. We will refer to its wind as the stronger wind. The values of the parameters of the winds in the simulations are given in the table in Appendix B. Both stars have a mass of 15 $M_{\odot}$ and the binary separation $a$ is 1 AU. The corresponding orbital period $P_{\rm orb}$ is 0.18 yr (67 days). The orbital velocity of the stars is $v_{orb}=81$ km s$^{-1}$. We only study circular orbits.

\subsection{2D and 3D simulations}
We perform our 2D simulations  in the orbital plane of the binary. We thus model the cylindrical $(r,\theta)$ plane instead of the $(r,z)$ plane as classically done (e.g.~\citealt{Stevens:1992on,1995MNRAS.277...53B,2006A&A...446.1001P}). This implies the density evolves $\propto r^{-1}$ instead of $\propto r^{-2}$ in a spherical geometry. For a given $\eta$ the structure of the colliding wind binary is thus different in 2D and 3D. However, as discussed in Paper I, the mapping $\sqrt{\eta_{\rm 3D}}\rightarrow \eta_{\rm 2D}$ captures most of the 3D structure in the 2D simulations. This point is re-discussed in \S5 where we compare the results of a 2D  simulation of WR 104 with a full 3D simulation including orbital motion. A major advantage of our 2D setup is the possibility to implement orbital motion for a modest computational cost, enabling the study of the flow structure up to scales currently inaccessible to full 3D calculations.

\section{Impact of orbital motion on the shock arms}\label{small_scale}

We carried out simulations of adiabatic colliding winds identical to those carried out in paper I except that they now include orbital motion to study its impact on the shock structure and development of the KHI. The simulations explore $\eta=1$ and $\eta=0.0625$ for different velocity ratios $\beta\equiv v_1/v_2=1,2,20$, all in a box of size $8a$. Briefly, the results without orbital motion were that (1) no instability is seen when $\beta$=1; (2) for $\beta \geqslant 2$, the instabilities affect the position of the contact discontinuity, for $\beta=20$ the KHI also affects the shocks positions; (3) for $\eta=0.0625$ the instabilities remain confined to the weaker wind. We present first the results of the simulations for $\beta$=1, where the KHI instability may be triggered (or not) by orbital shear (\S 3.1). We then discuss the simulations with $\beta\neq 1$. In these cases, the dominant wind is slower and much denser than the weaker wind. For $\eta=1$, there is no difference between simulations where $\beta=B$ and $\beta=1/B$. 

\begin{figure}
  \centering
  \includegraphics[width = .3\textwidth]{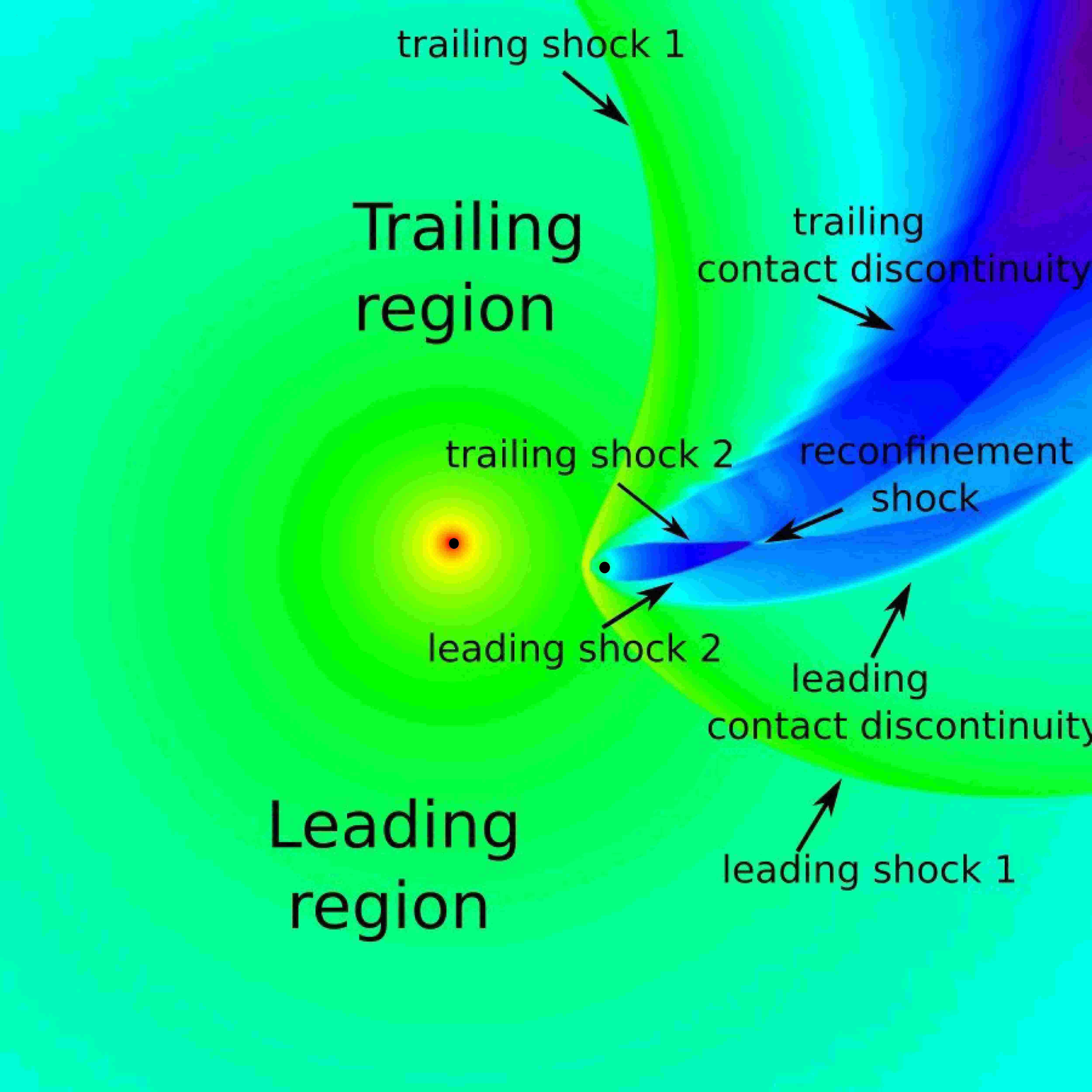}
  \caption{Geometry of a colliding wind binary including orbital motion ($\{\eta$=0.065,$\beta$=0.5$\}$). The stars are shown by the black circles. The spiral structure has a leading and trailing arm. Each arm is composed of one shock from each wind and a contact discontinuity. In this case both shocks from the wind of the second star intersect, totally confining the second wind.}
  \label{fig:geometry}
\end{figure}

A view of the overall colliding wind structure is given in Fig.~\ref{fig:geometry}. We define the leading arm as the arm preceding the second star, with respect to orbital motion (clockwise motion). The trailing arm is the second part of the spiral. Note that there is no dominant wind when $\eta=1$ so that the definition of leading/trailing is degenerate in this case. (The definition also has no link with the definition commonly used in galactic dynamics.) In each arm there is a shock in the wind from the first star and a shock in the wind from the second star, separated by a contact discontinuity. In 2D simulations, when $\eta <0.25$ the second wind is confined by the intersection of the shocks. In 3D simulations this occurs for $\eta \simeq 0.06$  (Paper I).

\begin{figure}
  \centering
  \includegraphics[width = .3\textwidth]{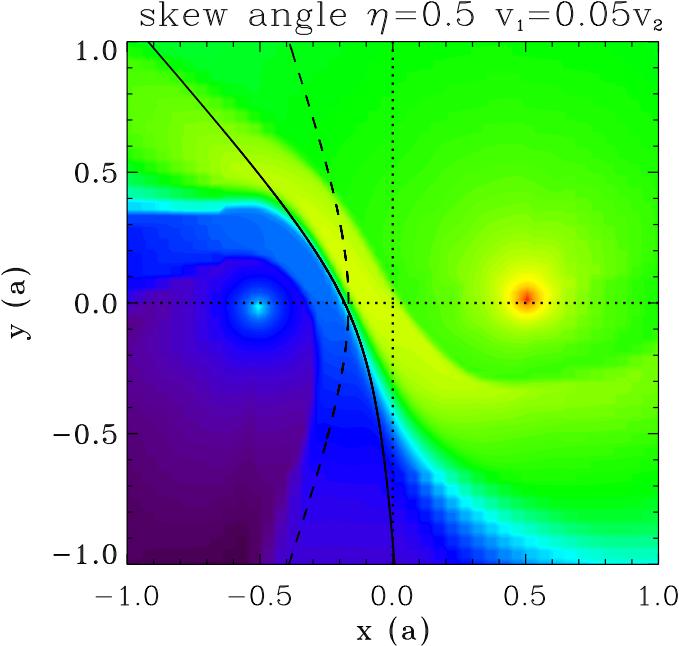}
  \caption{Determination of the skew angle $\mu$ for $\{\eta=0.5,\beta=0.05\}$. The dashed line showes the theoretical position of the contact discontinuity for $\mu=0$ (no orbital motion). The solid line fits  the contact discontinuity in the simultion.}
  \label{fig:skew}
\end{figure}

Close to the binary, relative motion of the stars creates an ``aberration'' \citet{2008MNRAS.388.1047P} of the shocked zone.\citet{2008MNRAS.388.1047P} introduce the skew angle $\mu$ which measures the offset between the line of centers of the stars and the symmetry axis of the shocked region. It is given by 

  \begin{equation}
    \label{eq:skew}
    \tan \mu =\frac{v_{orb}}{v}
  \end{equation}
where $v$ is taken as the speed of the slowest wind. This angle remains small unless the velocities of the winds are strongly reduced or orbital motion becomes important We measured $\mu$ in our simulations by finding the best fit of the analytic position of the contact discontinuity derived by \citet{Canto:1996jj}. An example is given on Fig. \ref{fig:skew}. We measured $\mu\simeq 22^{ \circ}$ in the simulation $\{\eta=0.0625,\beta=0.05\}$ while the theory predicts $\mu=21^{\circ}$. The simulation of WR 104 (see \S5) gives $\mu\simeq 9^{\circ}$ while the theory predicts $\mu=8^{\circ}$. $\mu$ is too small in our other simulations to allow correct measurements.

\subsection{Simulations  with $\beta=1$}
\begin{figure*}
  \centering
  \includegraphics[width = .3\textwidth]{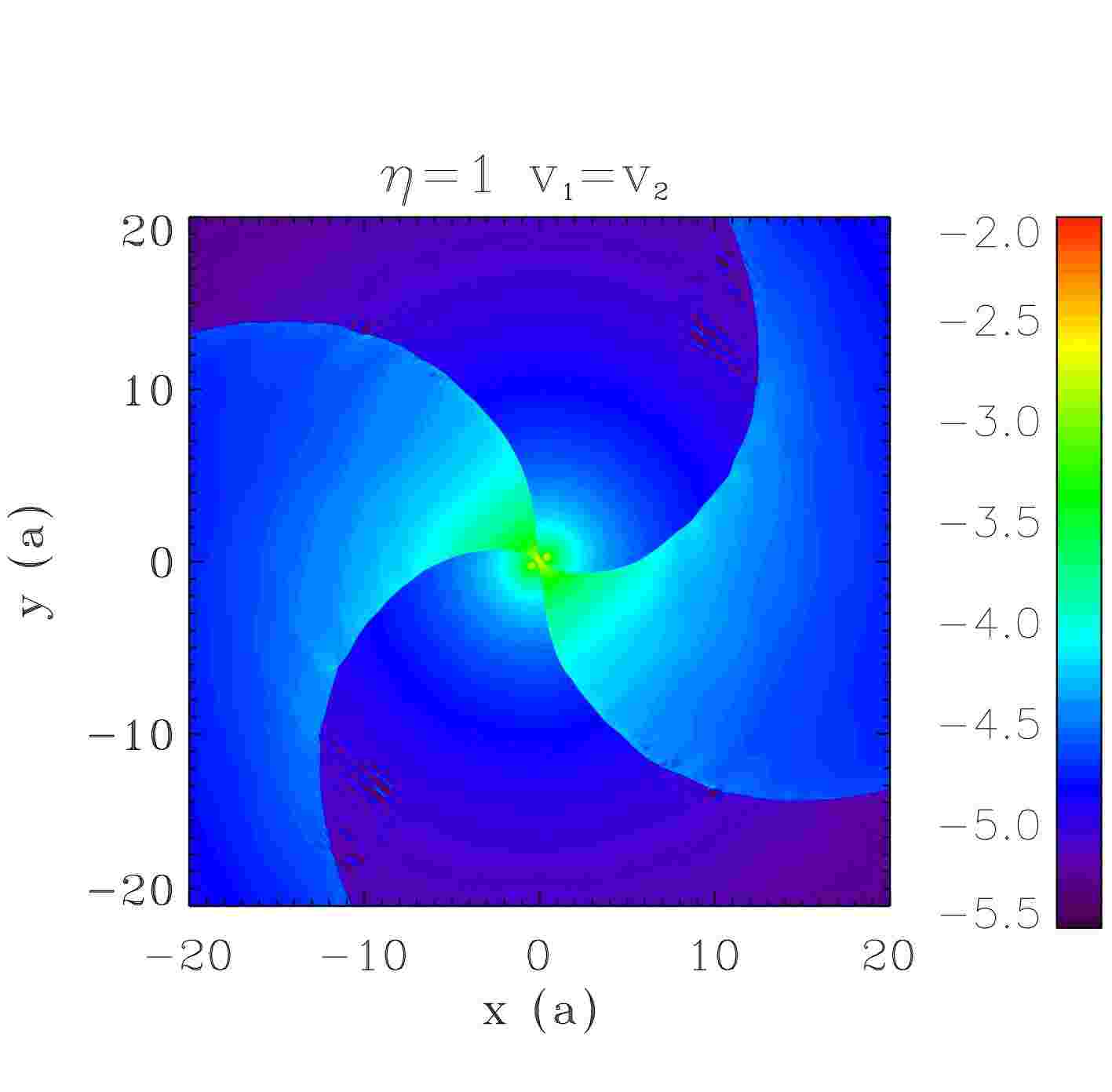}
  \includegraphics[width = .3\textwidth]{v_C_1_1_l}
  \includegraphics[width = .3\textwidth]{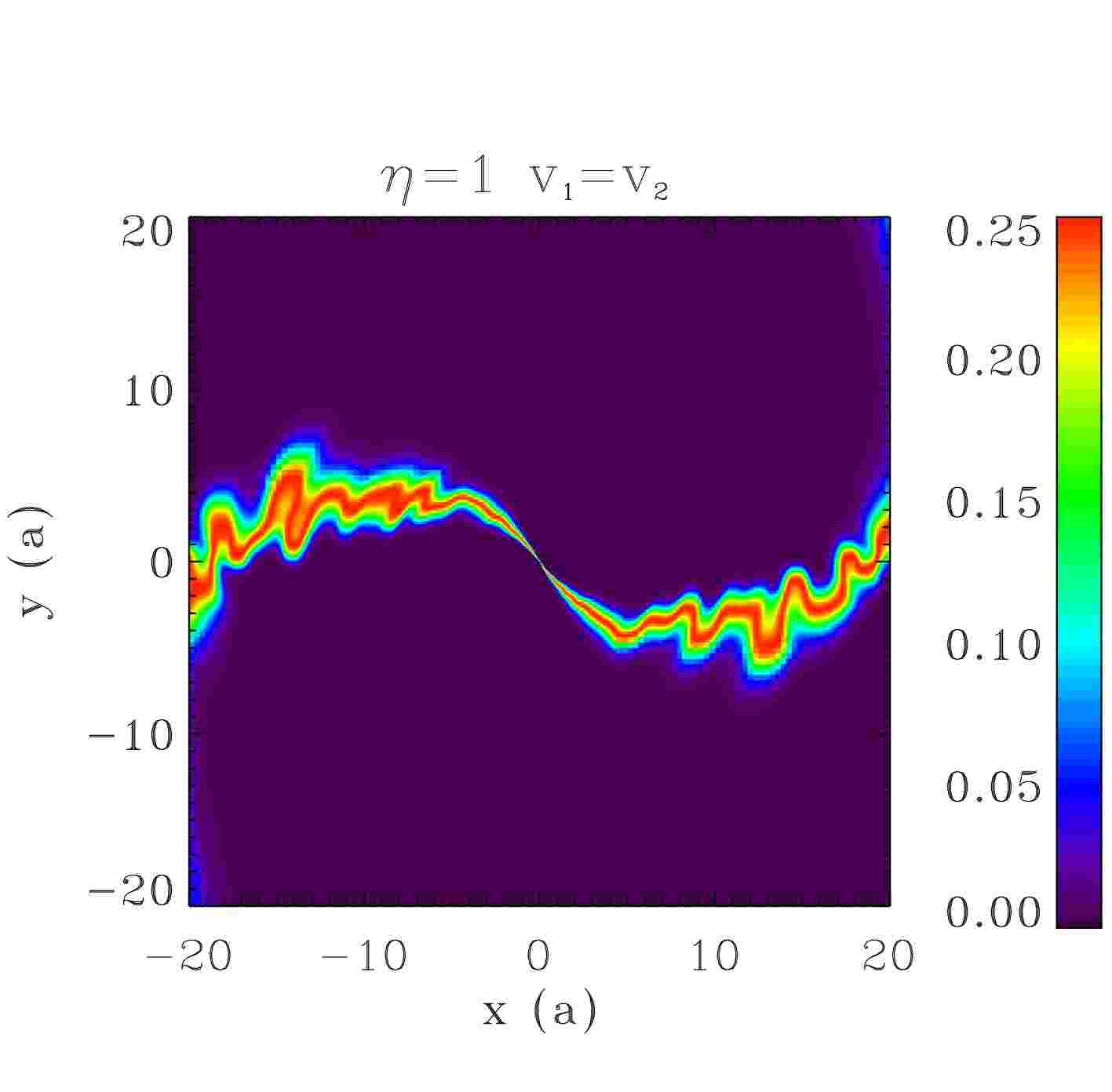}
  \caption{Small scale simulation: density, velocity, and mixing for a simulation with identical winds ($\eta=1,\beta=1$). The density is given in g cm$^{-2}$, the velocity is in km s$^{-1}$, the mixing of the winds is a dimensionless variable. The length scale is the binary separation $a$.}
  \label{fig:identical}
\end{figure*}

Fig.~\ref{fig:identical} shows the density, velocity, and mixing map for a simulation with identical winds $\{\eta=1,\beta=1\}$. We determine the mixing by the product of the passive scalars $s_1\times s_2$. The free (unshocked) winds correspond to the low density parts at the top and bottom. The denser parts are the shocked winds. The radial inhomogeneities visible in the unshocked wind region of the velocity map is a numerical artefact: it corresponds  to  minute anisotropies in the stellar wind due to the finite size of the masks. As shown in Appendix A the orbital period is much longer than the local shear timescale. The Coriolis force does not impact on the development of the KHI. However, the velocity map shows that a $\simeq 20 \%$ velocity difference develops in each arm at a distance $\simeq 20a$ from the binary. Orbital motion makes the shocked material leading the contact discontinuity (red region on the velocity map)  accelerate in the lower density free wind region while the shocked wind trailing the contact discontinuity (in green) moves into the denser, shocked material of the other wind. The resulting velocity difference is sufficient to trigger the KHI even though the original wind speeds are equal. The instability is clearly present in the mixing map. We therefore confirm the results of \citet{Lemaster:2007sl} that orbital motion triggers the KHI even when the winds are identical. We also observe, like they do, an artificial enhancement of the instabilities when the shocks align with the grid. Our simulations contradict the results from \citet{2011A&A...527A...3V} who find no KHI. Their simulations are performed with a Lax Friedrich Riemann solver. We ran a test simulation with the Lax Friedrich Riemann solver and observed no development of the KHI either, due to the important numerical diffusivity.

\begin{figure*}
  \centering
  \includegraphics[width = .3\textwidth]{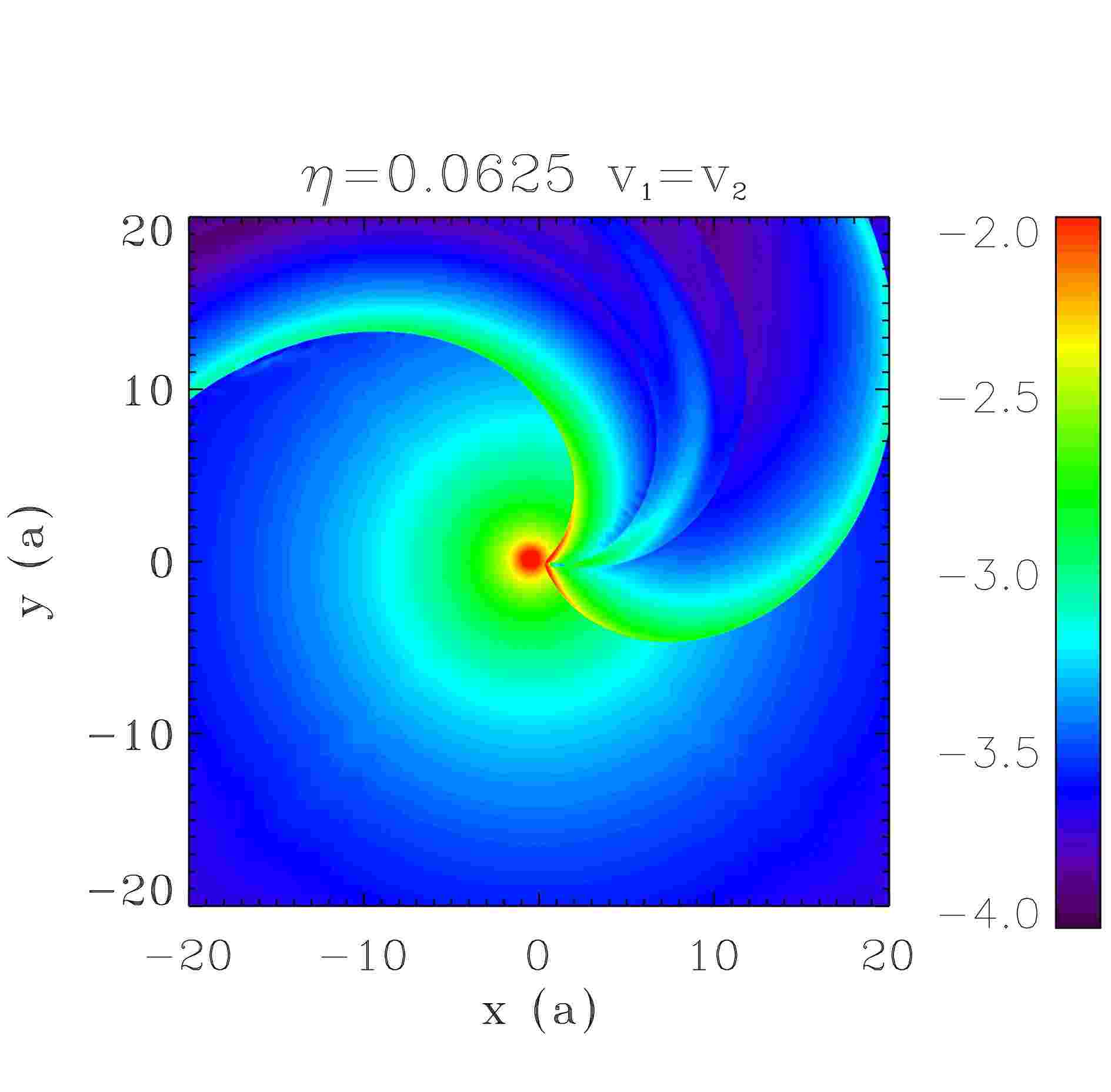}
  \includegraphics[width = .3\textwidth]{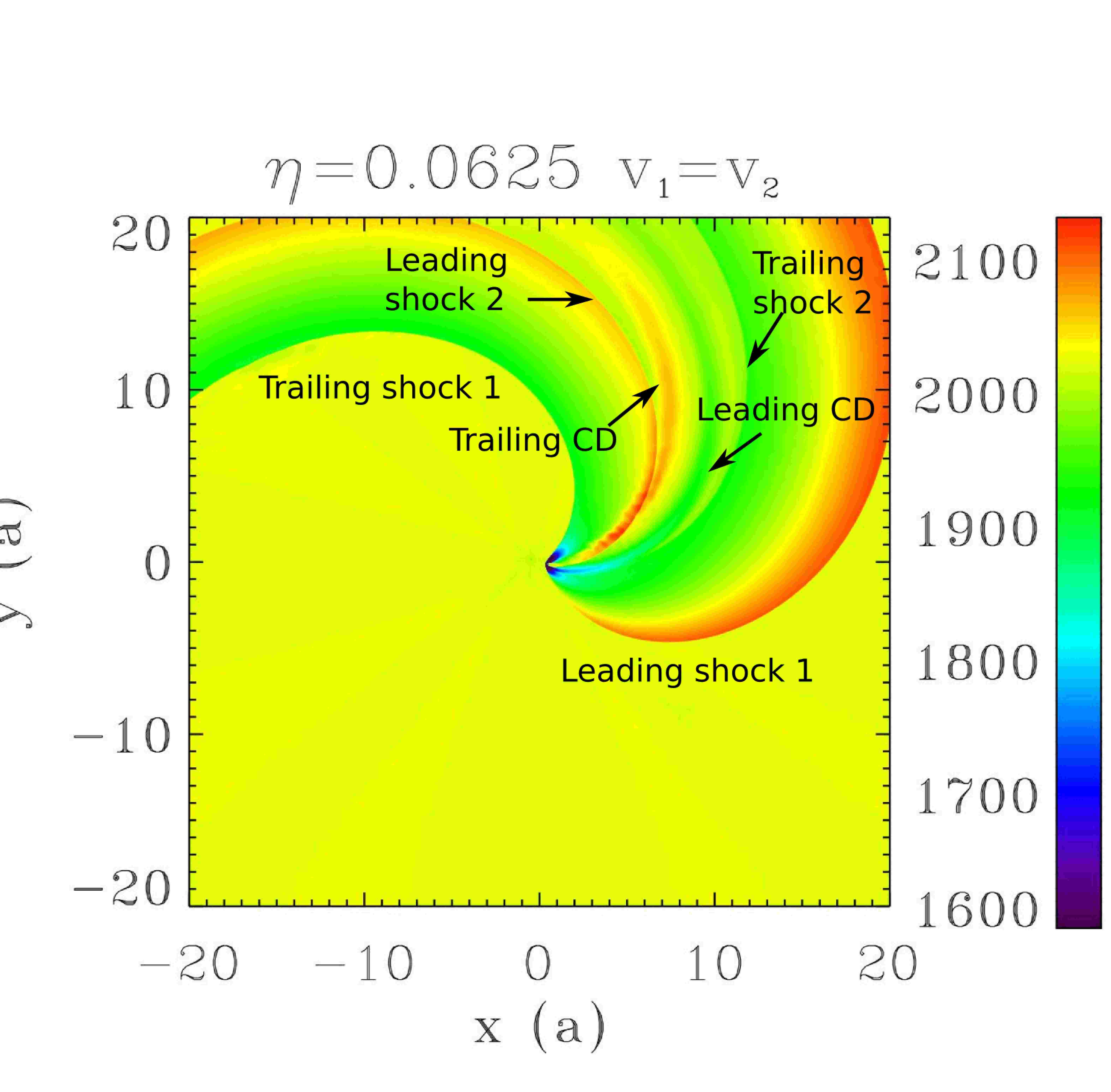}
  \includegraphics[width = .3\textwidth]{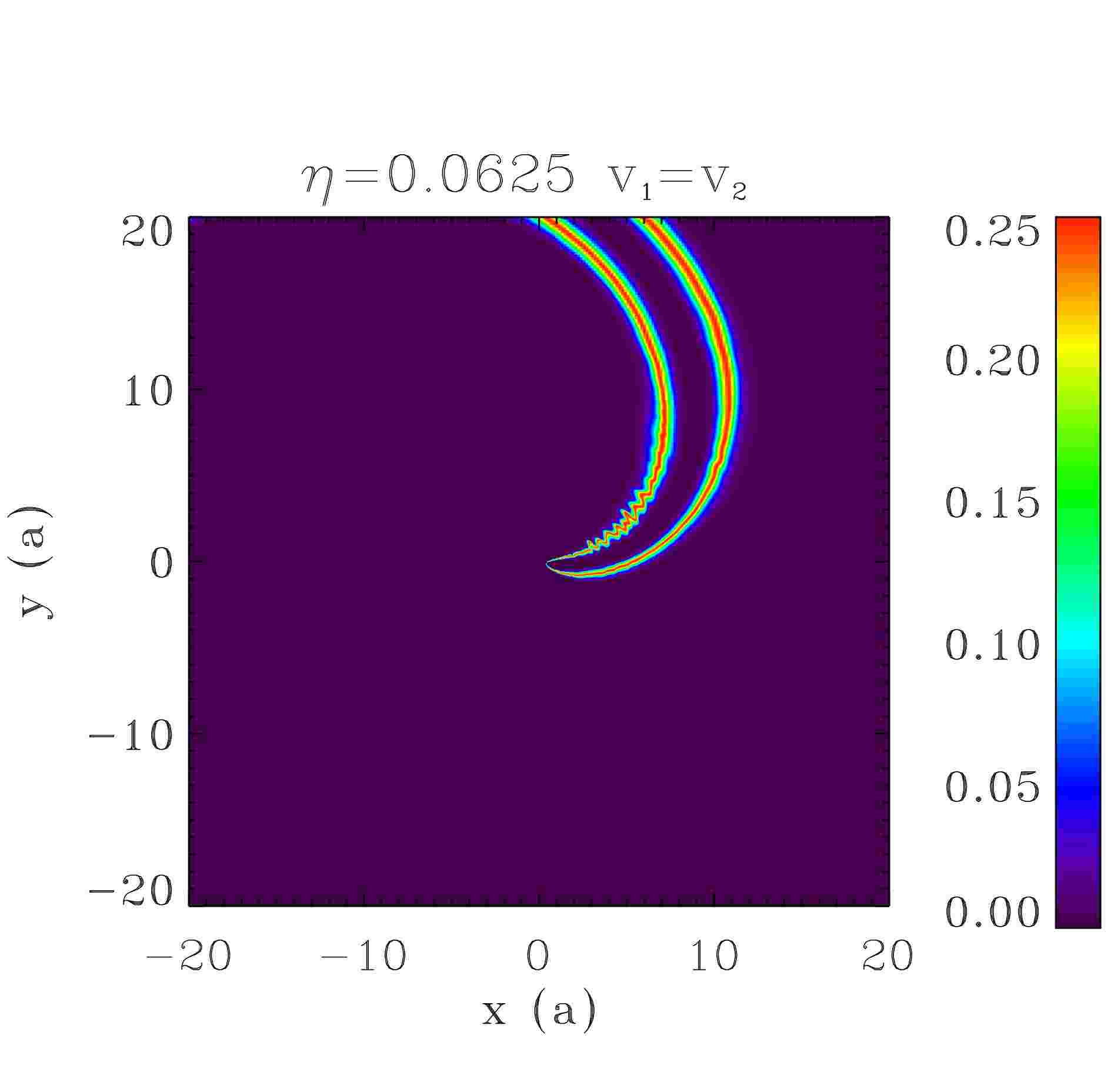}
  \caption{Small scale simulation: same as Fig.~\ref{fig:identical} for $\{\eta=0.0625,\beta=1\}$}. 
  \label{fig:small_scale_0625}
\end{figure*}

Fig.~\ref{fig:small_scale_0625} shows the  density, velocity, and mixing for $\{\eta=0.0625,\beta=1\}$. Because of the low value of $\eta$, there is a reconfinement shock (Paper I) behind the second star. The various discontinuities are indicated in the velocity map, to be compared with the simpler geometry shown previously in Fig.~\ref{fig:geometry}. Our simulations without orbital motion showed no KHI because the initial velocities are identical. Here, as in the $\eta=1$ case, orbital motion leads to velocity shear and mixing at the contact discontinuity. The KHI is confined to narrow regions close to the discontinuity because of the low $\eta$ and not because of orbital motion: we had found the same behaviour in the models explored in paper I.  We also see that complex velocity structures arise in the colliding wind region even in this {\em a priori} simple case where both winds have the same velocity, highlighting the possible difficulties in interpreting spectral line features arising from this region without guidance from numerical simulations.

\subsection{Simulations with $\beta \ne 1$}

\begin{figure}
  \centering
  \includegraphics[width = .23\textwidth]{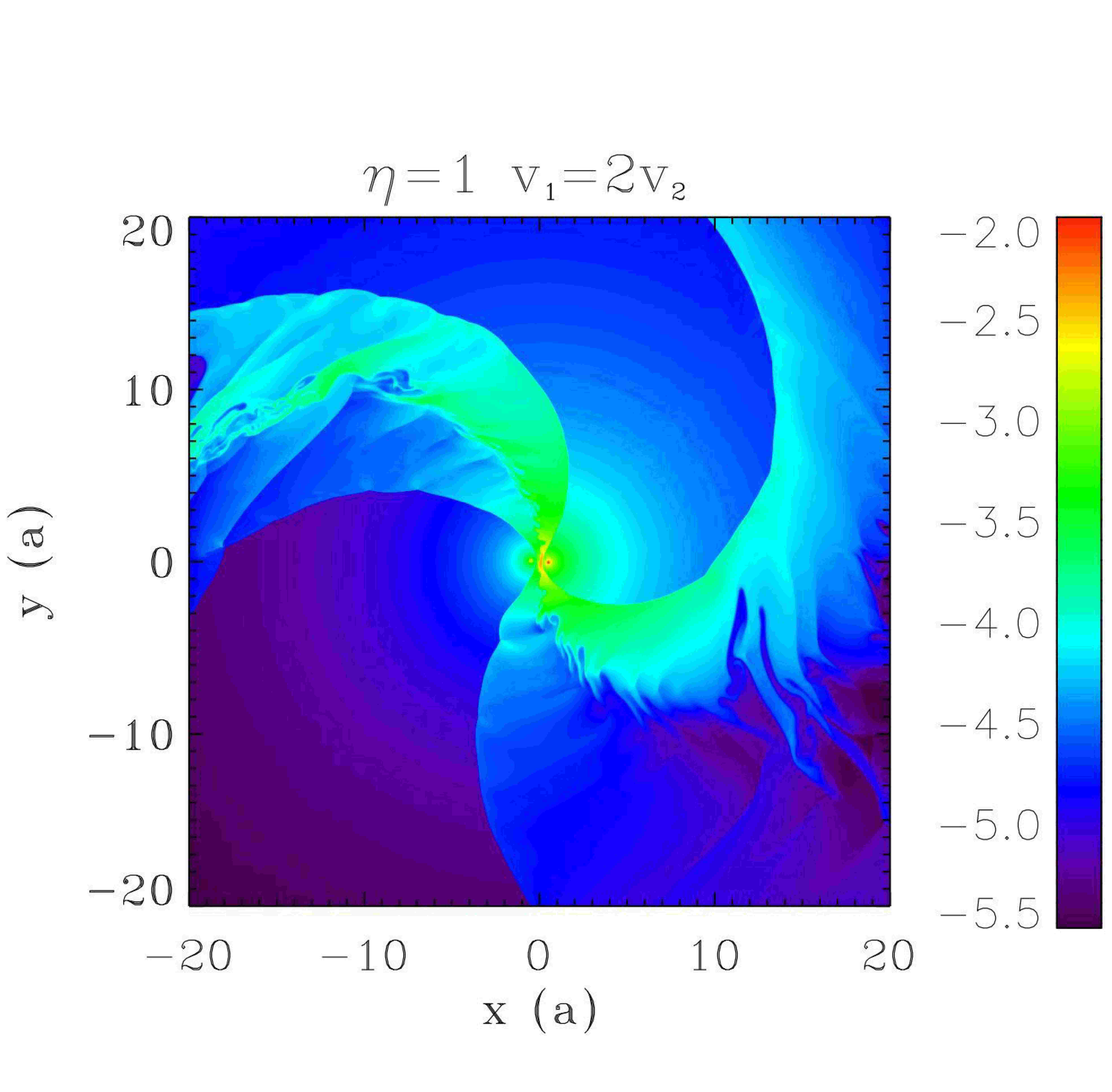}
  \includegraphics[width = .23\textwidth]{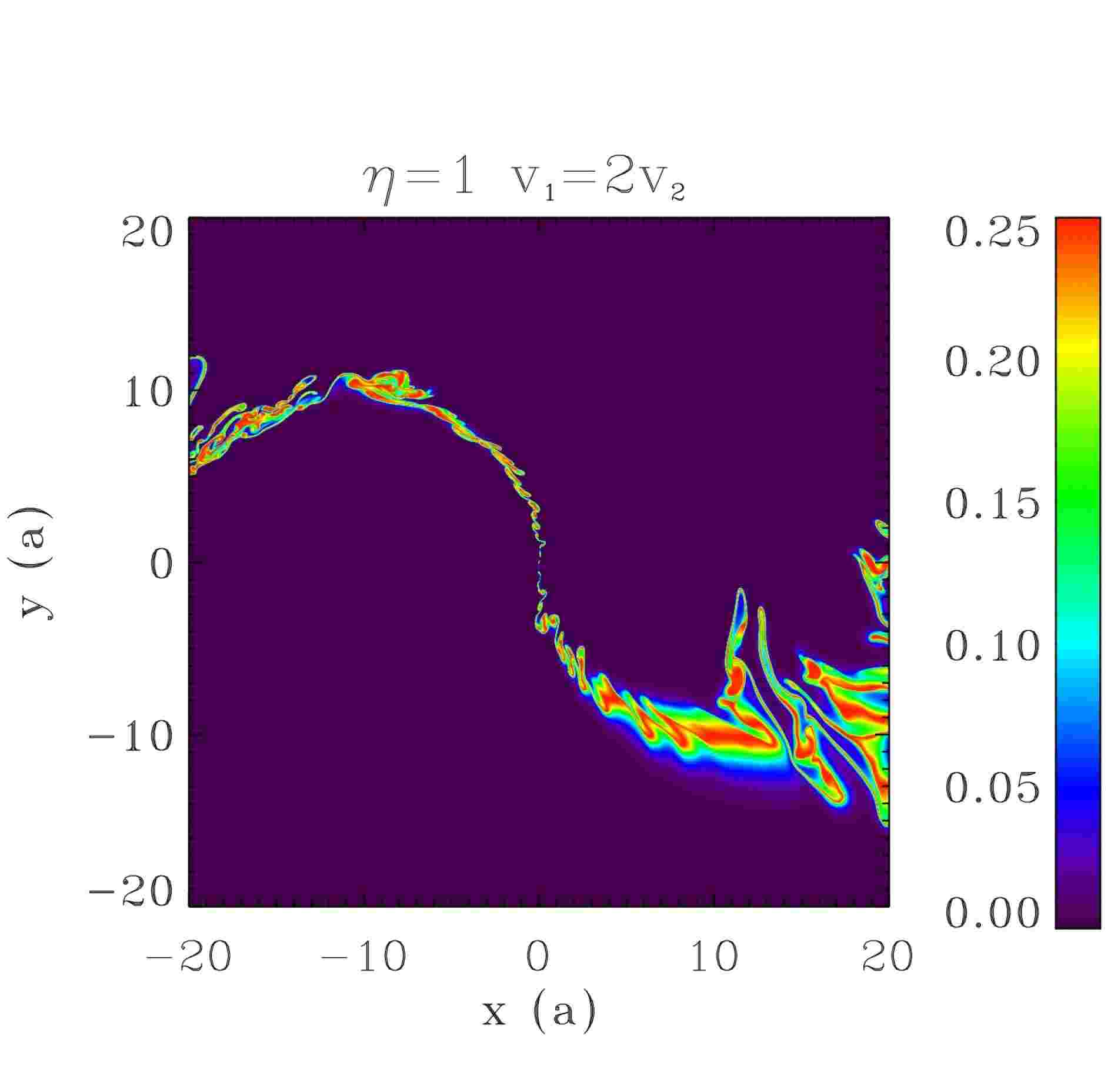} 
  \includegraphics[width = .23\textwidth]{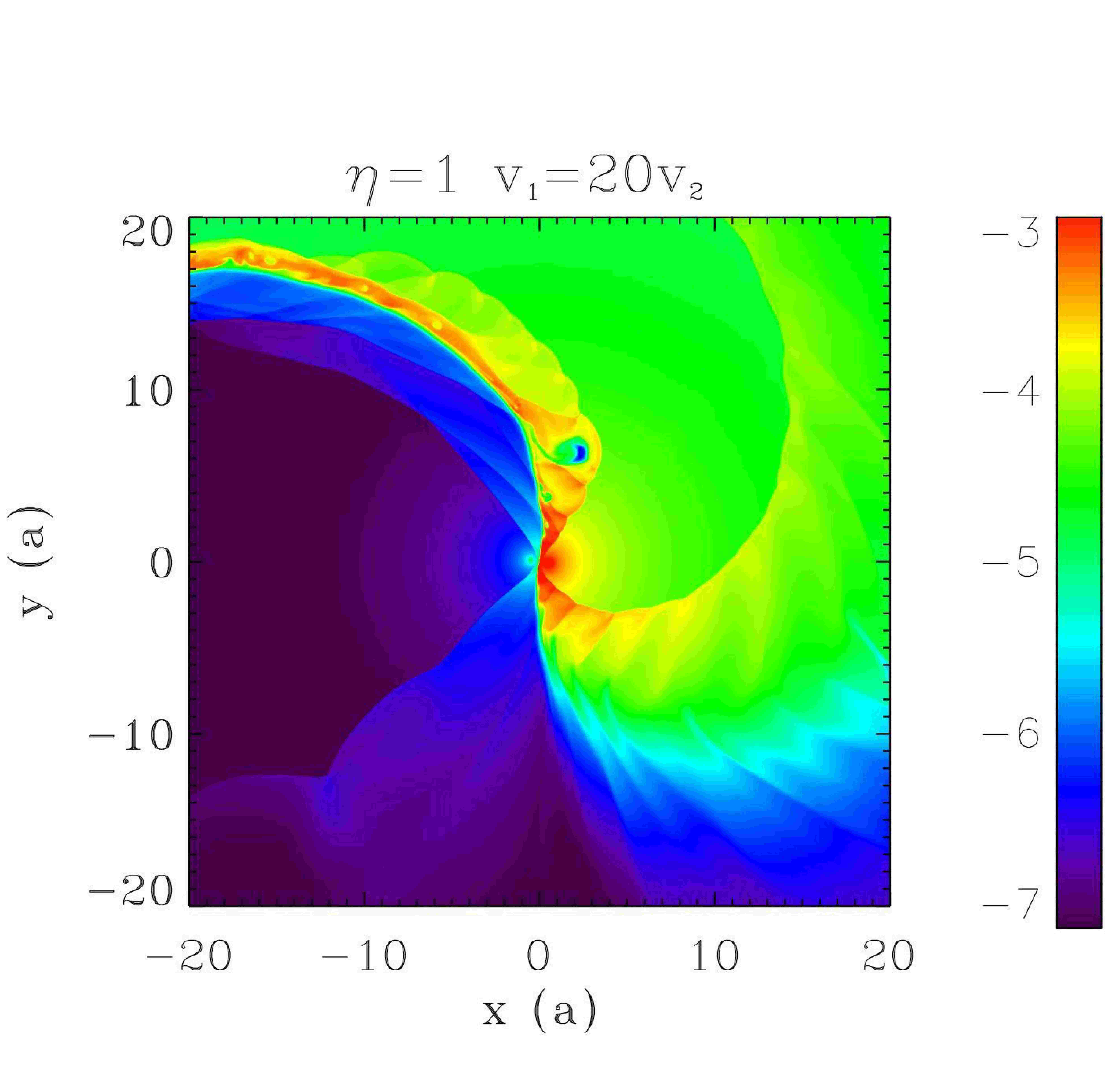}
  \includegraphics[width = .23\textwidth]{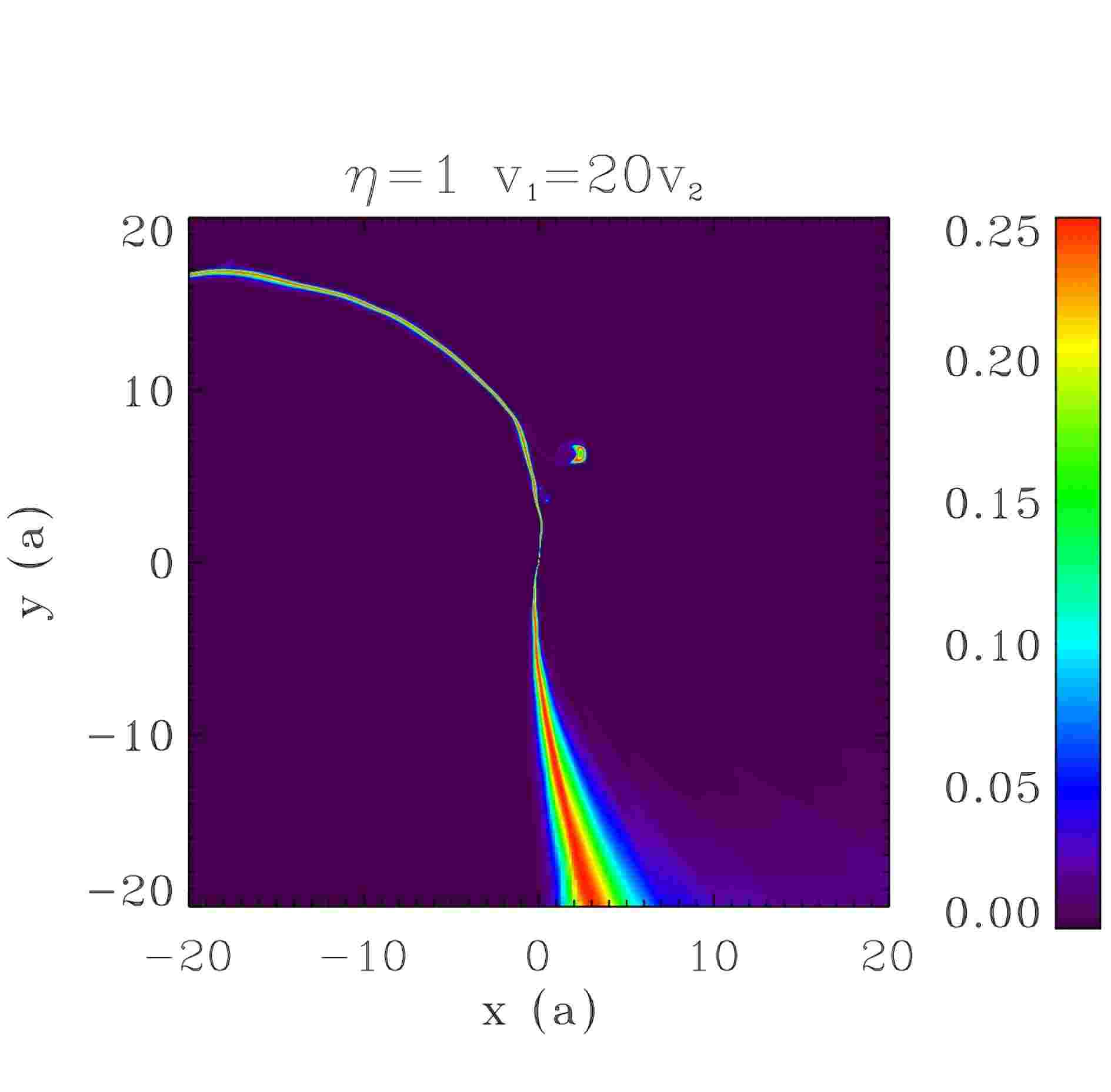}
  \includegraphics[width = .23\textwidth]{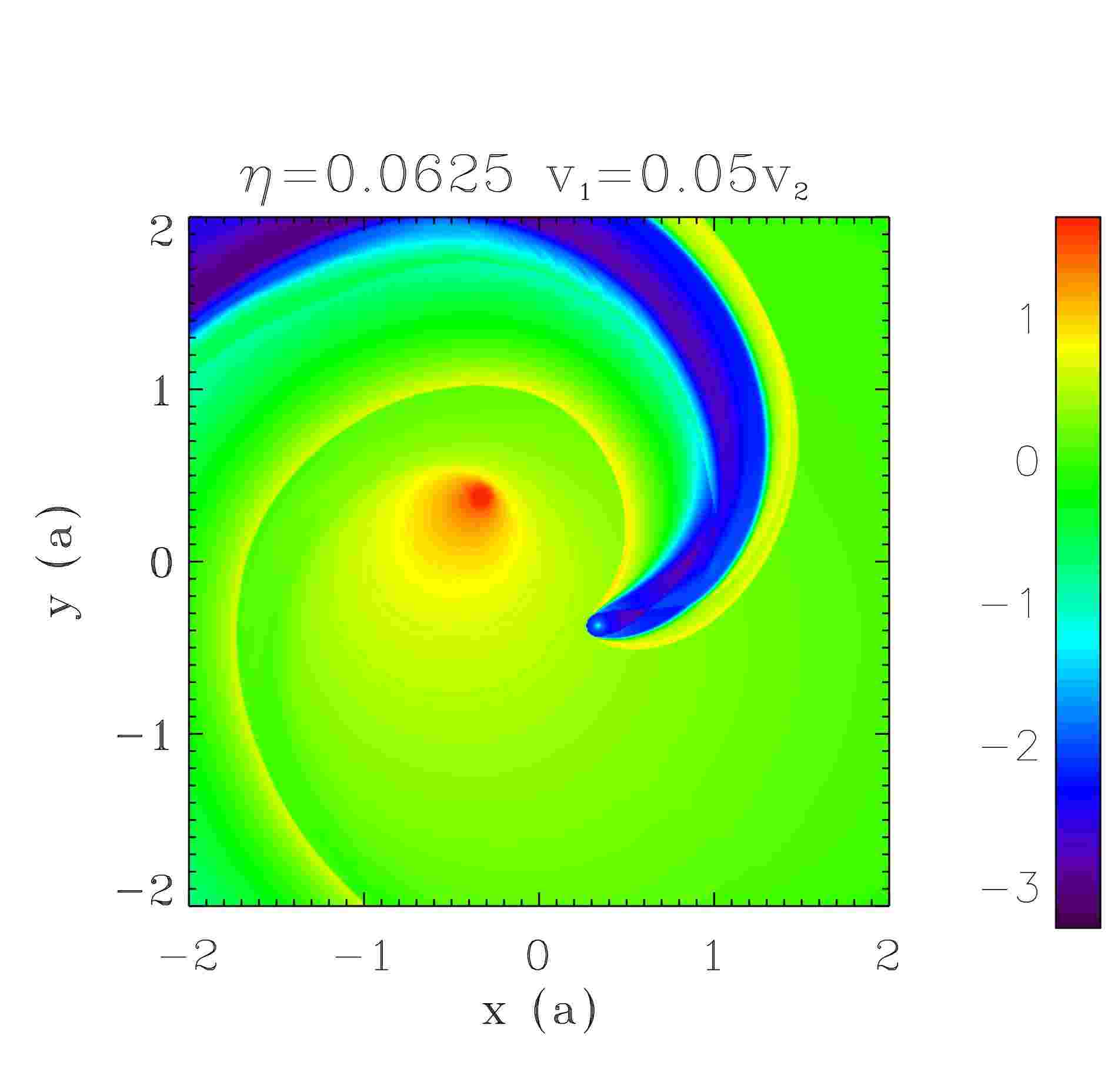}
  \includegraphics[width = .23\textwidth]{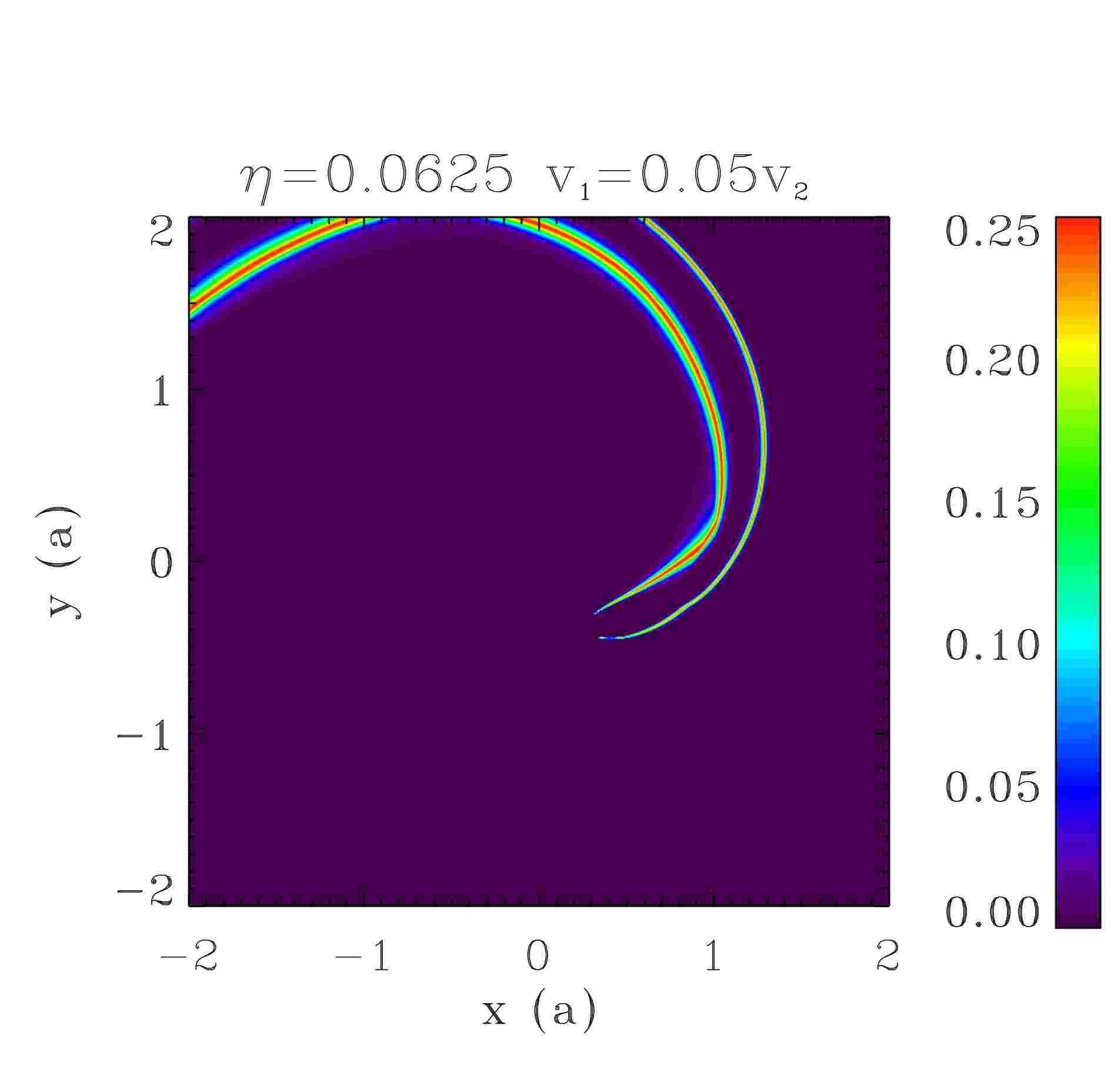}  
  \includegraphics[width = .23\textwidth]{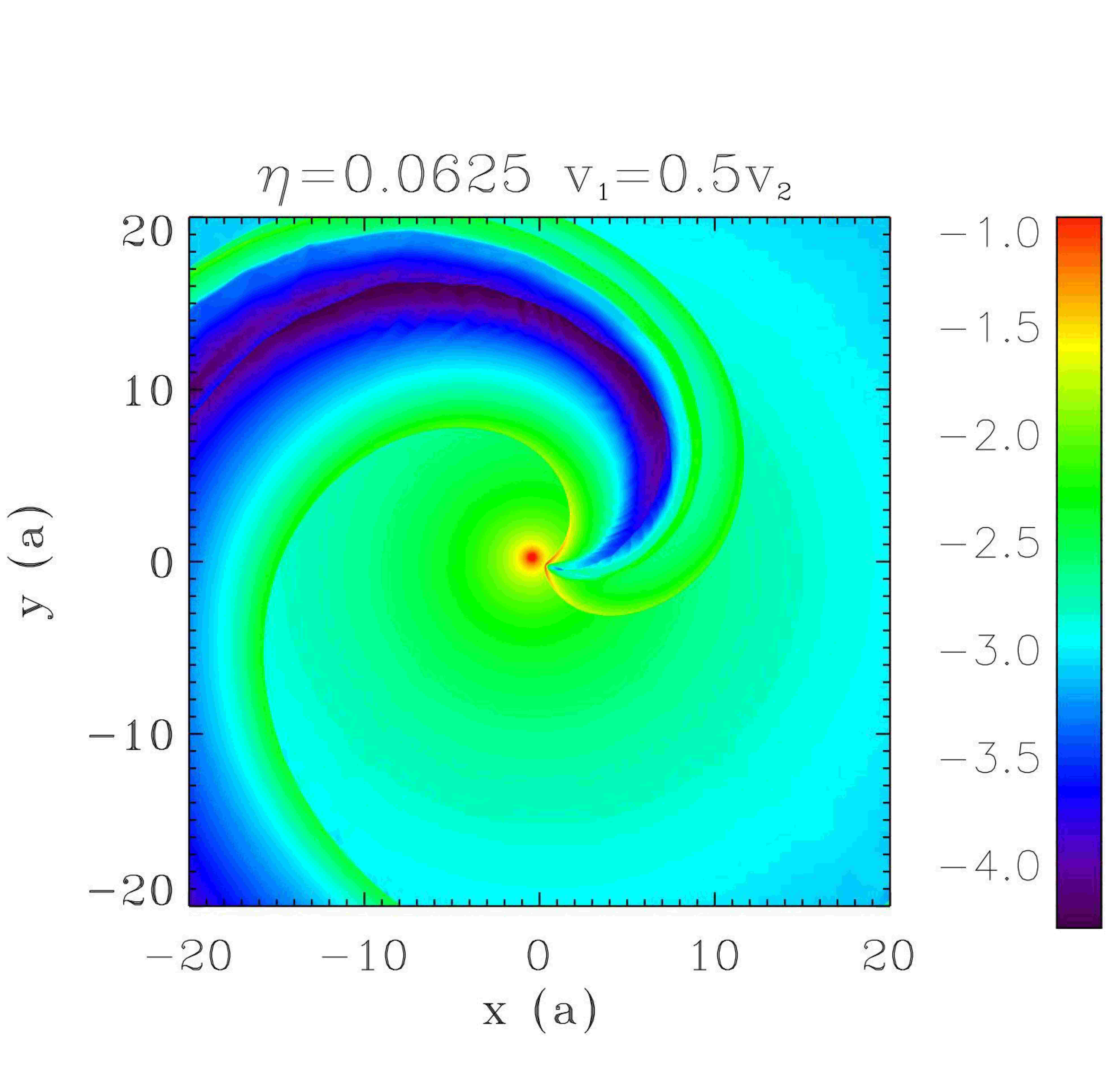}
  \includegraphics[width = .23\textwidth]{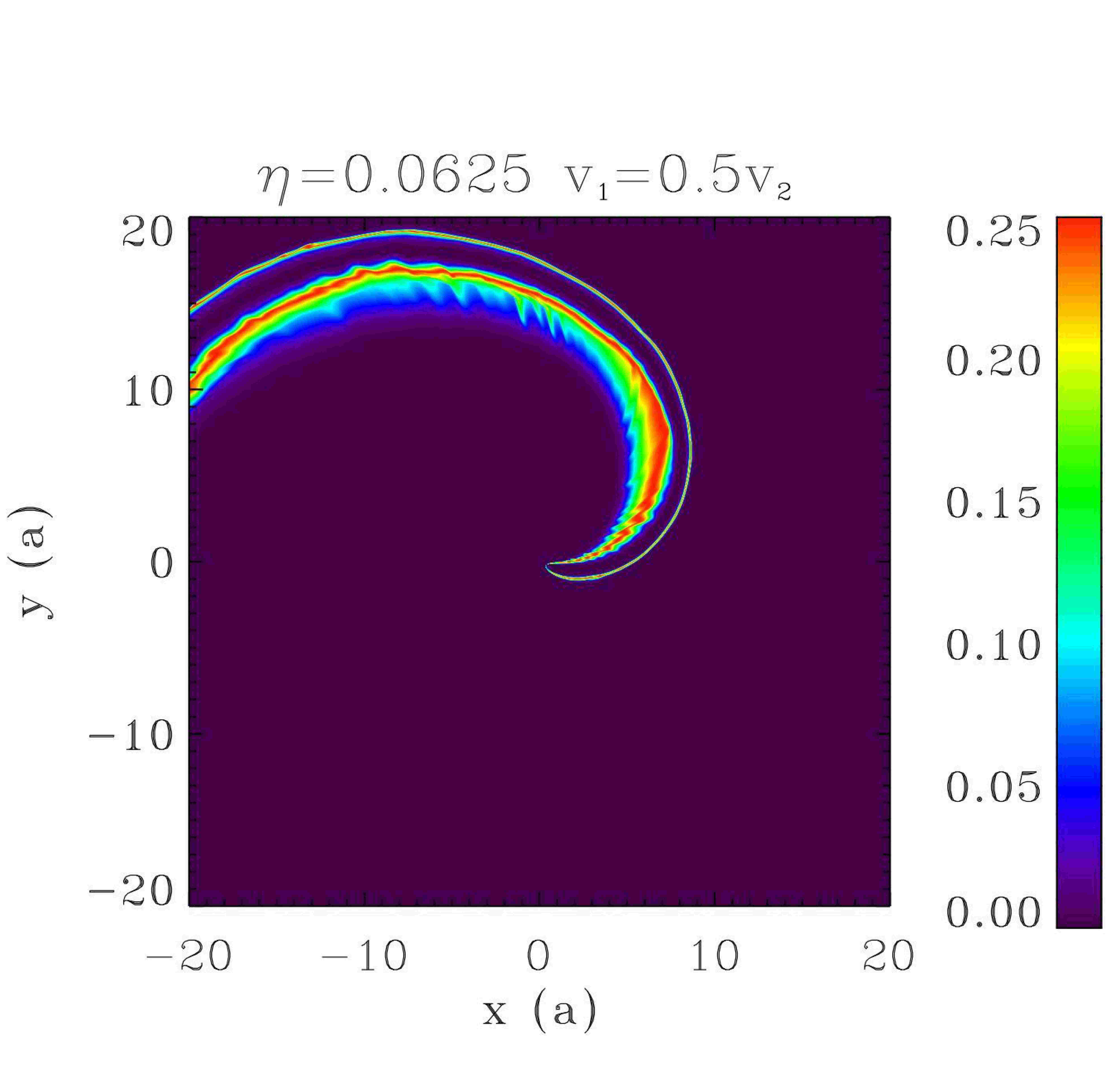}
  \includegraphics[width = .23\textwidth]{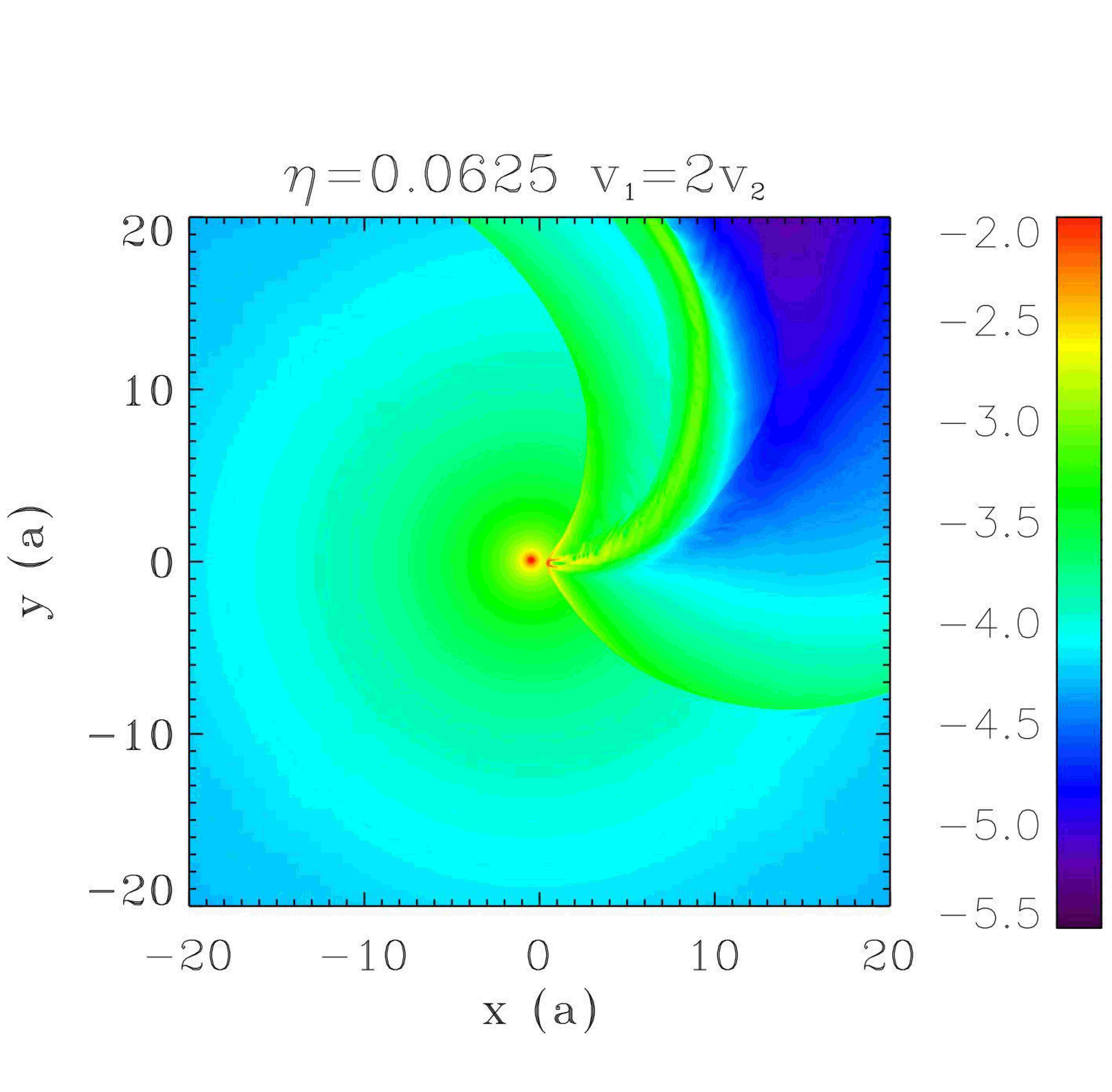}
  \includegraphics[width = .23\textwidth]{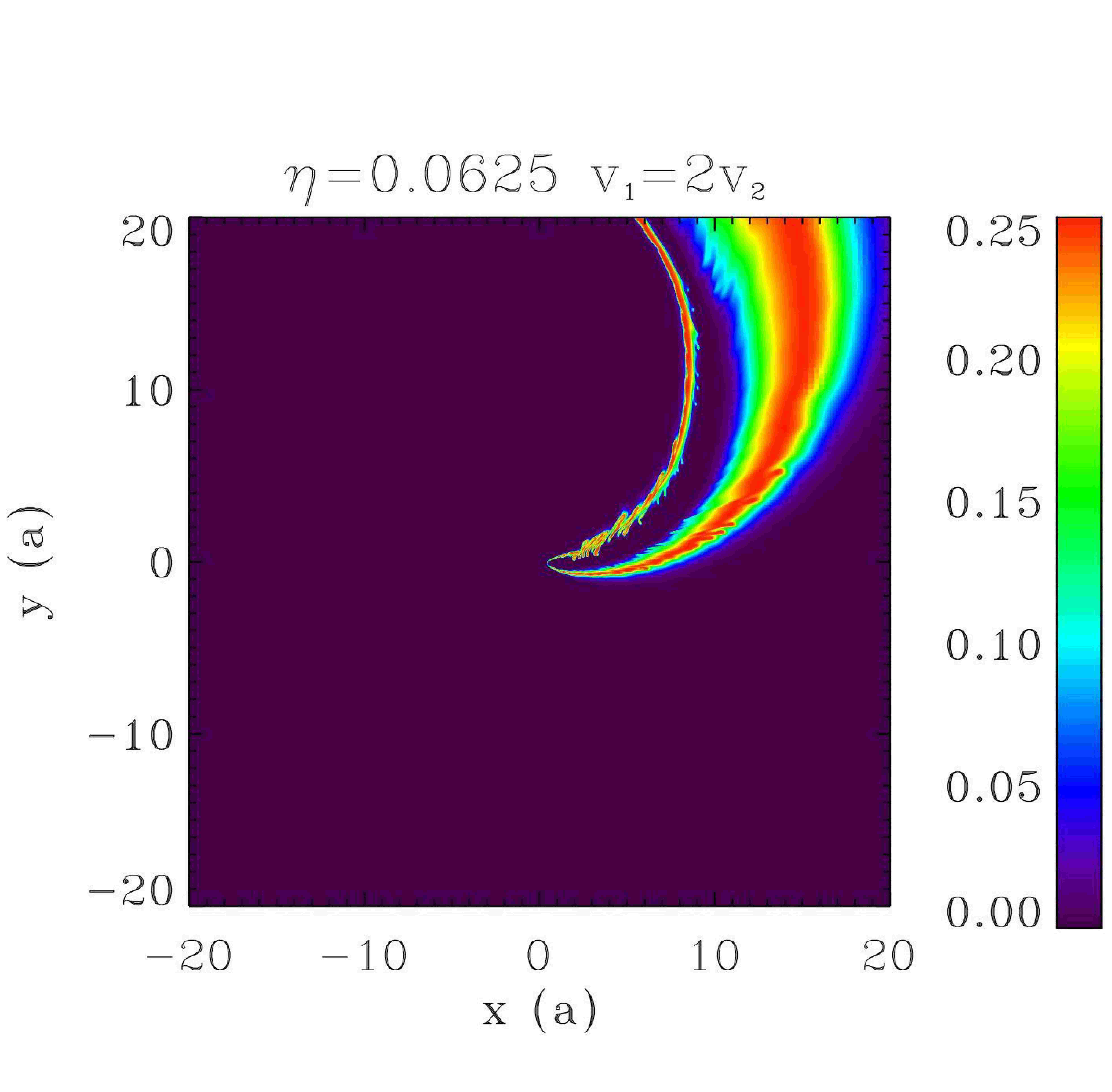}
  \includegraphics[width = .23\textwidth]{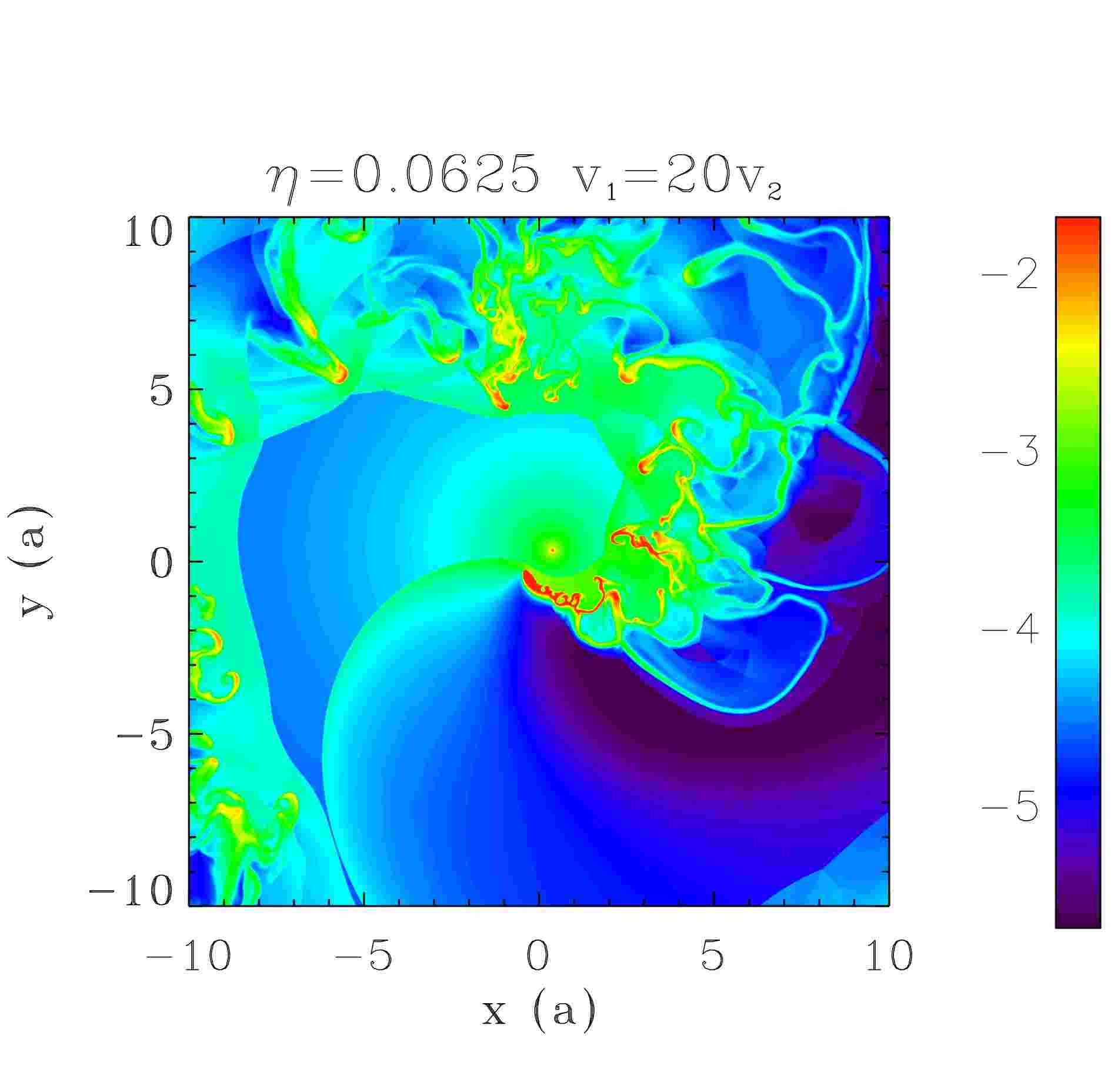}
  \includegraphics[width = .23\textwidth]{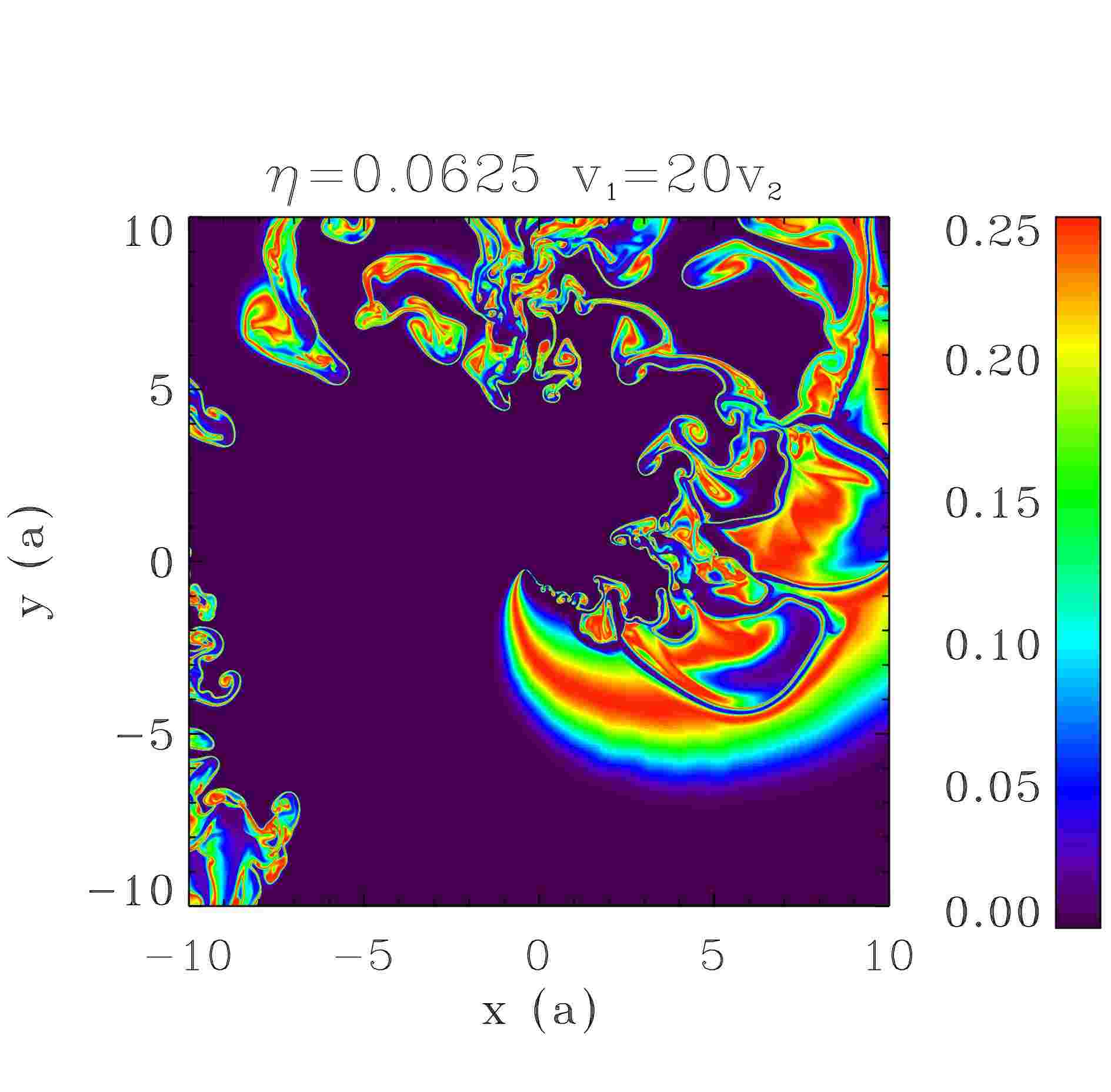}
  \caption{Small scale simulation: density (left column) and mixing (right column) for simulations with $\{\eta=1,\beta=2,20\} $ (two upper rows) and $\{\eta=0.0625,\beta=0.05,0.5,2,20\}$ (lower rows). %The density is given in g cm$^{-2}$ .
  The length scale is the binary separation $a$.} %The mixing of the winds is a dimensionless variable.}
  \label{fig:small_scale}
\end{figure}

Fig.~\ref{fig:small_scale} shows the density and mixing for simulations with $\{\eta=1,\beta=2,20\}$ and $\{\eta=0.0625,\beta=0.05,0.5,2,20\}$. These maps show the impact of rotation on arm geometry and the development of the instabilities. 

The leading and trailing arms become markedly different when the velocity difference increases, even when $\eta$=1. The shocked zone preceded by the unshocked wind with the higher velocity and lower density is larger than the zone preceded by the lower velocity, higher density wind. The latter shocked zone is compressed by the high velocity wind into a high density region \citep{2011A&A...527A...3V}. We verified that, as expected if this explanation is correct, for $\beta>1$ the compressed arm is the trailing arm while for $\beta<1$ the compressed arm is the leading arm (see Fig.~\ref{fig:small_scale}). For $\beta=20$, compression results in a rim of puffed up matter where the density increases by two orders of magnitude with respect to the simulation with $\beta=1$. The differentiation of both arms is independent of the instabilities in the winds but plays a role in their development.  

The KHI starts similarly in both arms close to the binary major axis as the velocity difference and density jump across the contact discontinuity are the same in both arms.  The symmetry between both arms is broken as the flow moves outwards. The compression of the shocked zone in the narrower arm results in a thin mixed zone with small scale structures, whereas the eddies are stretched out in the wider arm. Mixing covers a larger area in the wider zone. This is not just a geometrical effect. We show in Appendix A that when media have different densities ($\alpha\ne 0$, see Eq. A.10) mixing by the KHI occurs preferentially in the least dense medium. Both velocity and density profiles thus play a role in the development of the KHI in colliding wind binaries. They both impact the large scale outcome of the spiral structure as will be shown in \S \ref{large_scale}. 

We checked that the mixing is physical and not numerical in the narrow arm. If numerical, the physical size of the mixed zone increases with decreasing resolution. If mixing follows from instabilities, the physical size of the mixed zone is constant with resolution. We measure the width of the narrow mixed zone across the contact discontinuity at the upper edge of the simulation box for the case $\{\eta=0.0625,\beta=2\}$. The limit of the zone is determined by $s_1\times s_2=0.01$. For the simulations with the highest resolution (7 levels of refinement), we measure a width of 0.87$a$ while we measure a width of 0.93$a$ in a simulation with 6 levels of refinement. We conclude that numerical diffusion has a limited impact and that the mixing in our high resolution simulations is robust.

\section{Formation of a spiral structure}\label{large_scale}

\begin{figure}
  \centering
  \includegraphics[width = .23\textwidth]{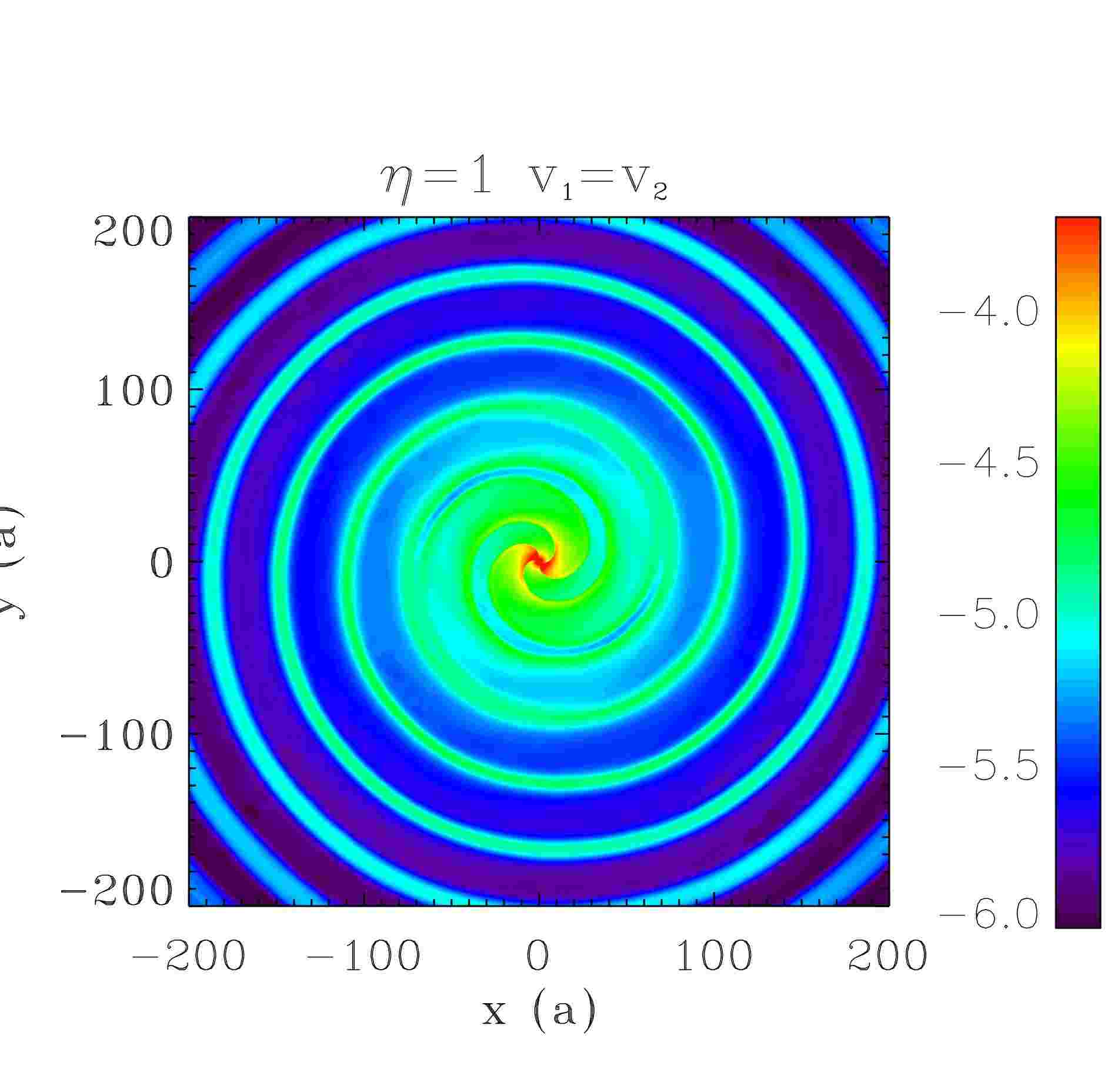}
  \includegraphics[width = .23\textwidth]{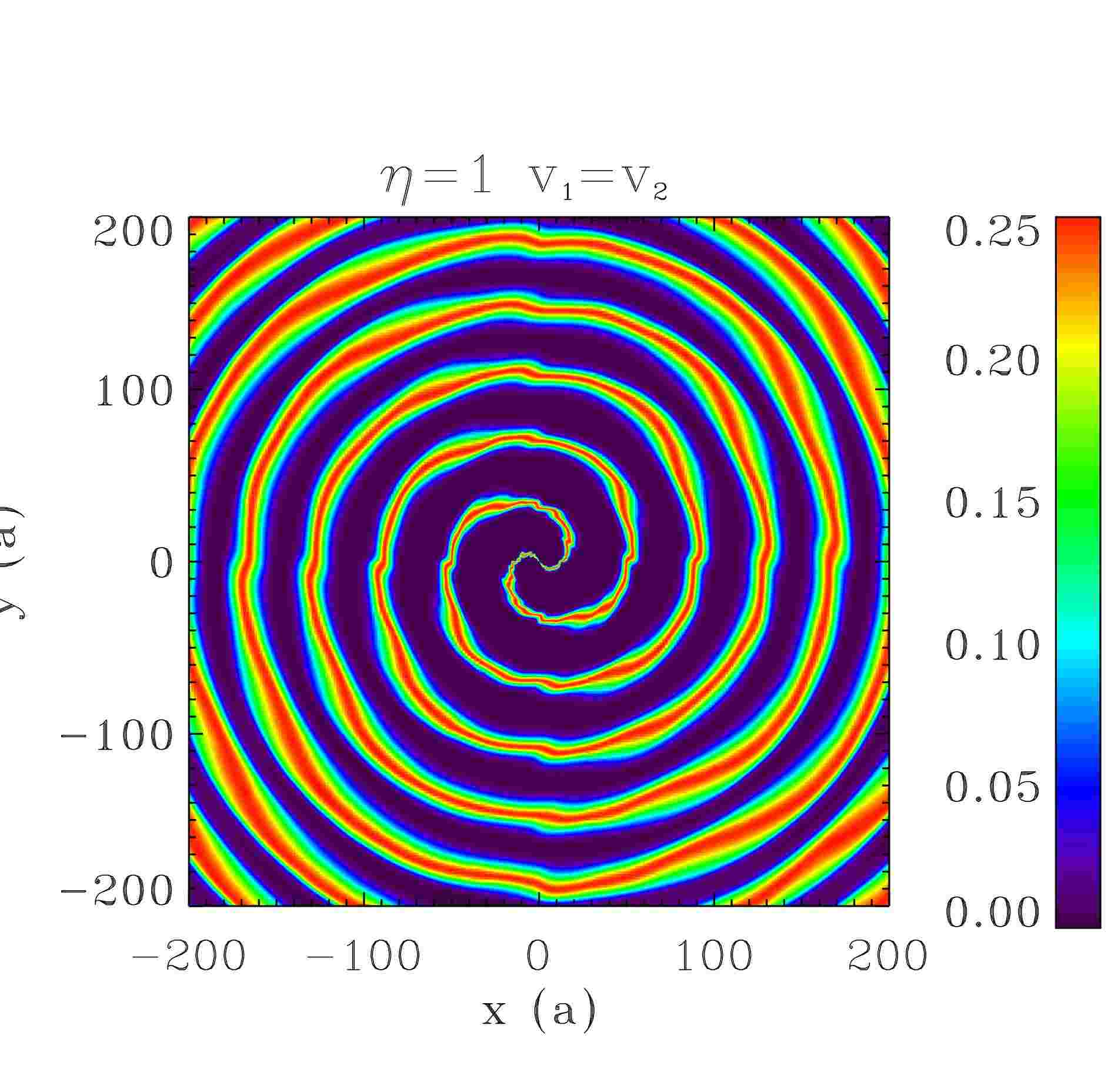}
  \includegraphics[width = .23\textwidth]{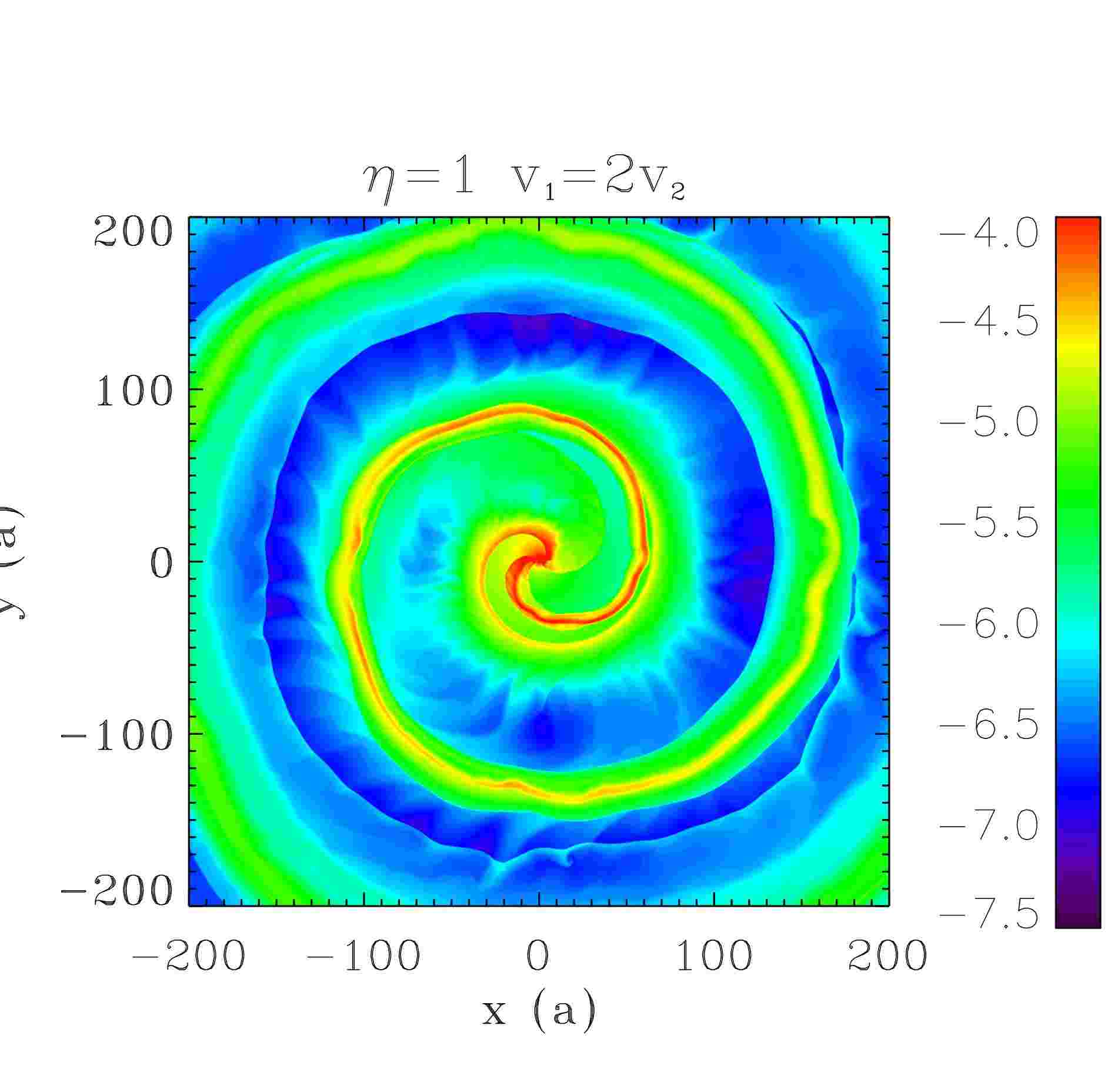}
  \includegraphics[width = .23\textwidth]{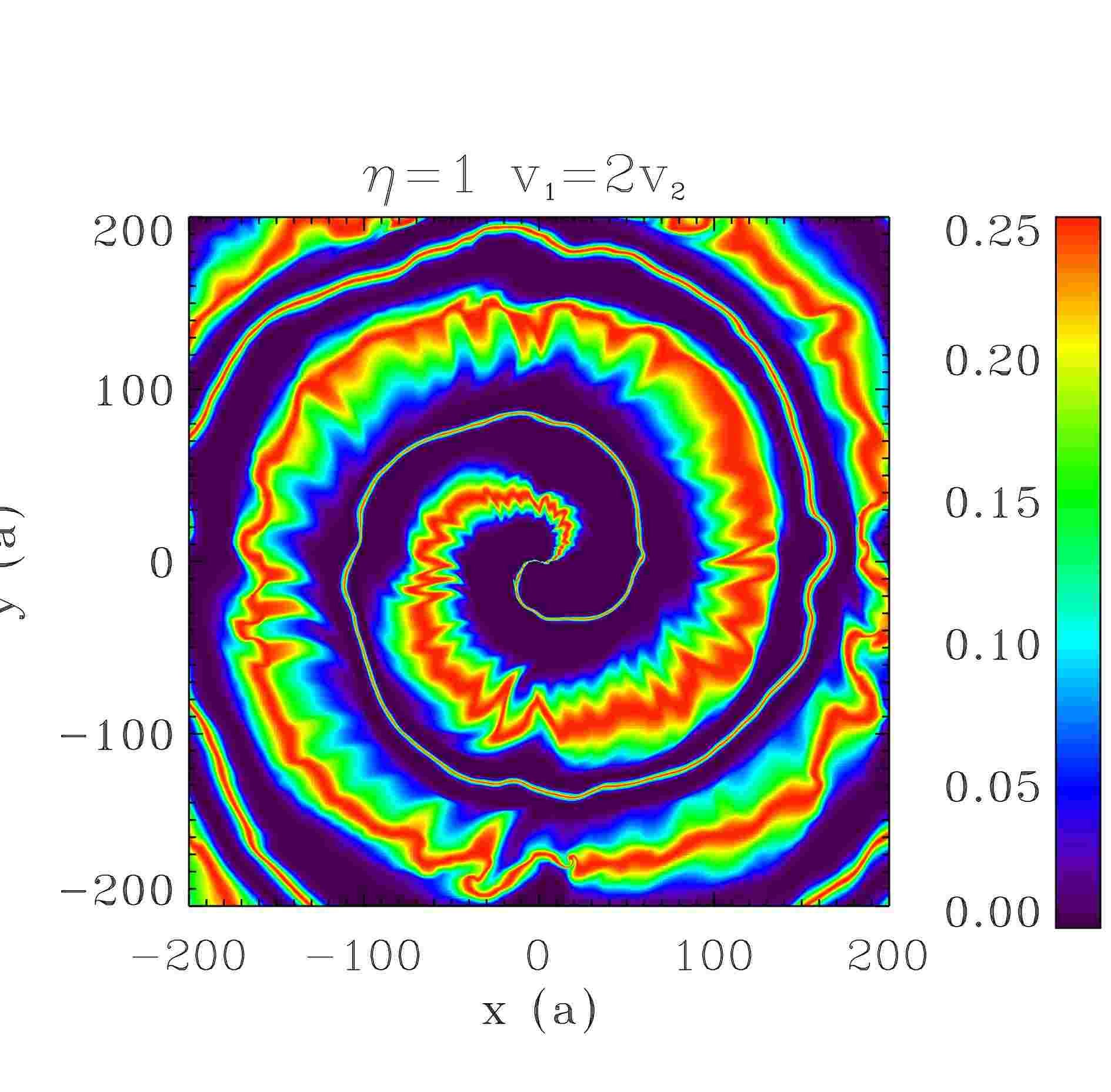}
  \includegraphics[width = .23\textwidth]{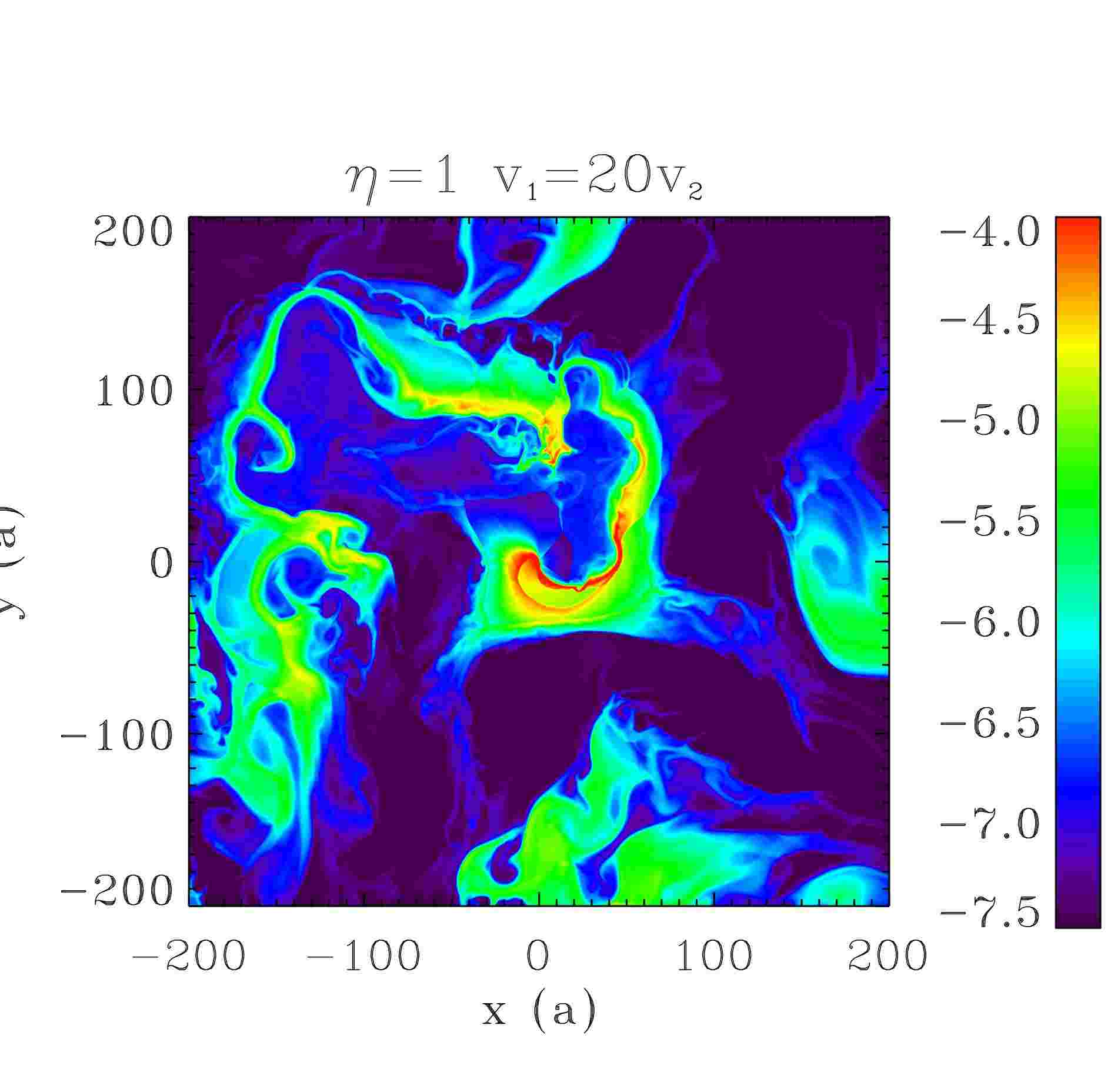}
  \includegraphics[width = .23\textwidth]{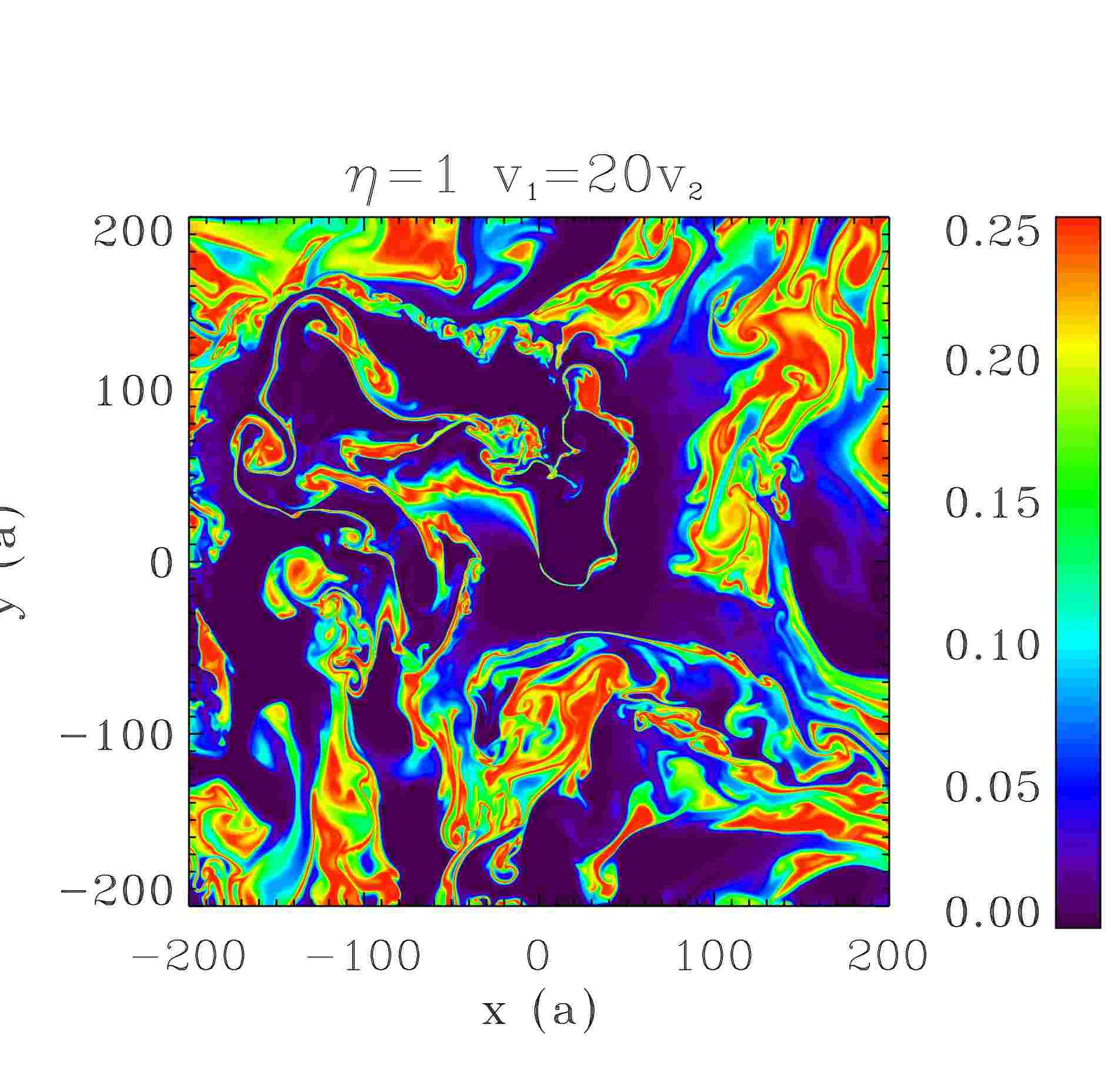}
  \caption{Large scale simulation : density (left column) and mixing (right column) for simulations with $\{\eta=1,\beta=1,2,20\} $ (from top to bottom) in the large boxes. The length scale is the binary separation $a$. The mixing of the winds is a dimensionless variable.}
  \label{fig:large_scale_1}
\end{figure}

\begin{figure}
  \centering
\includegraphics[width = .23\textwidth]{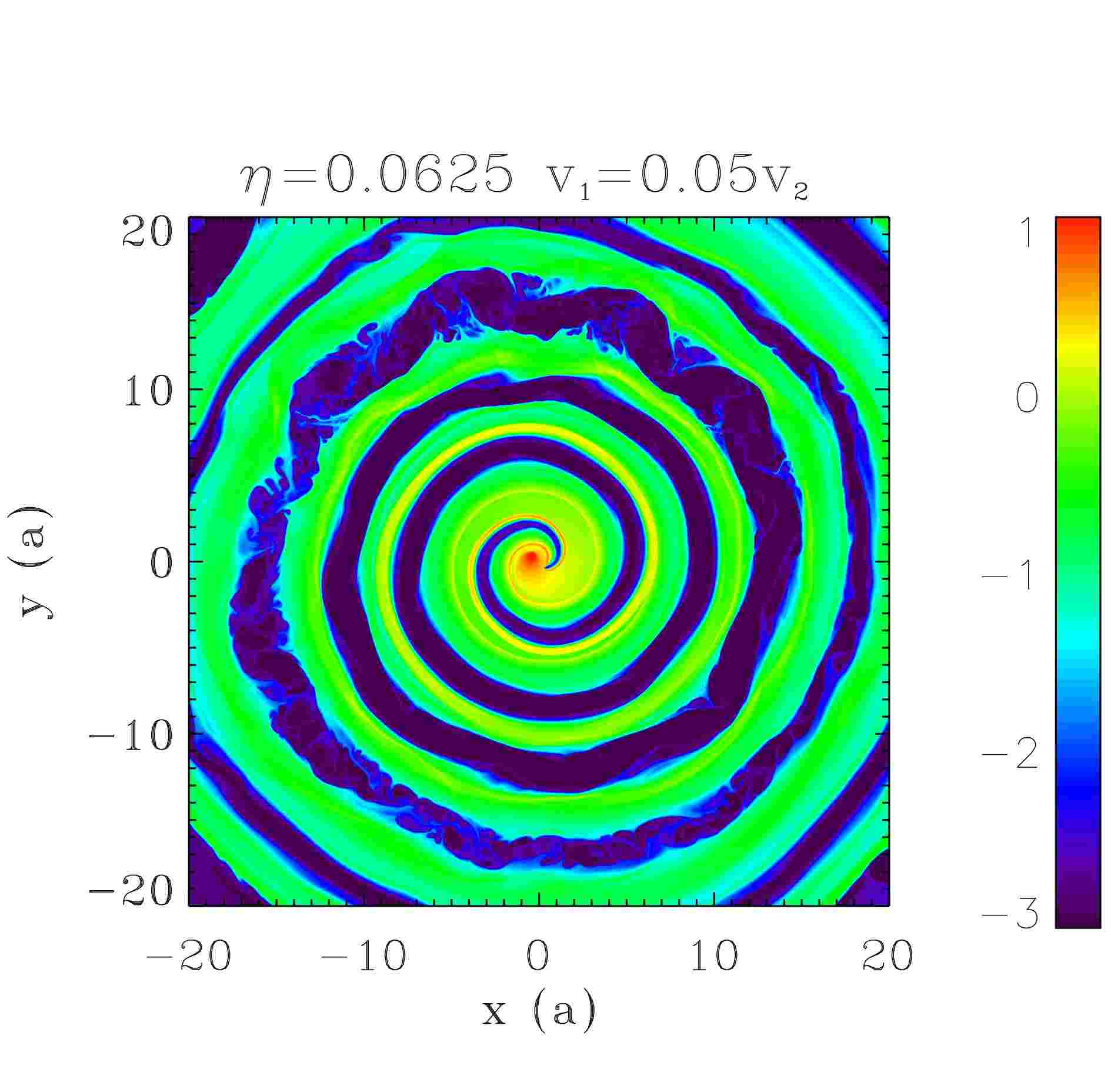}  
  \includegraphics[width = .23\textwidth]{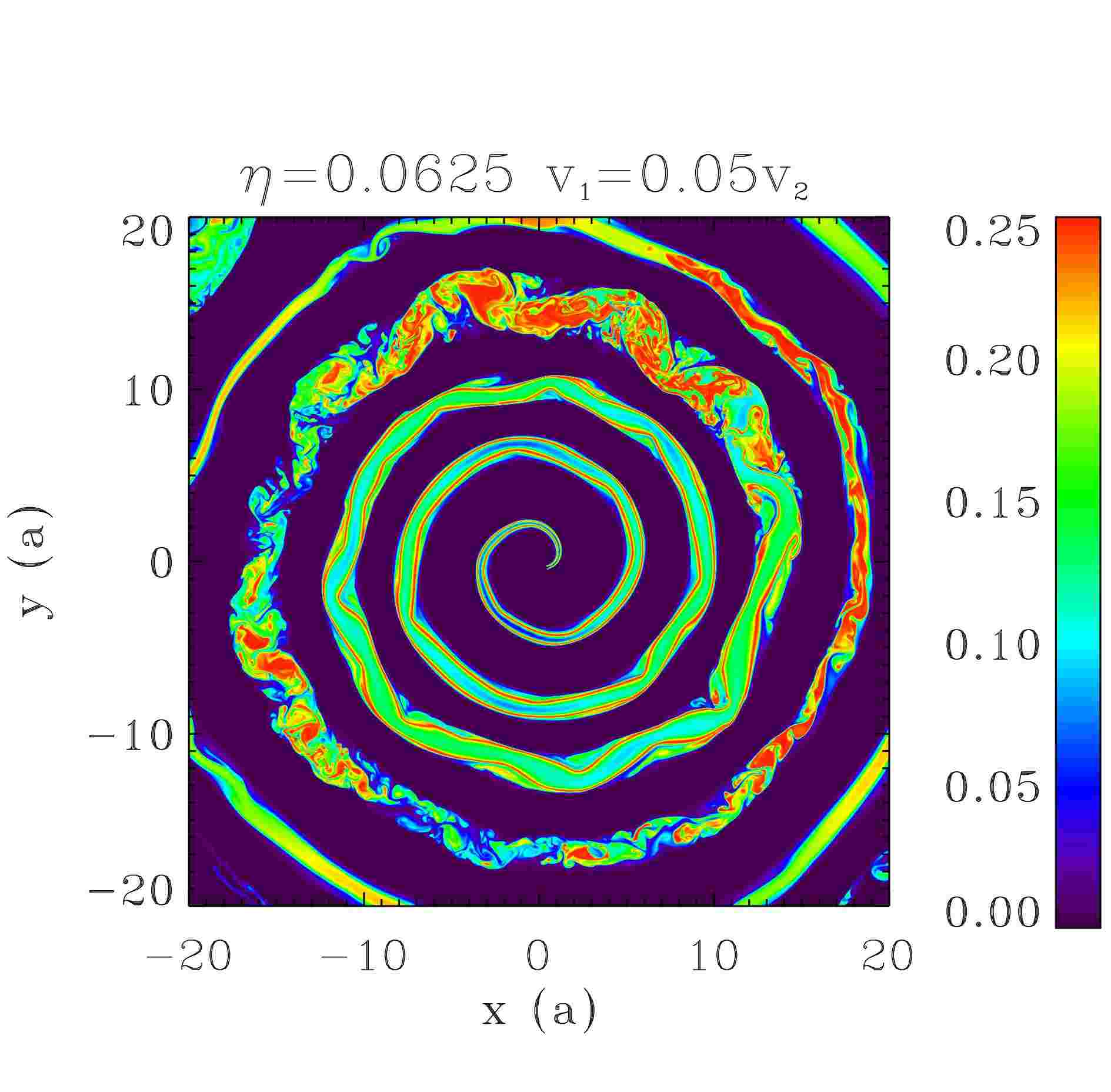}
  \includegraphics[width = .23\textwidth]{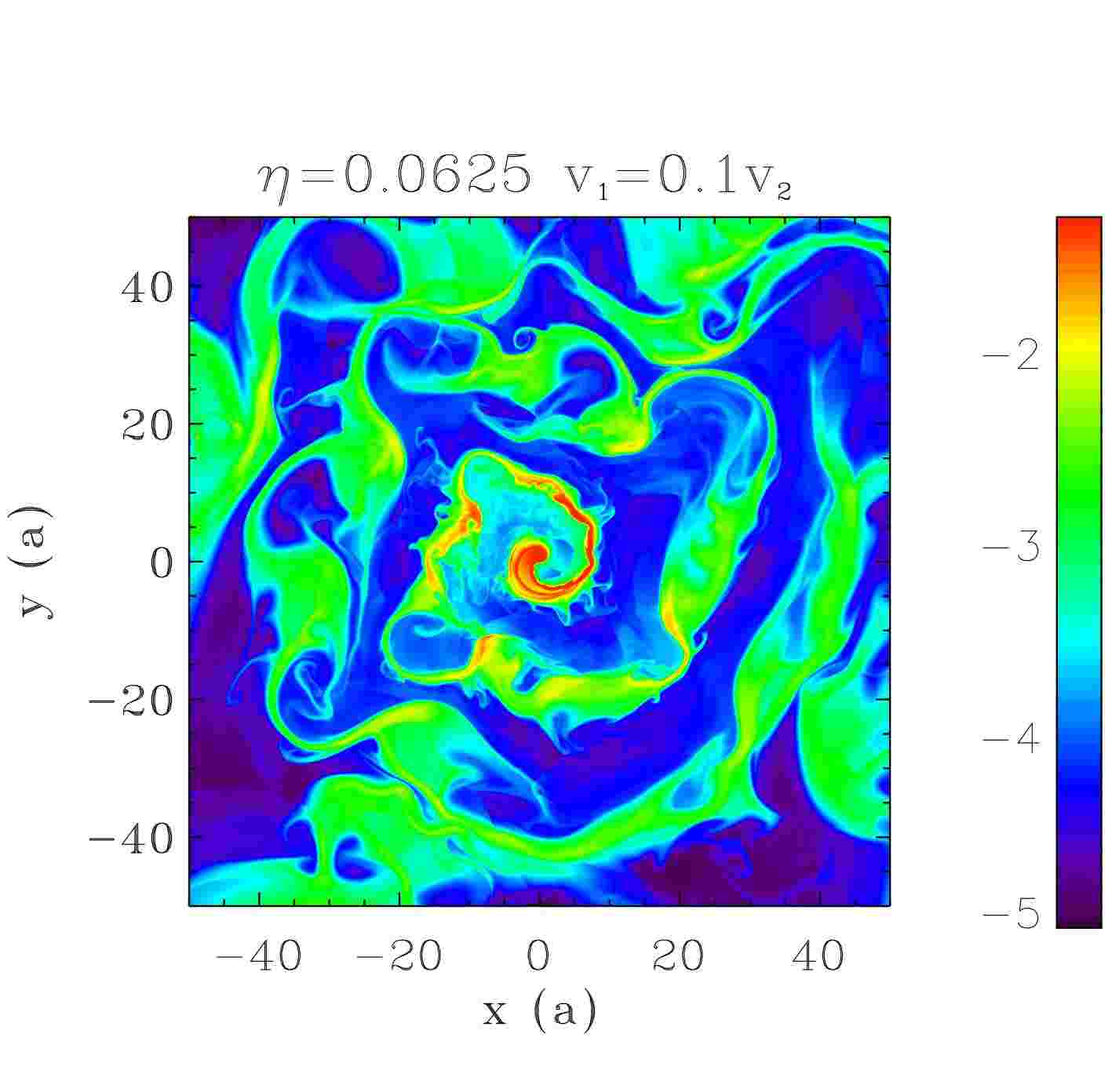}
  \includegraphics[width = .23\textwidth]{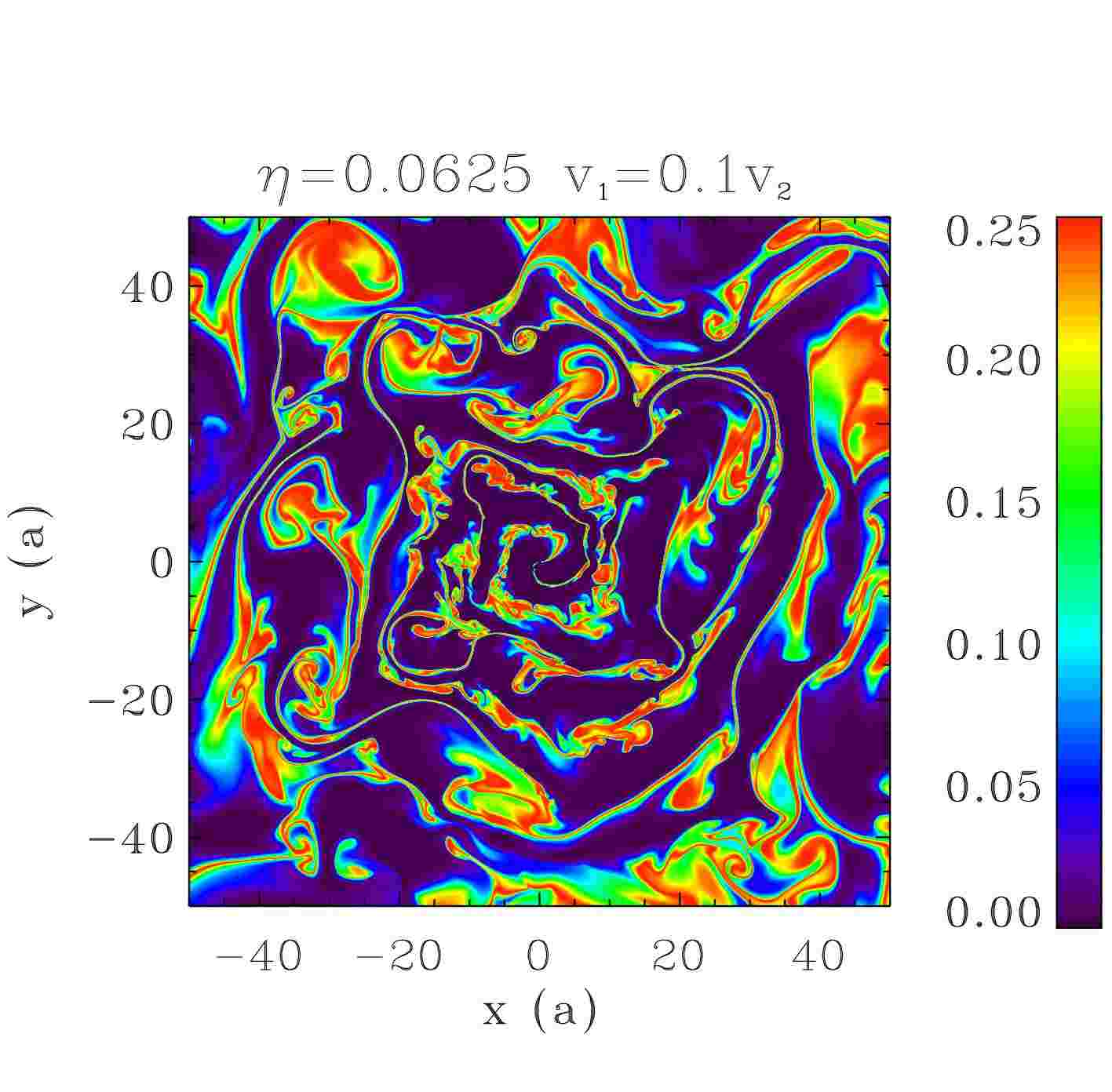}
  \includegraphics[width = .23\textwidth]{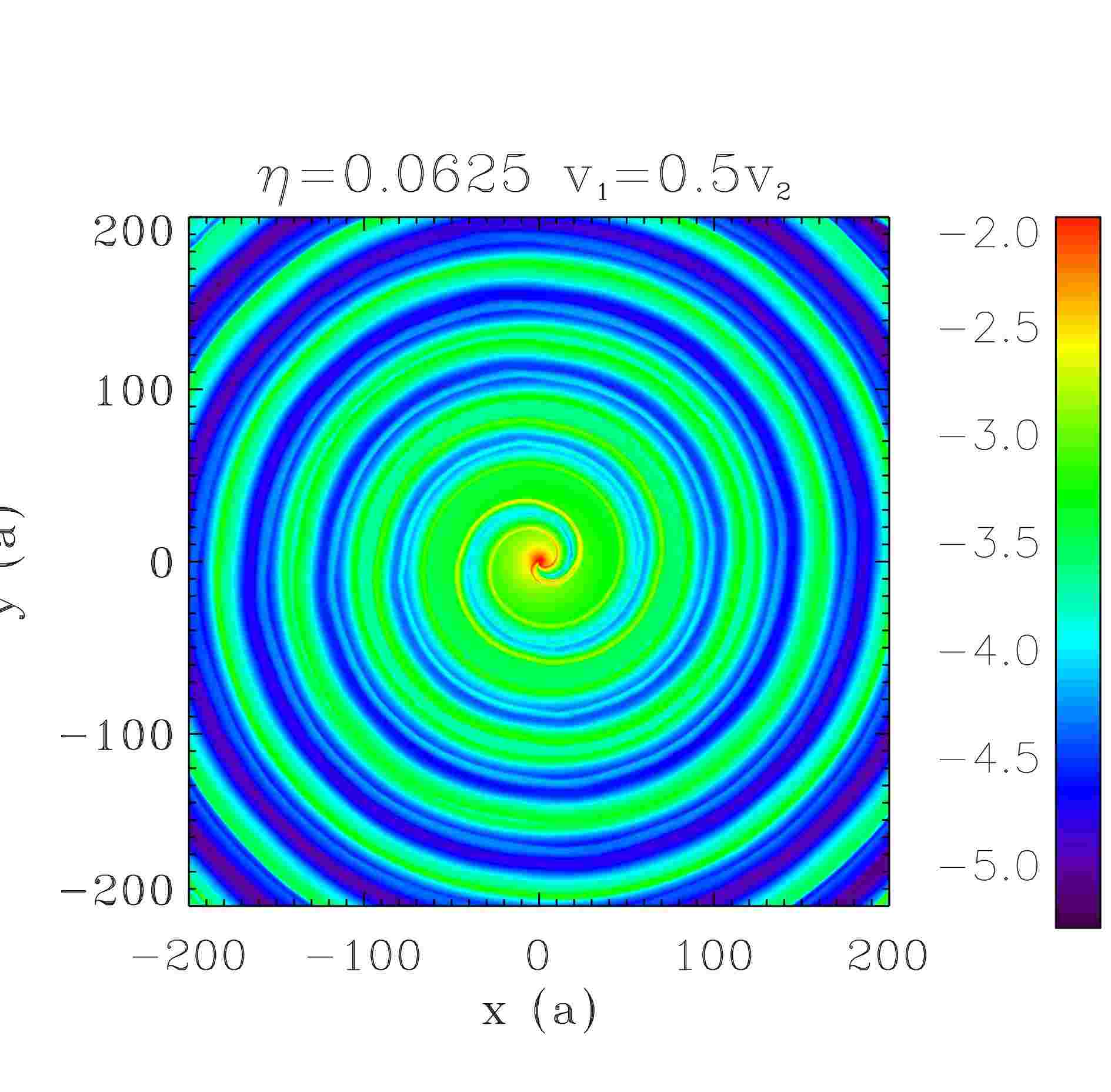}
  \includegraphics[width = .23\textwidth]{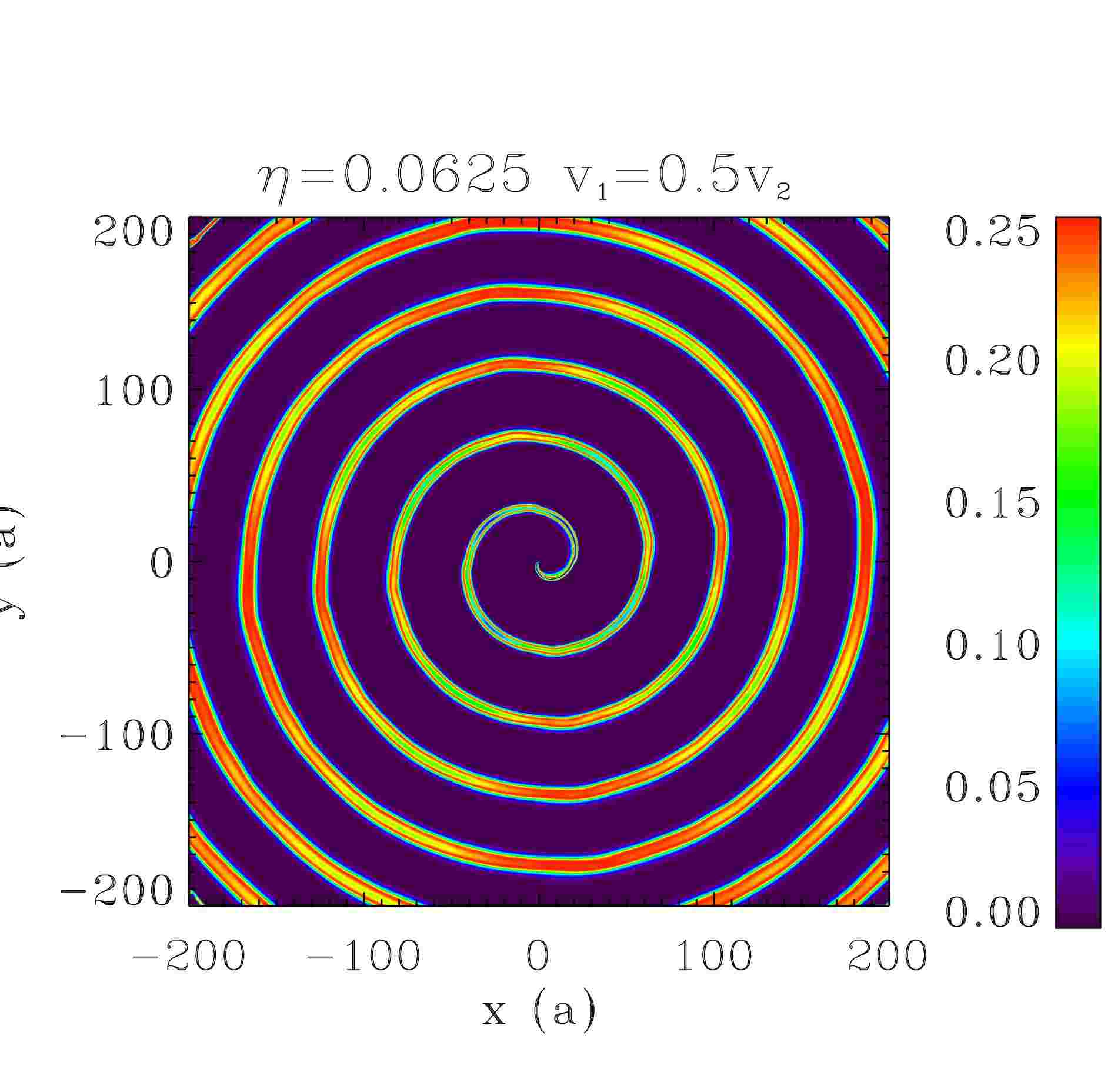}
  \includegraphics[width = .23\textwidth]{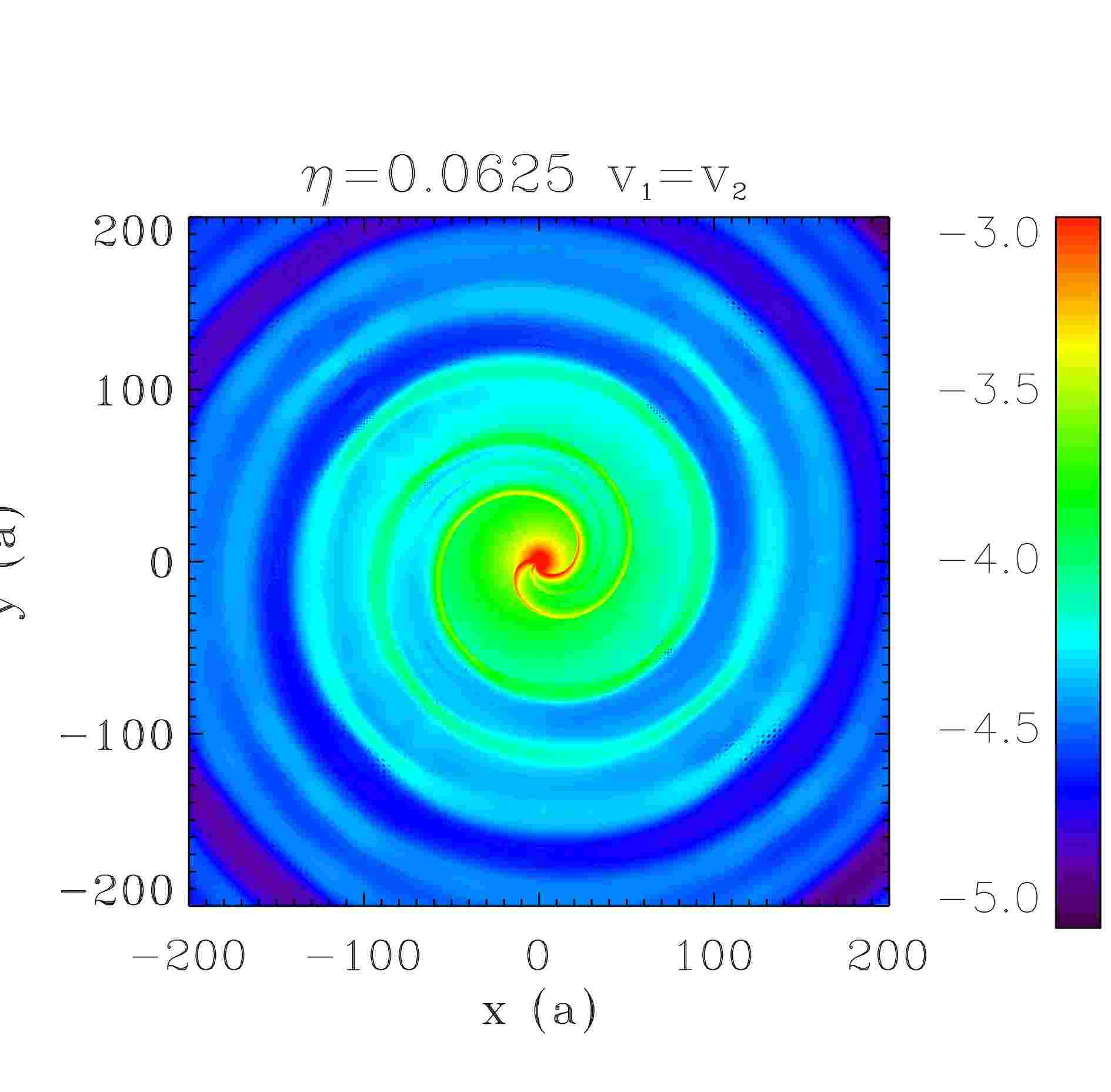}
  \includegraphics[width = .23\textwidth]{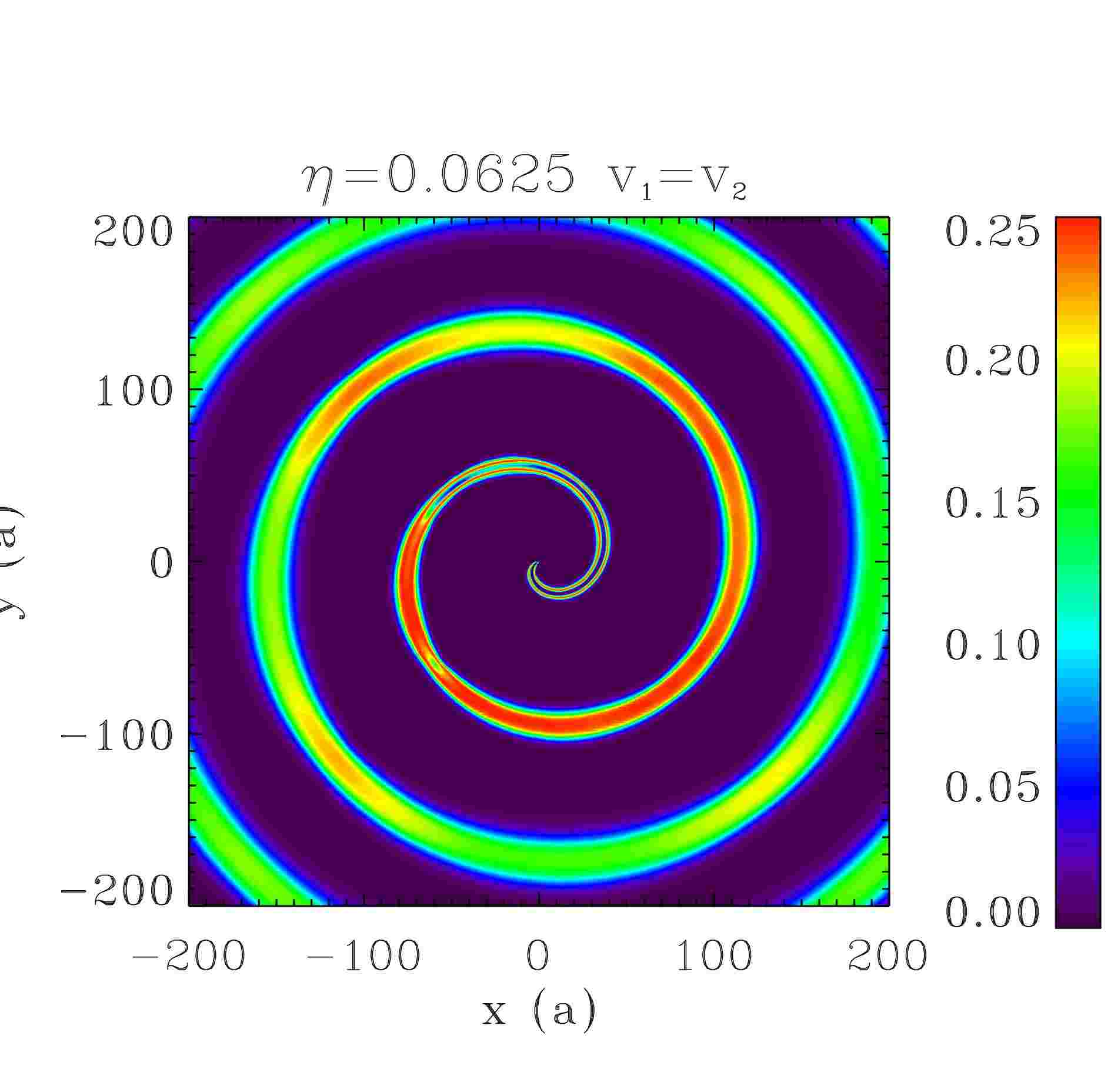}
  \includegraphics[width = .23\textwidth]{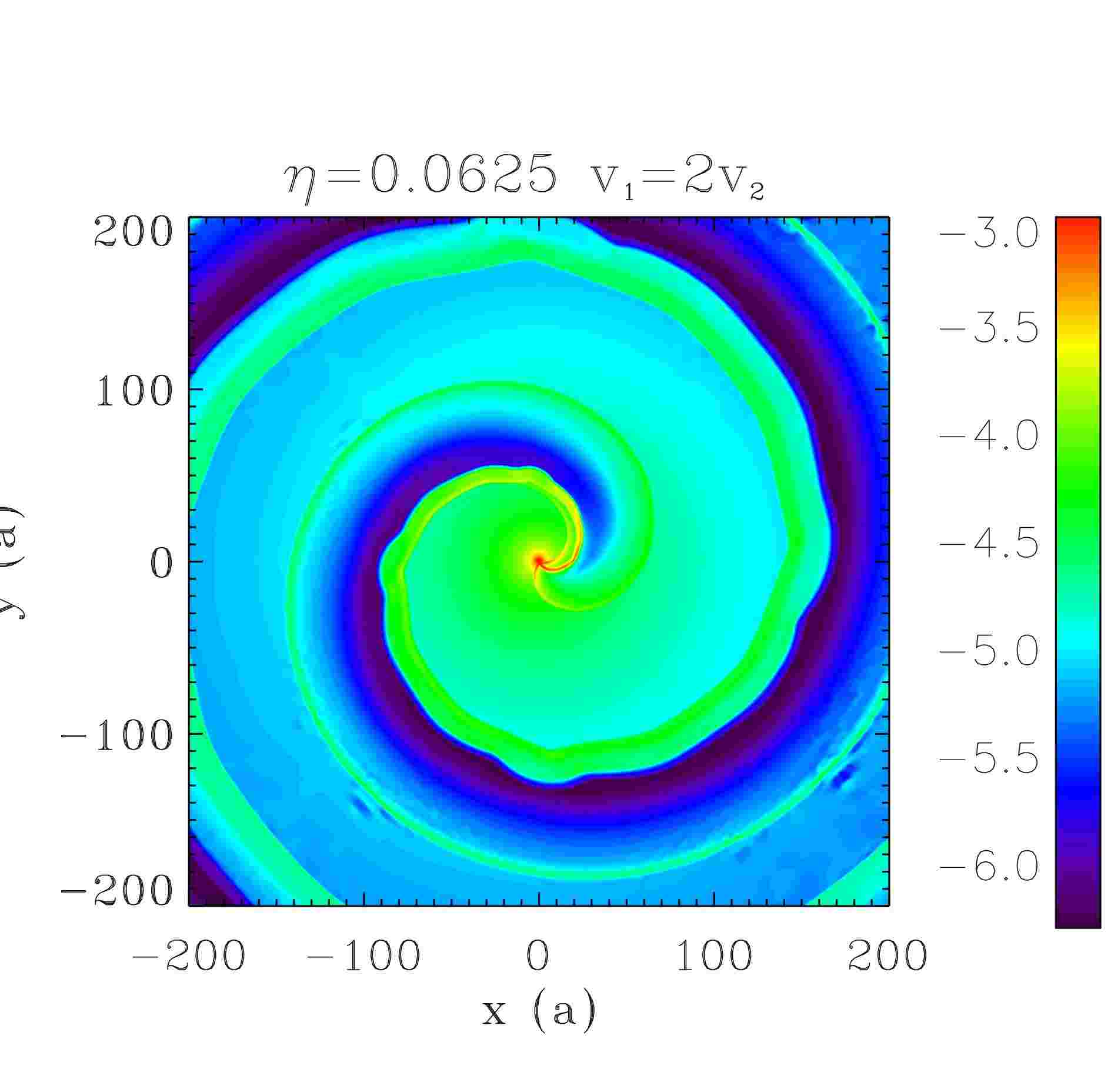}
  \includegraphics[width = .23\textwidth]{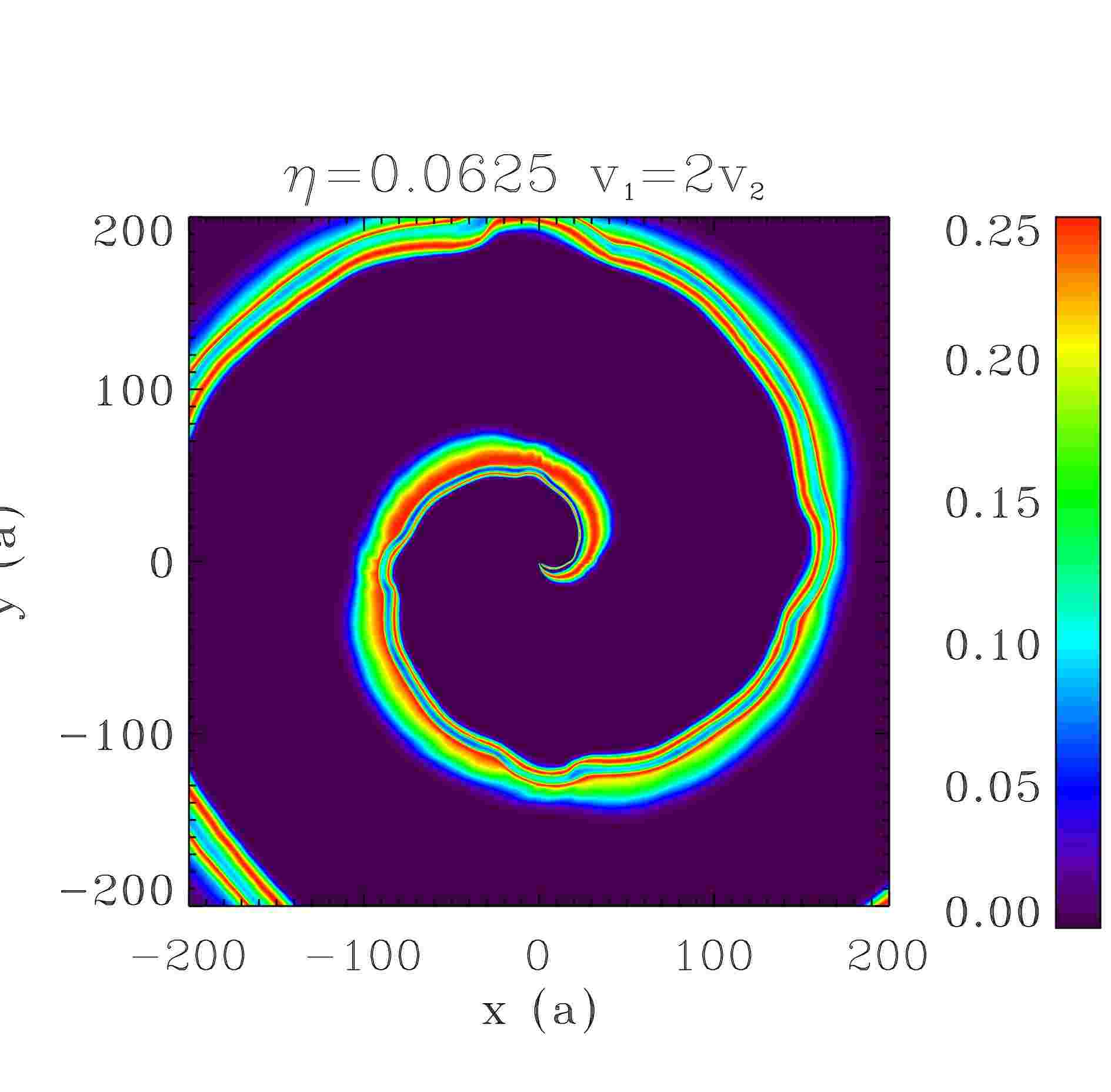}
  \includegraphics[width = .23\textwidth]{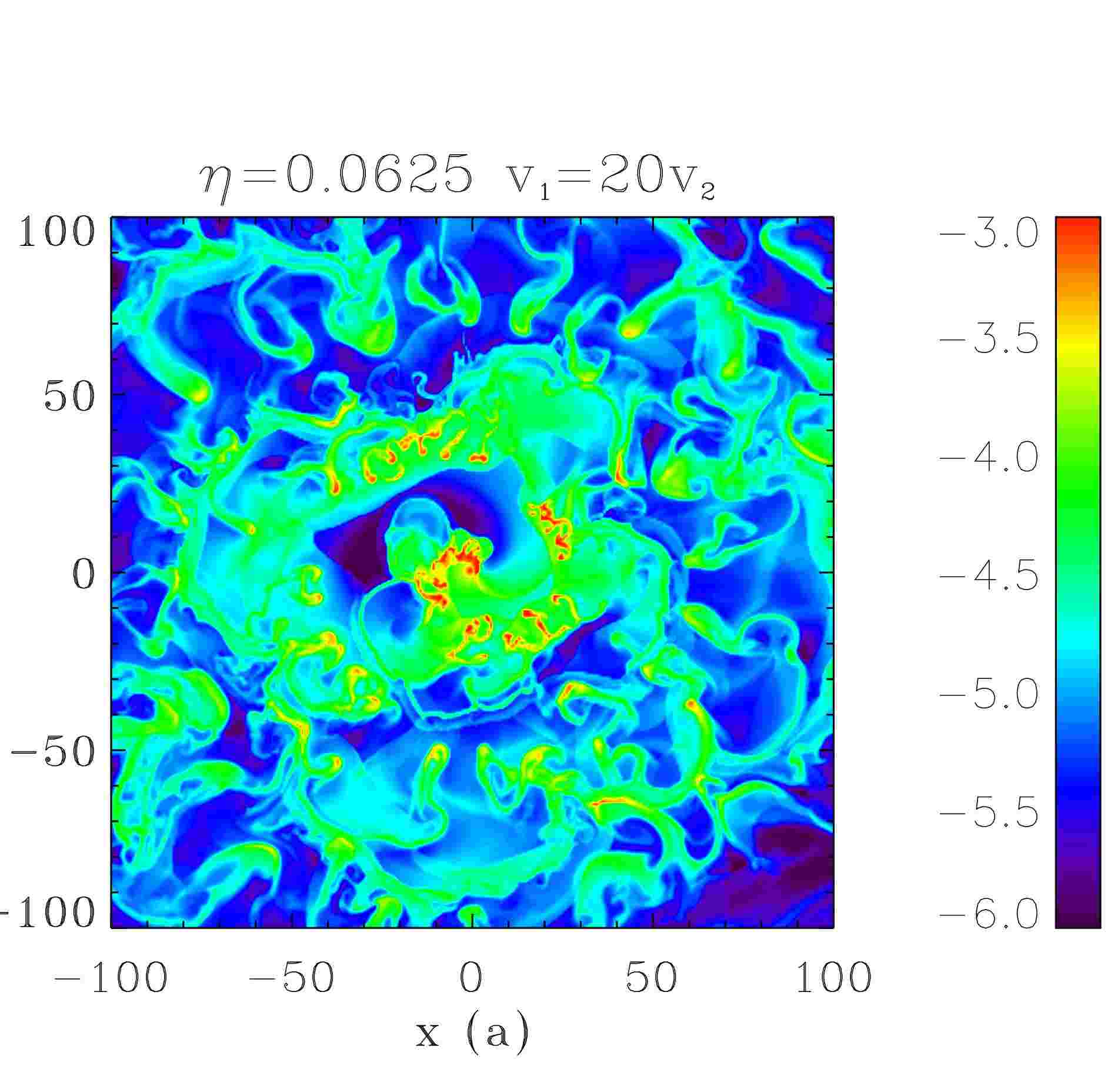}
  \includegraphics[width = .23\textwidth]{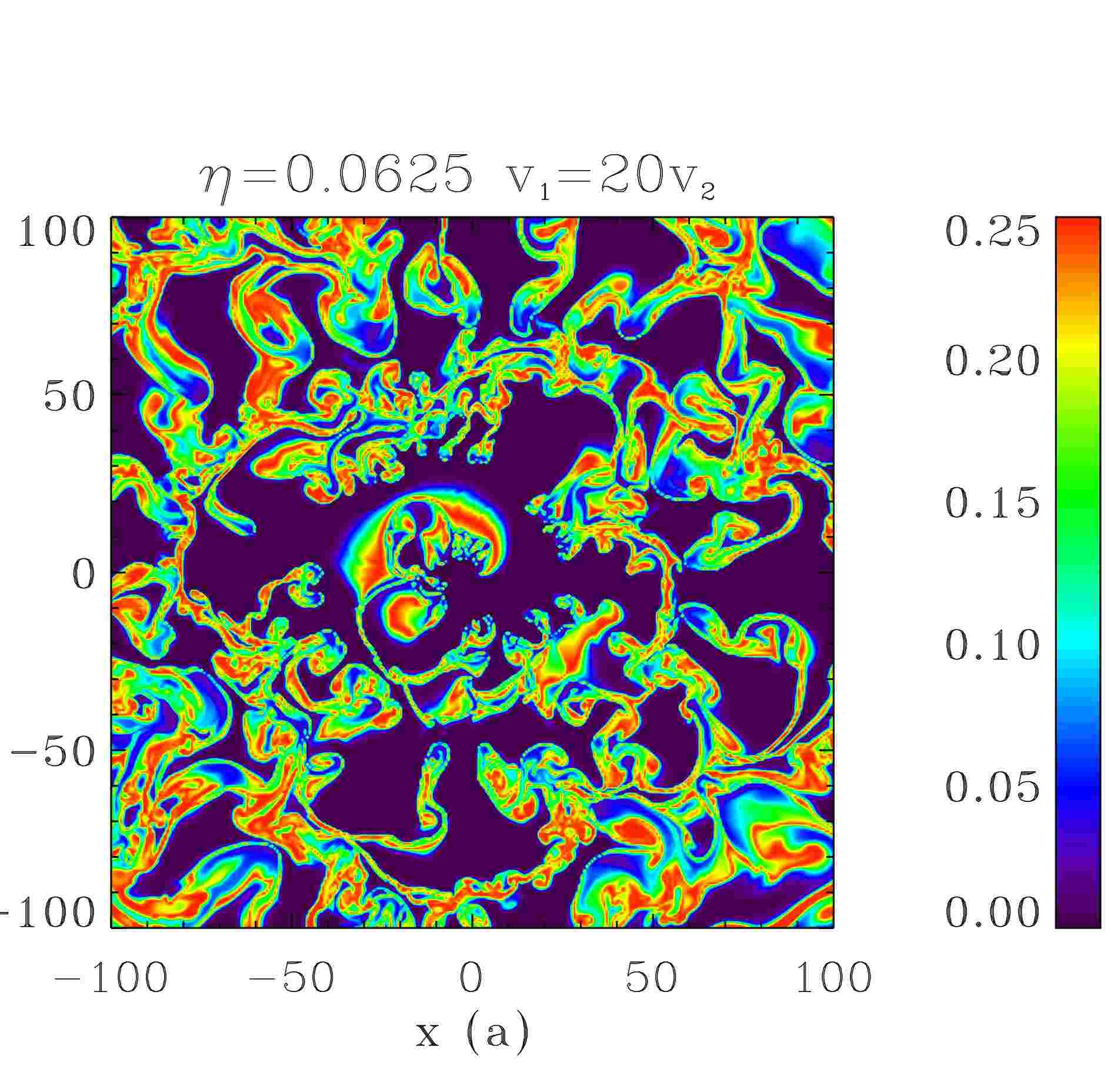}
  \caption{Large scale simulation : same as Fig.~\ref{fig:large_scale_1}  for $\{\eta=0.0625,\beta=0.05,0.5,0.1,1,2,20\} $ .}
  \label{fig:large_scale_00625}
\end{figure}

We now consider the large-scale evolution of the previous simulations (\S \ref{small_scale}) in a box of size  $l_{box}=400a$. Density and mixing maps are shown for $\eta=1$ in Fig.~\ref{fig:large_scale_1} and for $\eta=0.0625$ in Fig.~\ref{fig:large_scale_00625}, with $\beta$ increasing from top to bottom in both figures. The spatial scale is the same in all plots except for the top two panels of Fig.~\ref{fig:large_scale_00625}, where we reduced the size of the domain to avoid unnecessary computational costs. The different behaviour of mixing in both arms discussed in \S3 persists on the larger scales, eventually causing both contact discontinuities to merge into one single spiral in simulations with $\eta=0.0625$ (Fig.~\ref{fig:large_scale_00625}). We cannot exclude this merger results from numerical artefacts due to the use of a cartesian grid to describe an inherently spherical phenomenon. Merger should still occur naturally from inhomogeneities in the winds. Figures~\ref{fig:large_scale_1}-\ref{fig:large_scale_00625} also show that the colliding wind region does not always turn into a stable spiral and that, for given $\eta$ the appearance depends strongly on the velocity ratio $\beta$. We discuss below the step size that we measure when a steady spiral forms before addressing the issue of the stability of the pattern.

\subsection{The step of the spiral}

\begin{figure}[h]
  \centering
     \includegraphics[width=0.75\linewidth]{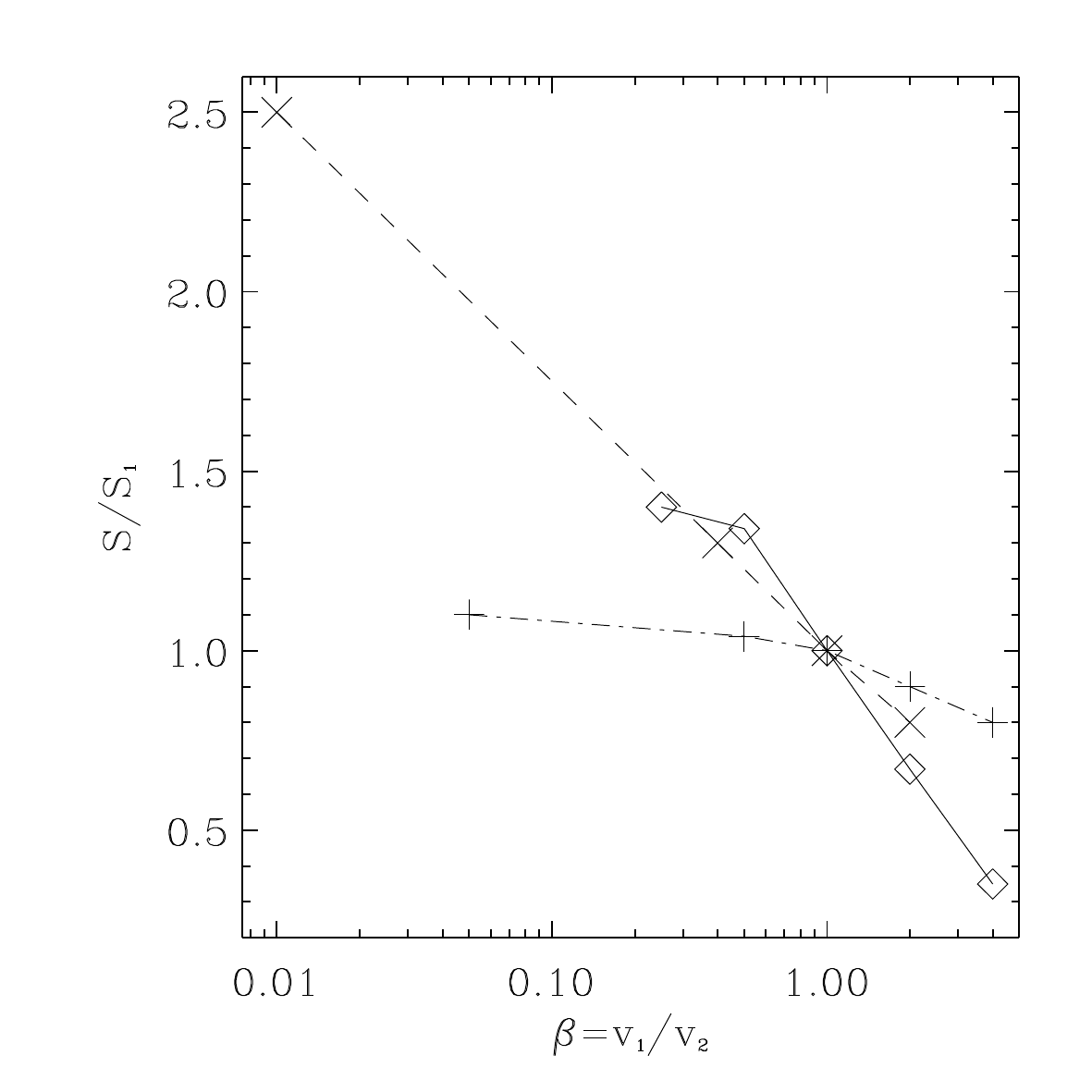}
  \caption{Step of the spiral $S/S_1$ as a function of $\beta$ for $\eta$ =1 (diamonds), 0.5 (diagonal crosses), and 0.0625 (crosses). The symbols are respectively joined by a solid, dashed, and dash-dotted line for easier identification.}
\label{fig:pas_spirales}
\end{figure}

We found that, when a stable spiral structure is formed, an Archimedean spiral with a step size $S$ provides a good fit to the results of our simulations. As in \citet{2009MNRAS.396.1743P}, we find that the fit with an Archimedean spiral is not perfect at the apex. However, the deviation is small and limited to a region $\simeq 10a$. Mixing follows closely the contact discontinuities in the arms so we used the mixing maps to trace the spiral. In fact, the spiral is not always clearly apparent in the density maps ({\em e.g.} $\{\eta=0.0625,\beta=1\}$ in Fig.~\ref{fig:large_scale_00625}), especially when a complex flow is established by the presence of a reconfinement shock behind the weaker star (Fig. 1). The fitted $S$ for various values of $\eta$ and $\beta$ is compared to the theoretical estimate $S_1$ in Fig.~\ref{fig:pas_spirales}. $S_1$ assumes the velocity of the stronger wind controls the structure scale so that $S_1=P_{\rm orb}v_1$ (e.g. T2008). When $v_1=v_2$, there is no ambiguity in the velocity that sets the step size and we verify that, in this case, $S=S_1$ for all $\eta$. This also rules out any significant numerical issue with the way the spiral develops. There are significant deviations from $S_1$ in all the other cases, except when $\eta\ll 1$ {\em i.e.} when the first wind largely dominates momentum balance. For more balanced ratios $\eta$, the spiral step is smaller than expected when the weaker wind is slower than the stronger wind, and vice-versa when the weaker wind is the fastest. Using the slowest wind speed instead of $v_1$ \citep[e.g.][]{2008MNRAS.388.1047P} does not work better. The results do not suggest a straightforward analytical correction using $\eta$ and $\beta$ that could be used to interpret observations of pinwheel nebulae without requiring hydrodynamical simulations.

\subsection{Stability of the spiral structure \label{spiral}}
The complete disruption of the spiral structure for large velocity ratios (Figs.~\ref{fig:large_scale_1}-\ref{fig:large_scale_00625}) was unexpected. The breakdown can be traced to the KHI: we found that the structure was stabilised when we tested a setup with a Lax-Friedrich Riemann solver and a smaller resolution, in which case the KHI is artificially suppressed (Paper I). The amplitude of the KHI appears to destroy the spiral when strong velocity gradients are present, resulting in widespread turbulence and important mixing throughout the domain. Curiously, for $\eta=0.625$, the structure is unstable when $v_1=20v_2$ while it is stable for the opposite velocity gradient $v_1=0.05v_2$: the strong density gradient ($\dot{M_1}/\dot{M_2}=320$) is also likely to play a role in the stability. Indeed, we measure $\alpha\simeq 0.95$ in the mixed region at $r\simeq 50 a$, which implies a strong reduction of the KHI growth rate.  

Table 1 summarises the presence or absence of spiral structures, for all the simulations we performed. We compared these results against the KHI growth rate, normalised by the rate at which eddies propagate (see Appendix A for details). Although the growth rate gives no indication on the saturation in the non-linear regime, we suspect that the spiral is destabilized most easily when the eddies grow quickly before propagating further away.

\begin{figure}[h]
  \centering
   \includegraphics[width=0.85\linewidth]{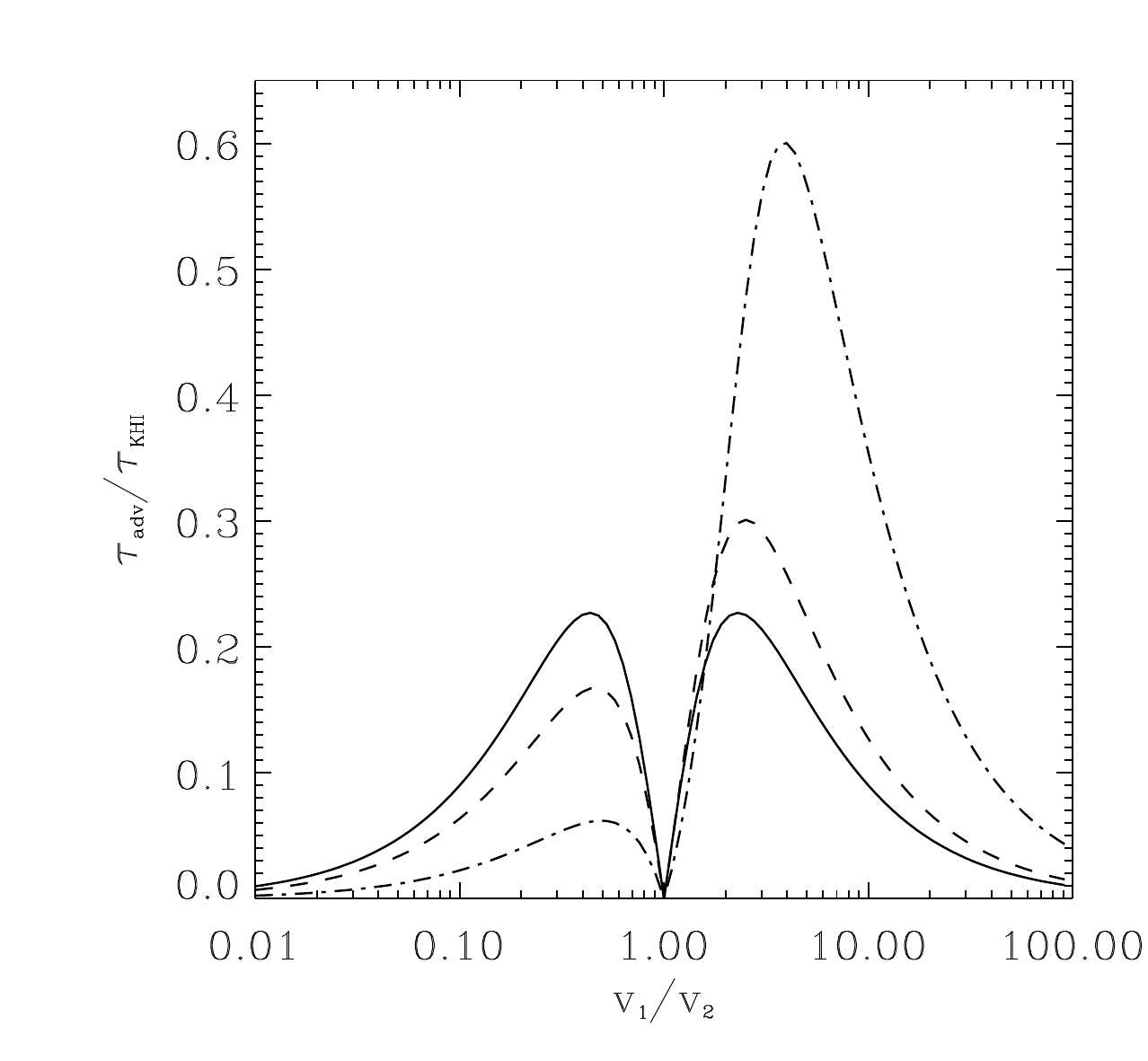}
  \caption{Theoretical 2D growth rate of the KHI in colliding wind binaries as a function of the velocity ratio $\beta=v_{1}/v_{2}$ of the winds. The solid, dashed and dash-dotted lines correspond to $\eta=1, 0.5 , 0.0625$ respectively.}
  \label{fig:growth}
\end{figure}

Figure~\ref{fig:growth} shows the normalised growth rate (see Eq. A.12) as a function of $\beta$ for $\eta=1, 0.5$, and 0.0625. For $\eta=1$ the curve is  symmetric. The KHI does not develop when there is no velocity difference, peaks at $\beta_{{\rm max}}\simeq 2$ and drops for higher values of $\beta$ because the density gradient dampens the growth rate. The symmetry with respect to $\beta=1$ is broken when $\eta< 1$: the normalised growth rate is weaker for $\beta<1$ and stronger for $\beta>1$. The lower the value of $\eta$, the stronger this asymmetry. Hence, stable structures are expected near $\beta=1$, for $\beta\gg 1$, and for $\beta \ll 1$. In addition, when $\eta\neq 1$,  structures with $\beta<1$ should be more stable than with $\beta>1$.

The results of the simulations are in qualitative agreement with these expectations. When $\eta=1$, there is a spiral for $\beta=1,2,4$ but not for $\beta=8,20$ (Tab. 1), which is consistent with the faster growth of the KHI when $\beta$ increases from 1. However, the transition from stable to unstable spirals occurs further away than expected from Fig~\ref{fig:growth} ($\beta_{\rm max}\simeq 2$). Also, we were not able to recover a stable final structure for very high $\beta$ (or, equivalently in the case $\eta=1$, very low $\beta$), up to $\beta=200$  (Fig.~\ref{fig:growth}). Tests at higher $\beta$ are computationally too expensive (and may not have much astrophysical relevance). For $\eta=0.5$, we find that the spiral is maintained for values close to $\beta=1$ but is quickly destroyed for higher/lower values of $\beta$. In this case, a stable spiral is recovered when $\beta \leq  0.01$, consistent with the lower growth rate, while the spiral remains destroyed for the symmetric value of $\beta=20$ (higher growth rate). We observe a similar behaviour for $\eta=0.0625$.  Stabilisation is possible for a higher $\beta=0.05$, which is consistent with the lower growth rate of the instability for $\beta<1$ as $\eta$ decreases. We conclude the presence of a  spiral depends on $\eta$ and $\beta$ in a way that is consistent with having stable structures for near-equal velocity winds ($v_{1}\simeq v_{2}$) or when the weaker wind is much faster ($\dot{M}_{1}v_{1}\geq \dot{M}_{2}v_{2}$ and $v_{2}\gg v_{1}$).

\begin{table}
\caption{Presence (S) or absence (X) of a spiral structure in simulations for various wind momentum ($\eta$) and velocity ($\beta$) ratio.}
\label{tab:spirals}
\centering
\begin{tabular}{c c c c c c c c c c c}
\hline
\noalign{\smallskip}
 $\eta \backslash \beta$  &  .01  &   .05& .1      &   .5 & 1& 2 & 4 &8  & 20 & 200 \\ 
\noalign{\smallskip}
\hline
\noalign{\smallskip}
 1                          &      &   X  &         &  S   & S &S&  S & X & X  & X                   \\
  0.5                     &   S &  X   & X       &  S   & S &S&    & X & X  &  X                 \\
  0.0625              &      &  S   &  X      &  S   & S &S&  S/X & X & X  &                       \\
\noalign{\smallskip}
\hline
\end{tabular}
\end{table}

\section{The pinwheel nebula WR 104}\label{wr104}

WR 104 is a binary composed of an early type star and a WR star. The system shows an excess of IR emission related to dust production. The IR emission has been resolved into a spiral structure with several steps imaged (T2008). The high temperatures and low densities in WR winds are difficult to reconcile with dust formation, which requires a temperature around $1000$ K and a number density range between  $10^{6}$ cm$^{-3}$ and $10^{13}$ cm$^{-3}$ \citep{1995IAUS..163..346C}. An additional constraint for dust formation arises from the absence of hydrogen in the WR wind, leading to uncommon chemical processes \citep{2000A&A...357..572C}. Dust production appears closely-related to binarity and the presence of dense colliding wind structures: in eccentric systems, such as WR 48 or WR 112, dust production is limited to orbital phases close to periastron while it is continuous in systems with circular orbits. Systems viewed pole-on show an extended spiral structure in infrared. WR 104 is the prototype system of these pinwheel nebulae. Our aim is to determine whether a hydrodynamical model with adiabatic winds reproduces the observed large-scale structure of WR 104, study mixing and identify regions where dust production may be possible. Detailed modelling of dust formation and growth in colliding wind binaries is beyond the scope of this study. 
%These observations suggests that colliding wind structures favour dust production. They create a higher density region and allow for mixing between hydrogen from the early-type star and  carbon from the WR star.% Although the growth of amorphous carbon grains in WR stars seems understood~\citep{1998MNRAS.295..109Z}, no theory explains the nucleation of the initial dust seeds. 

% and measure the impact of mixing of the winds through the KHI.

\subsection{Simulation parameters}
 
\begin{table}\label{tab:param_wr104}
\begin{center}
\caption[]{System parameters for WR 104}
\begin{tabular}{c c c}
\hline
\noalign{\smallskip}
 & WR & OB \\
\noalign{\smallskip}
\hline
 \noalign{\smallskip}
 $v$ (km s$^{-1}$)  & 1200 (a) & 2000 (b)\\
$\dot{M}$ ($M_{\odot}$ yr$^{-1}$) &  $0.8\times 10^{-5}-3 \times 10^{-5}$ (c)& $6\times10^{-8}$ (d) \\
%R$_*$ ($R_{\odot}$)& - & 10 (e)\\
\noalign{\smallskip}
\hline
\end{tabular}
\end{center}
(a) \citealt{1992A&A...261..503H}, (b) estimate according to spectral type \citep{Harries2004}, (c) \citealt{1997MNRAS.290L..59C}, (d) using the mass-loss luminosity relation by \citealt{1989ApJS...69..527H}.
\end{table}

Table 2 has the wind parameters for the binary system WR 104. The characteristics of the companion to the WR star are not well constrained \citep{2001NewAR..45..135V} and, like T2008, we will refer to the companion star as the ``OB" star. The orbital period (241.5$\pm 0.5$ days), eccentricity $e<0.06$, inclination ($i<16\degr$), and angular outflow velocity of the spiral 0.28 mas day$^{-1}$ in WR 104 were found by fitting an Archimedean spiral to the IR maps (T2008). The orbital separation $a$ is about 2.1-2.8 {\sc au} for a total mass of 20-50 M$_\odot$. We took $e=0$ and $a=2.34$ {\sc au}. Given the uncertainties on mass loss rate and velocities $\eta$ varies between $0.0125=1/80$ and $0.0033=1/300$. Assuming a constant velocity for the OB wind and $R_{\rm OB}=10$ R$_\odot$ \citep{Harries2004}, the second shock forms at $2.7 R_\mathrm{OB}<r<5.1 R_\mathrm{OB}$ depending on $\eta$. 

The shock position can be influenced by additional physical processes. The OB wind is accelerated on distances of $\simeq 2-3$ stellar radii and has not necessarily reached its final velocity at the shock, which modifies the effective momentum flux ratio of the collision. The shock position moves to $2.2R_\mathrm{OB} <r<4.7 R_\mathrm{OB}$ if acceleration is taken into account by using the velocity law $v=v_{\infty}(1-R_\mathrm{OB}/r)$. Radiative braking of the WR wind by the OB radiation field  \citep{1997ApJ...475..786G} can also play a role in WR 104 (T2008). A slower WR wind moves the shock away from the OB star (up to 12 $R_{\rm OB}$ if radiative braking is able to stop the WR wind completely, which is only marginally possible in WR 104, see T2008). The magnitude of both effects, their compensating influence, and the uncertainties in the wind parameters did not justify including these processes. We adopted constant velocity winds and $\eta=0.0033$ to ease comparison with T2008.

Radiative cooling can significantly change the shock structure. The ratio $\chi$ of the cooling timescale $t_{cool}$ over the dynamical timescale $t_{esc}$ provides an estimate of its importance \citep{Stevens:1992on}
\begin{equation}
  \label{eq:chi}
  \chi=\frac{t_{cool}}{t_{esc}}=\frac{k_BT_s}{4 n_w\Lambda(T_s)}\frac{c_s}{a}
\end{equation}
where $\Lambda\approx 2\times 10^{-23}$ erg\,cm$^{3}$ s$^{-1}$  is the emission rate, $n_{w}$ the number density of the unshocked wind, $k_{B} T_{s}=(3/16) \mu m_{p} v_{w}^{2} $ the shock temperature and $c_{s}$ the associated sound speed. The system is adiabatic if $\chi >1$ and isothermal if $\chi \ll 1$. We find $3<\chi_\mathrm{OB}<15$ for the OB star and $0.3<\chi_\mathrm{WR}<1.2$ for the WR star. The system is at the transition between the two regimes. The escape timescale is assumed to be $\simeq a/c_s$ in Eq.~\ref{eq:chi} but could be as short as $2.7-5.1 R_\mathrm{OB}/c_s$ (increasing $\chi$ by a factor 10-20) if one takes the distance from the OB star ($\sim$ shock curvature radius, \citealt{Stevens:1992on}).  In the following, we neglected radiative cooling in the energy equation and assumed an adiabatic shock.

The low value of $\eta$ is challenging for numerical simulations (see discussion in paper I). The mask of the star needs to be as small as possible so that the shocks can form properly. A minimum length of 8 computational cells per direction is needed to obtain spherical symmetry of the winds. Numerical resolution on scales much smaller than a stellar radius (0.05 {\sc au}) is thus required close to the binary. Further away, we need to maintain a high resolution in order to properly study the instabilities, while following a spiral step requires a box size $\geq 200$ {\sc au}. We carried out two complementary simulations:  a 3D simulation covering scales up to $12a$ and a 2D simulation to model a whole step of the spiral structure.

We use the large scale 2D simulation to determine the step of the spiral and the impact of mixing. As explained in \S2.4, we use the mapping $\sqrt{\eta_{\rm 3D}}\rightarrow \eta_{\rm 2D}$ to obtain comparable 2D and 3D results. We took $\eta_{\rm 2D}=0.0625$ to help comparisons with the results in \S3-4, which is close enough to $\eta_{\rm 2D WR104}\simeq 0.057$ derived from a straight application of the mapping. It is important to have the right velocity difference for the Kelvin-Helmholtz instability. We thus adapted $\dot{M}_{WR}$ in order to have $\eta_{\rm 2D}=0.0625$ for the 2D simulation. We use a 200$a\approx 500$ {\sc au} simulation box with $n_x=128$ and 12 levels of refinement. This gives an equivalent resolution equal to $2^{19}\simeq  5\times 10^5$ cells. We use nested grids to slowly decrease the maximum allowed resolution away from the binary.

We use the smaller scale 3D simulations for quantitative results on the density and temperature in the winds. The 3D simulation follows $1/8$th of an orbit of WR 104 in a 12$a\approx 30$ {\sc au} simulation box, large enough to see the impact of orbital motion. The orbital plane is the mid-plane of the box and the centre of mass of the binary is placed in a corner of the box to maximise the use of the simulated volume. We use adaptive mesh refinement with a maximal equivalent resolution of 4096$^3$. We limit the high resolution to a narrow zone of $3a$ close to the binary where the instabilities develop. It corresponds to the same equivalent resolution as in our 2D model.  We model only $\simeq 20$ layers at this high resolution in the $z$ direction  and we gradually reduce the resolution when going away from the orbital plane.

\subsection{Global structure} 
\begin{figure*}
  \centering
  \includegraphics[width = .23\textwidth]{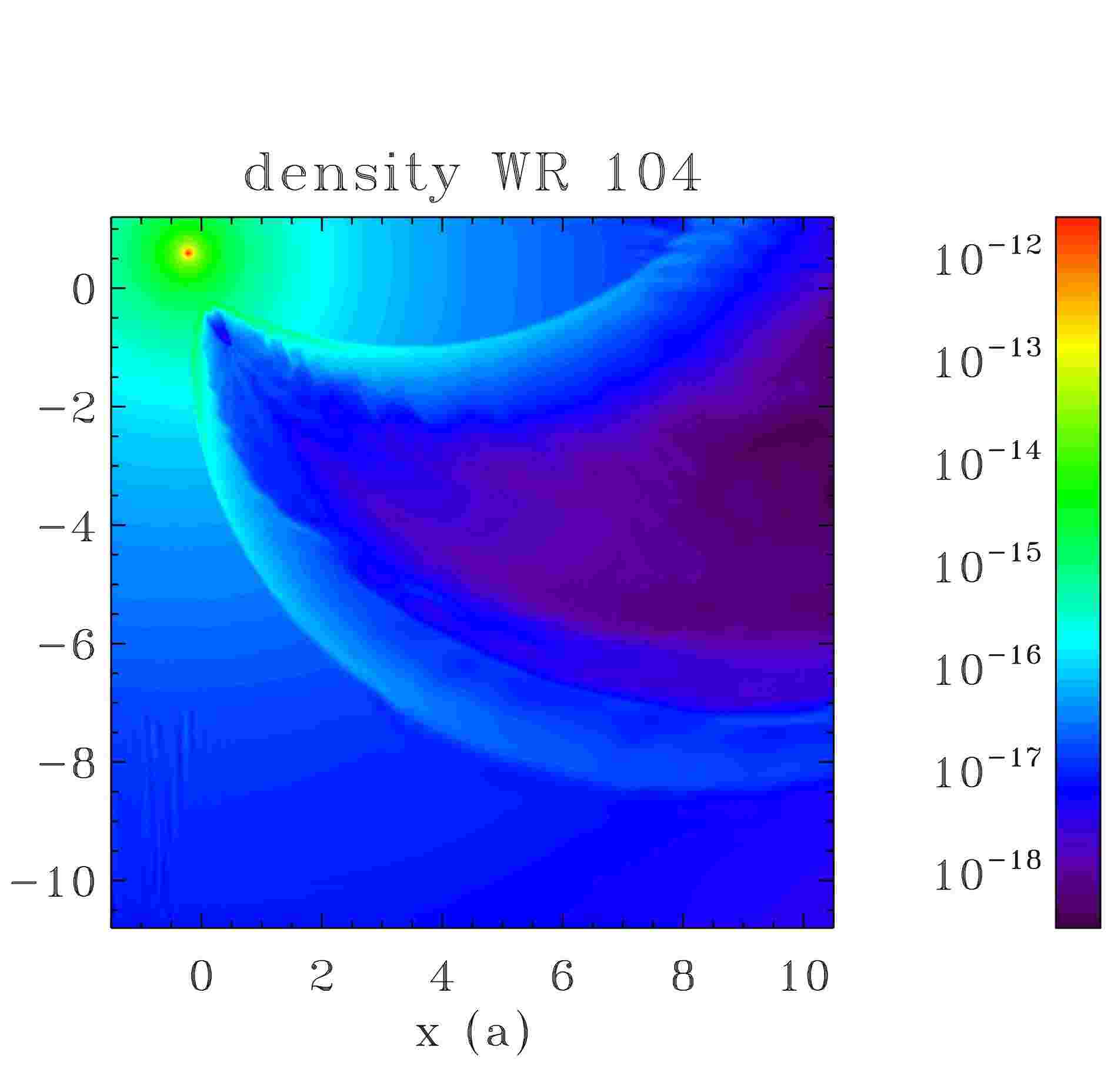}
  \includegraphics[width = .23\textwidth]{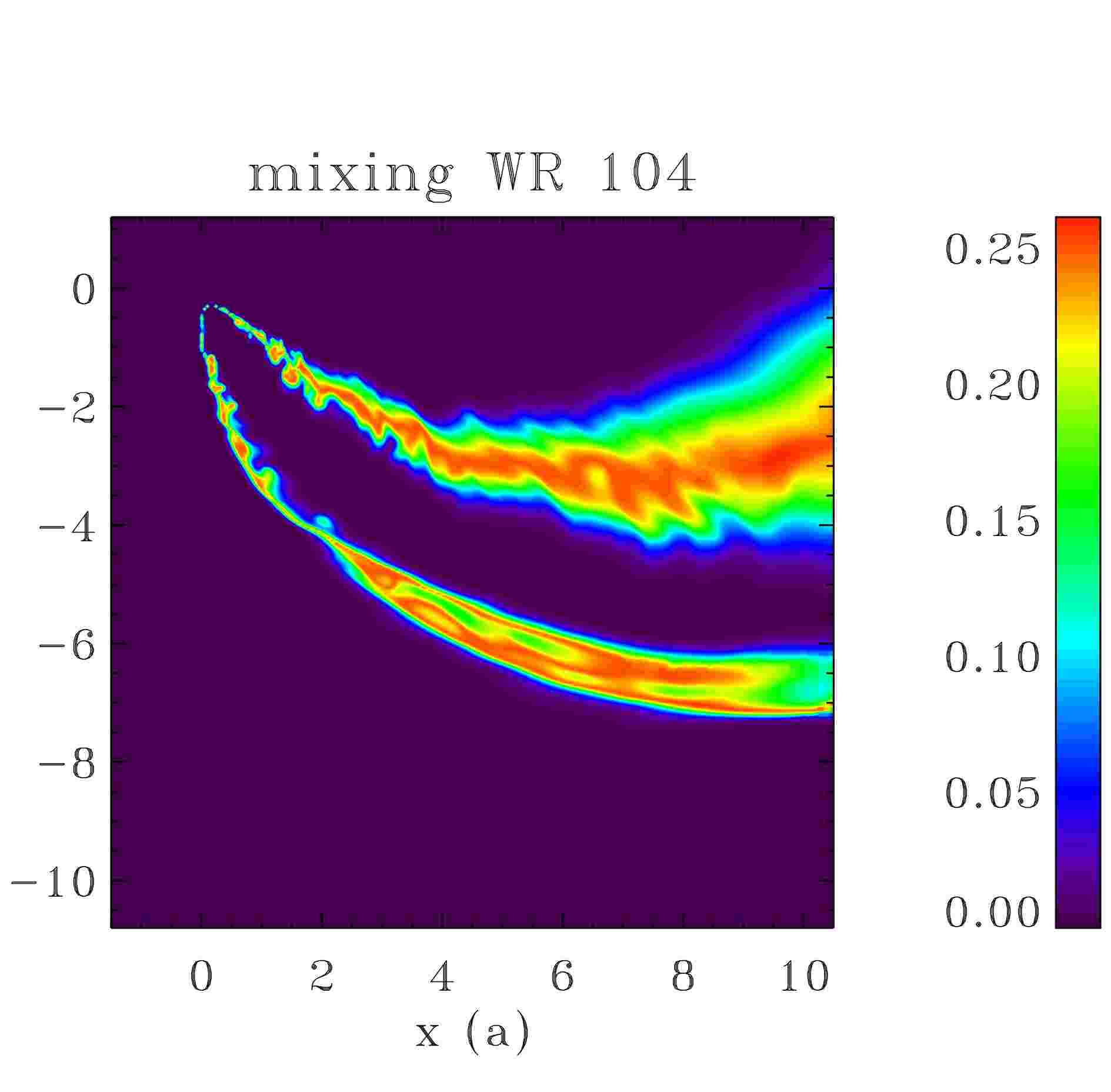}
  \includegraphics[width = .23\textwidth]{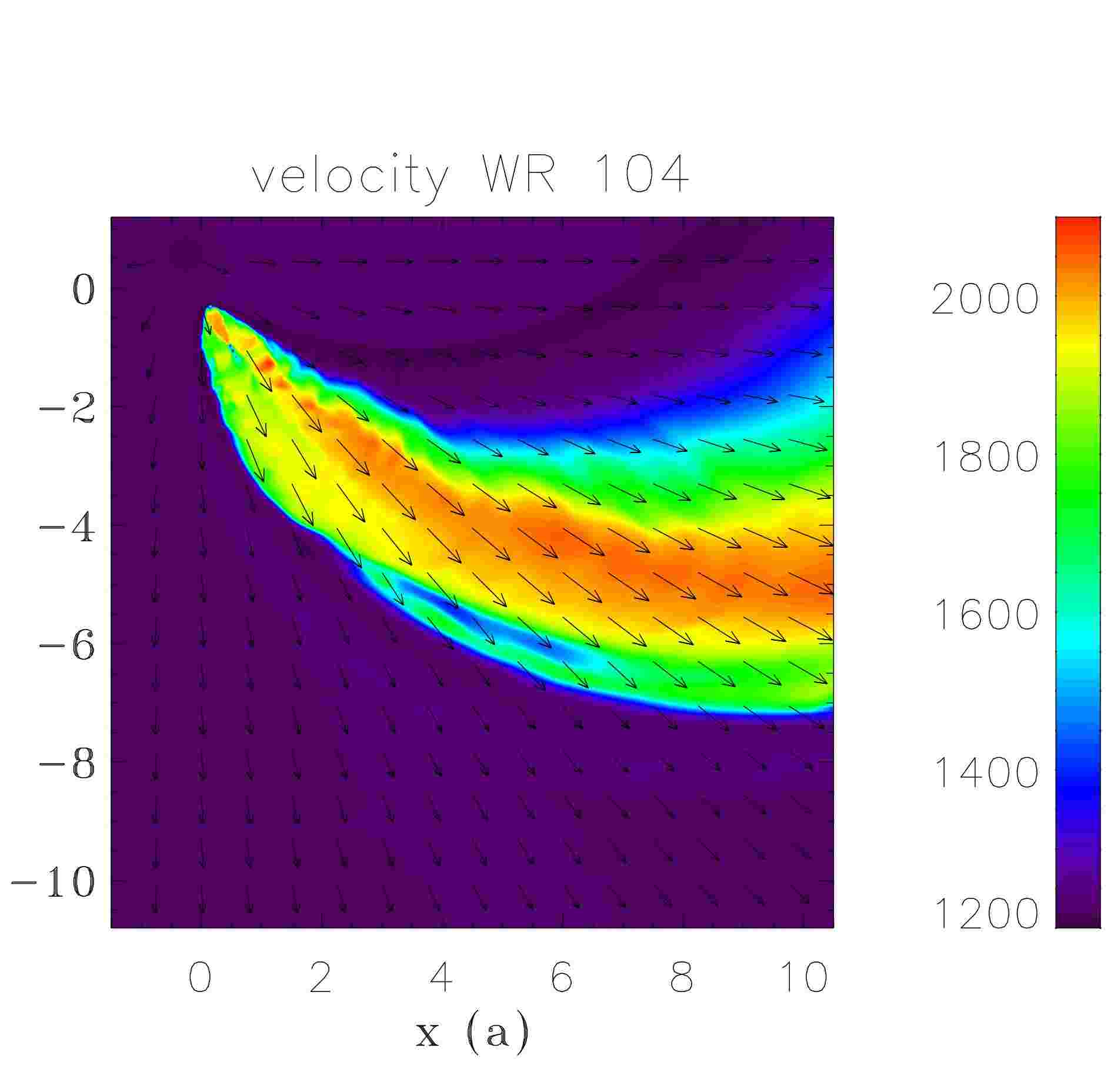}
  \includegraphics[width = .23\textwidth]{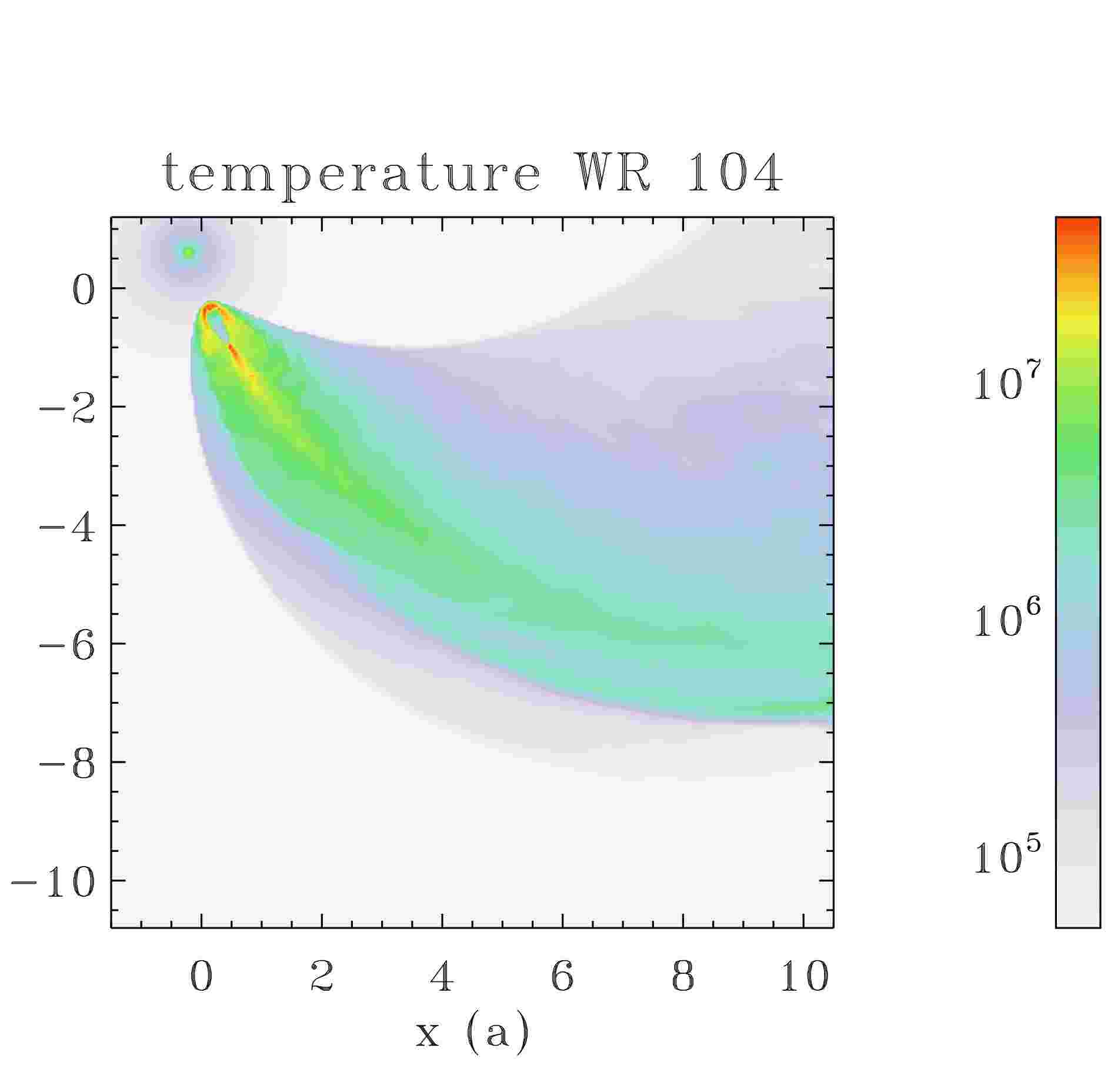}
  
  \includegraphics[width = .23\textwidth]{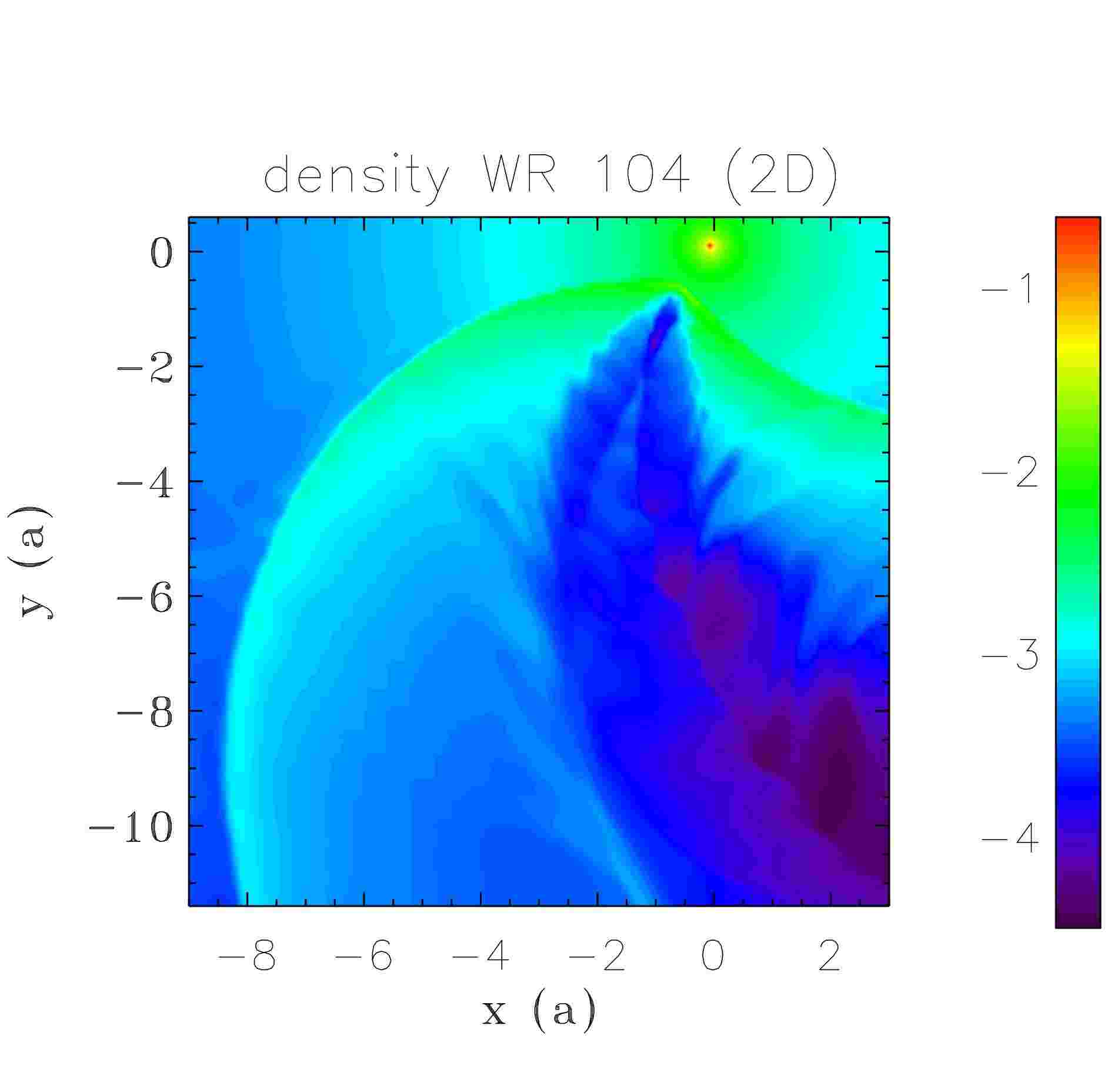}
  \includegraphics[width = .23\textwidth]{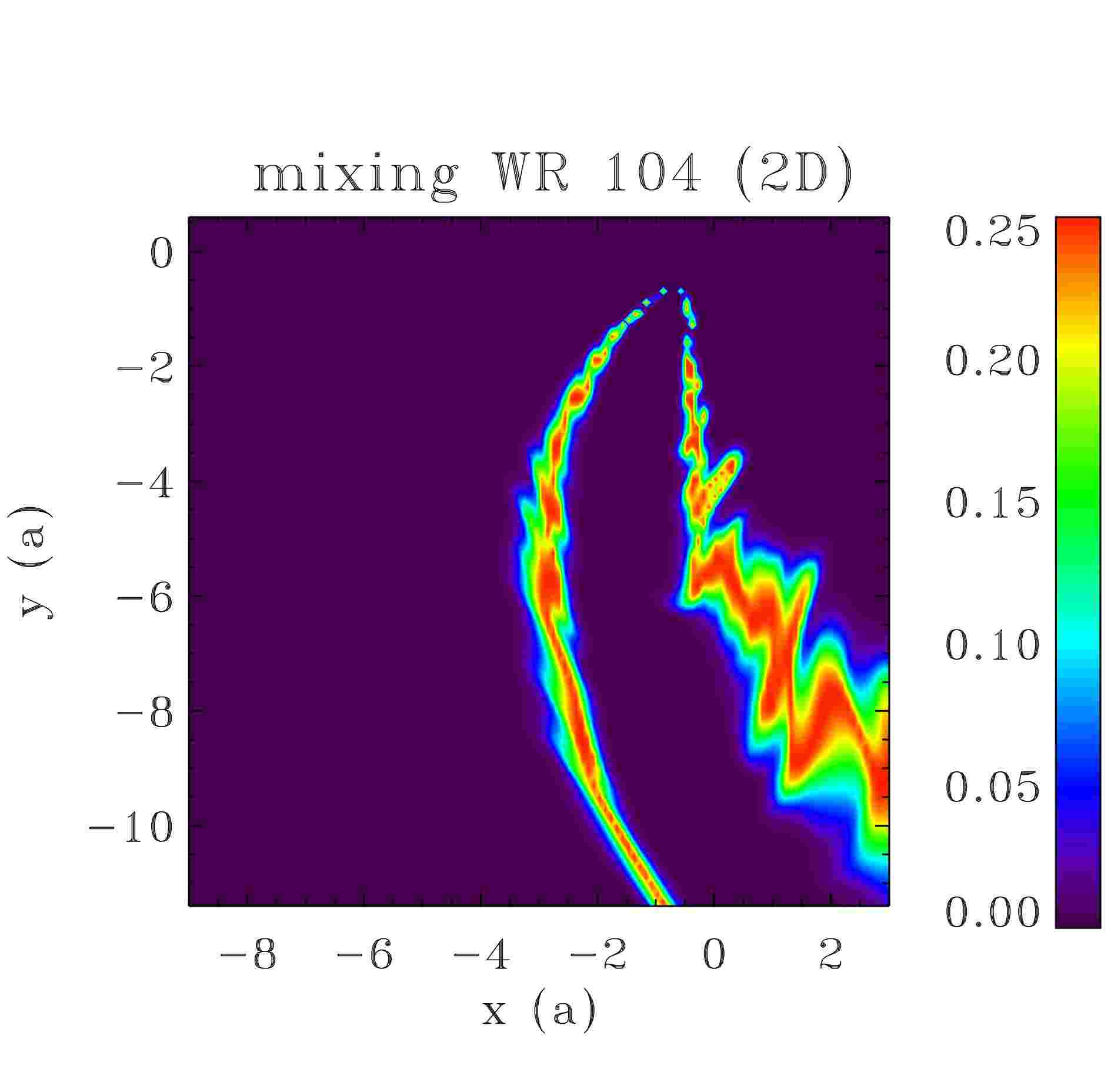}
  \includegraphics[width = .23\textwidth]{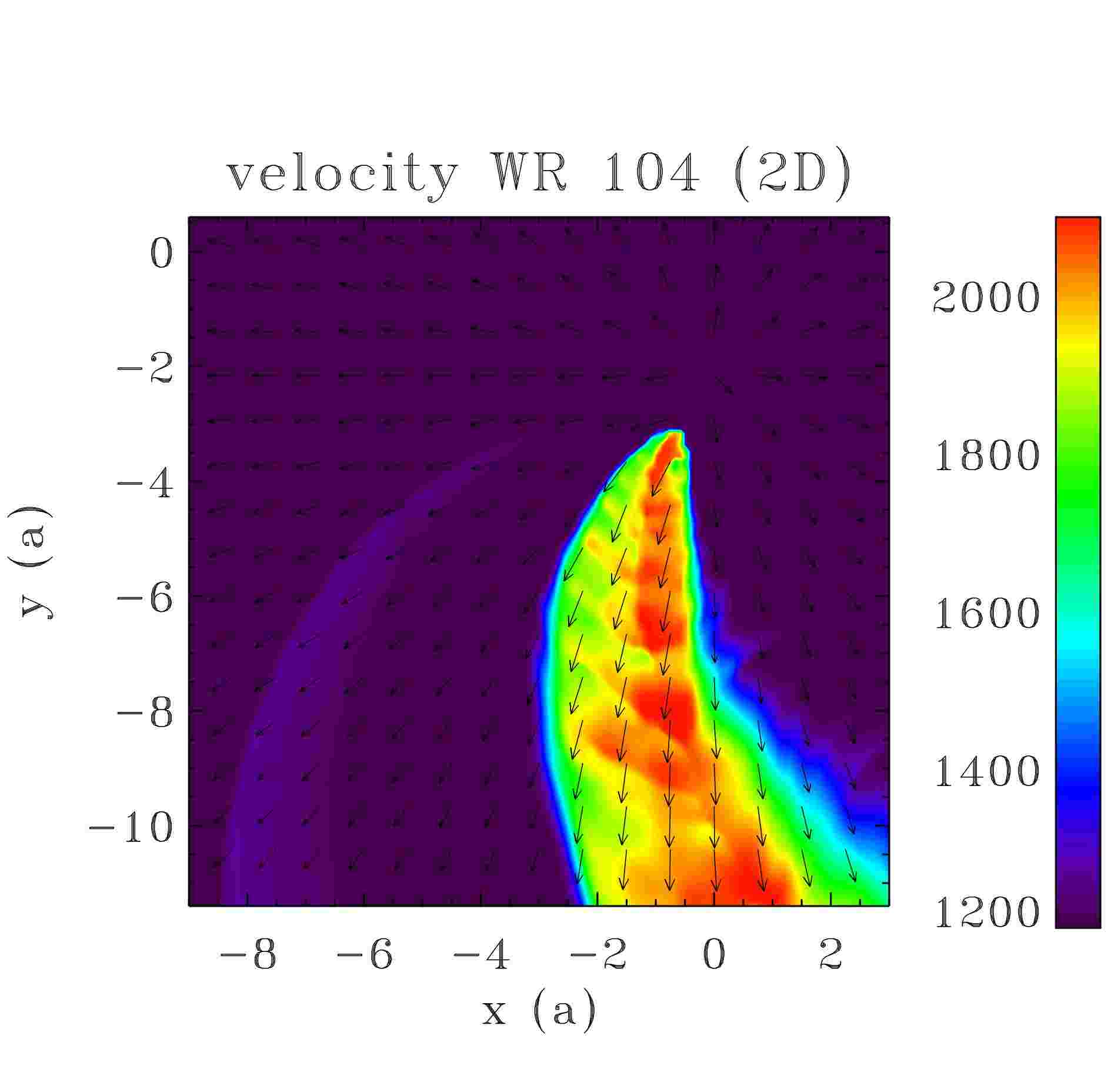}
  \includegraphics[width = .23\textwidth]{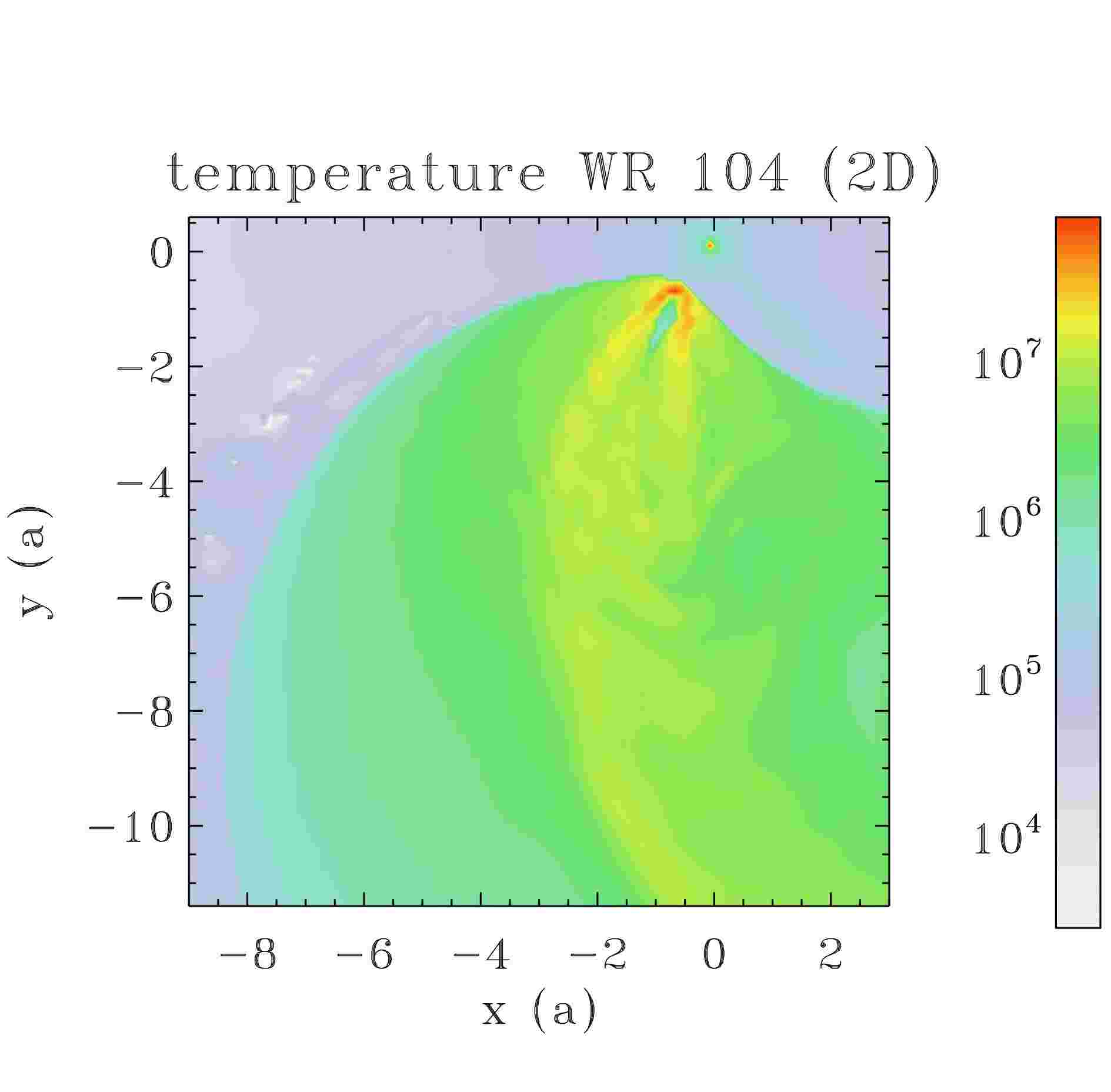}
  \caption{Density (g cm$^{-3}$), velocity (km s$^{-1}$), mixing and temperature (K) in the orbital plane of the 3D simulation of WR 104 (top row). Corresponding 2D maps on the same scale (bottom row). The length scale is the binary separation $a$.}
  \label{fig:wr104_3D}
\end{figure*}

\begin{figure*}
  \centering
 \includegraphics[width = .3\textwidth]{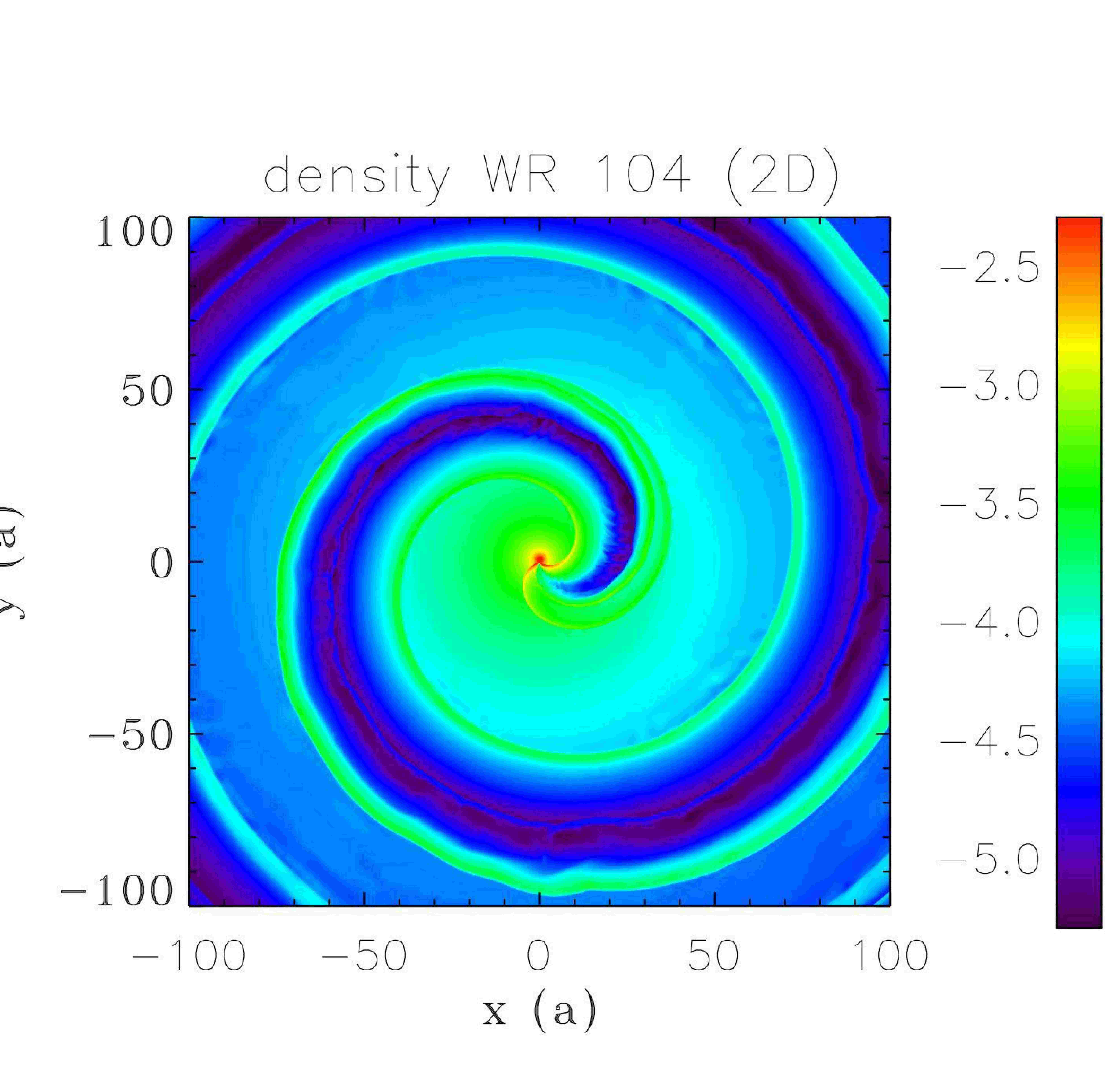}
 \includegraphics[width = .3\textwidth]{s_WR104_2D}
 \includegraphics[width = .3\textwidth]{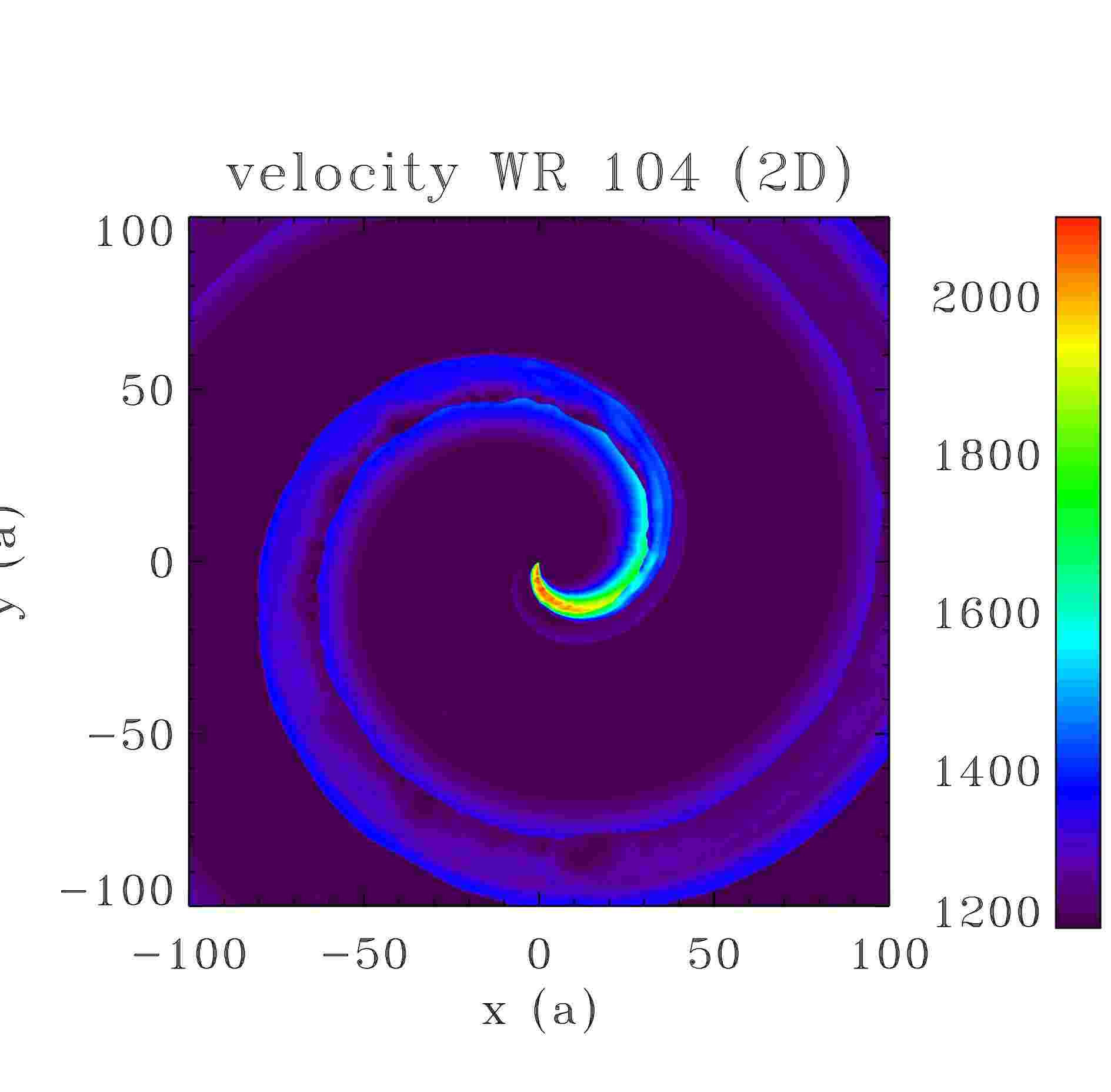}
  \caption{Density (g cm$^{-2}$), velocity (km s$^{-1}$) and mixing in 2D simulations of WR 104. The length scale is the binary separation.}
  \label{fig:wr104_2D}
\end{figure*}

Figure~\ref{fig:wr104_3D} shows the density, velocity, mixing, and temperature in the binary orbital plane of the 3D simulation (top row). The bottom row has the  corresponding 2D map on the same scale. The comparison confirms the mapping in $\eta$ captures adequately the 3D structure in the 2D simulation. The positions of the shocks and contact discontinuity along the line of centres are similar in both 2D and 3D simulation and match the analytic solutions (Paper I). The opening angle defined by the contact discontinuities, well traced in the mixing map, is $15 \pm 1^{\circ}$. This angle is consistent with the analytical estimates that have been used (\citealt{Harries2004}, T2008). However, the opening angle defined by the location of the shocks is wider in 2D than in 3D, which may have some influence on the density structure at larger scales (see \S5.3). Because of the low $\eta$, there is a reconfinement shock behind the OB wind at a distance $\simeq 0.75a$ in the 3D simulation (1.5$a$ in the 2D simulation). All of the OB wind is involved in the collision and no fraction escapes freely to infinity.

Material piles up in both arms of the spiral. T2008 suggest different strengths of the shock can change the conditions for dust formation in each arm. In the 3D simulation, the Mach number of the trailing arm at $r\simeq 12a$ is 13$\%$ higher than in the leading arm, in agreement with the results in \citet{Lemaster:2007sl}. The small temperature difference is unlikely to affect dust formation. A more significant effect is that compression keeps a hotter temperature in the leading arm than in the trailing arm. Material in the mixing zone of the trailing arm experiences a temperature an order-of-magnitude cooler than in the mixing zone of the leading arm. Dust formation may be favoured in this arm, seeding the spiral structure when the contact discontinuities merge farther out (see below). 

The amount of mixing increases with the distance to the binary. Integrating in spheres of increasing radii, the ratio of mixed material to the total amount of material within $r$ increases from $\simeq 0.01\%$ at $r=a$ to $0.4\%$ at $r=10a$. These values are constant during the last stages of the simulation (lasting $\simeq 0.1 P_{\rm orb}$), indicating the development of the instabilities has reached a steady state. As can be seen on Fig.~\ref{fig:wr104_3D} and as expected from theory (Appendix A), mixing occurs mostly in the lower density regions of the colliding wind zone. The velocity map shows that the velocity is mostly radial and that matter is accelerated on a distance of a few times the binary separation. After  substraction of the radial component, we find the velocity of the flow along the spiral in the 3D simulation reaches a maximal value of $\simeq 800$ km s$^{-1}$.  This corresponds to the low density region in the center of the spiral. In the outer regions of the spiral, the velocity along the spiral reaches $\simeq$ 500 km s$^{-1}$.

Figure~\ref{fig:wr104_2D} shows the 2D simulation on the largest scale (200$a$ or about 470 {\sc au}). A stable spiral structure forms as expected for $\beta=0.6$ and $\eta=0.0625$ (Tab. 1). The collimated OB wind generates a low density spiral bounded on each side by walls of material where the density is $\sim$100 times larger. The initially different mixing in both arms blurs at a distance of $\simeq 50 a$. The mixing zones more or less merge and follow the leading arm, overlapping slightly with the density enhancement of the arm. The step of the spiral is 1.05$S_{\rm WR}$ where $S_{\rm WR}=v_{\rm WR} P_{\rm orb}=170$ {\sc au}$=77a$. T2008 assumed $S_{\rm WR}$ to determine a distance of 2.6 kpc from the observed step size. The 5\% correction to this distance due to the intrinsically larger spiral step is smaller than the uncertainty on the measured WR velocity and observed angular step size.

The single-armed spiral observed in infrared is more reminiscent of the mixing region than the double spiral in the density map. A double-armed spiral  structure, separated by a very under-dense region of angular size $\simeq 27$ mas (at 2.6 kpc), would have been resolved if the IR emission correlated with density ({\em i.e.} for a constant gas-to-dust ratio). However, we caution that the width of the low density zone may be overestimated in this 2D simulation since it is likely to be related to the opening angle of the shocks, which Fig.~\ref{fig:wr104_3D} shows to be wider in 2D than in 3D. Including dust radiative transfer is required for a closer comparison of the observations with the hydrodynamical simulation.

\subsection{Conditions for dust formation}
One criterion for dust formation is a high enough density. \citet{1995IAUS..163..346C} indicate different paths towards the formation of amorphous carbon for number densities $n$ ranging from 10$^6$ to 10$^{13}$ cm$^{-3}$ and give a detailed study for $n=10^{10}$cm$^{-3}$. This gives $\rho=1.4 \times 10^{-14}$ g cm$^{-3}$ assuming a mean molecular weight $\mu=1.4$, typical for a ionized WC wind \citep{Stevens:1992on}. Such a density is only present in our 3D simulation at the edge of the spiral, up to a distance  $\simeq$ 2$a$ from the WR star. In the 2D simulation, along the walls we find that the density drops as $\rho \propto r^{-1}$. Using this as guidance, we expect $\rho \propto r^{-2}$ in 3D on large scales. The minimum value $n$=10$^{6}$ cm$^{-3}$ considered by \citet{1995IAUS..163..346C} is reached at {$r\simeq 25 $ $a$ at the inner wall of the spiral. This is equivalent to 1/3 of a turn} along the spiral. The density is too low for dust formation beyond this distance so that any dust present far away has been advected out.

Temperature is another criterion. Models show that dust condensation is possible for 1000\,K $<T<6000$\,K and does not vary much within that temperature range \citep{2000A&A...357..572C}. A strong temperature gradient remains between leading and trailing regions of the shocked zone, with the more compressed material maintaining a very high temperature. Still, the temperature falls from $\simeq 10^7$ K close to the stars to $\simeq 10^5$ K in parts of the mixing zone at the outer edge of the 3D simulation box ($\simeq 10a$). The temperature is expected to fall below 6000 K at half a turn of the spiral by extrapolating using $T \propto r^{-4/3}$ (from $P\propto \rho^{5/3}\propto \rho T$ and $\rho\propto r^{-2}$ in the arms). This can be taken as an upper limit to the dust condensation distance as this is calculated in the adiabatic approximation. It is consistent with the infrared observations of a quarter-orbit shift between the maximum infrared emission and the binary centre. Radiative cooling and photoionization heating of the wind would have to be included to have an accurate determination of the temperature and of the impact of shielding from the stellar radiation fields.

\section{Discussion}

\subsection{Asymmetries due to orbital motion}
Orbital motion breaks the symmetry around the binary axis and introduces significant differences from the stationary case with adiabatic winds. It causes a velocity difference that triggers the KHI even when both winds are strictly identical. It results in differentiation of the two arms flanking the weaker star. The arm moving into the densest unshocked wind is compressed, dampening the KHI while the other arm expands and sees Kelvin-Helmholtz eddies of larger size.  The density difference between the inner cavity and the bracketing walls can reach two orders of magnitude. According to \citet{2011ApJ...726..105P} radiation pressure can have a similar effect, either enhancing or reducing the initial difference.
 \citet{2004MNRAS.351.1307V} modelled the variations in emission line profiles of \object{WR 140} by a rotating cone with dense edges, allowing them to constrain the opening angle of the colliding wind region. They found a wider opening angle than expected using the analytic formula for the opening angle of the contact discontinuity \citep{Canto:1996jj} with the standard value of $\eta$ for this system. Our simulations also show that matter accumulates at the shock rather than at the contact discontinuity. The observed opening angle thus corresponds to the opening angle of the shocks, which is wider than the opening angle of the contact discontinuity. This also increases the fraction of the WR wind involved in the collision compared to estimates using the contact discontinuity. Some of the spectral line features that are not explained by models where both arms have equal emission (absorption) could be due to differences between leading and trailing arm \citep{1999MNRAS.302..549S}.
The skew angle that we measured in the simulations matches the theoretical value given by Eq. \ref{eq:skew}. However, as we found that the step of the spiral is mostly determined by the speed of the stronger wind (and not the speed of the slower wind), we wonder whether this could also be the case for the skew angle. This is implicitely assumed by \citet{2007ApJ...662.1231K}. Our simulations do not allow us to answer this question.

\subsection{To spiral or not to spiral}
The simulations presented here are the first including at least one step of the spiral. We have shown that a structure is maintained on these scales when the two winds have nearly equal velocities ($\beta\simeq 1$). This is consistent with the observations of pinwheel nebulae in several WR + O star binaries \citep{2006Sci...313..935T,2007ApJ...655.1033M,2009A&A...506L..49M}, since their winds do have comparable velocities. The spiral is destabilised when the stronger wind has a velocity between 10-50\% of the weaker wind (Tab. 1). For example, the episodic ejection of large amounts of (initially) slow-moving material could have temporarily destabilised any spiral structure in the luminous blue variable (LBV) / WR binary \object{HD 5980} \citep{2007ApJ...658L..25N,2011AJ....142..191G}.  \object{Eta Carina} may be a case where any large-scale structure generated near apastron (when the system is closer to being adiabatic) is destroyed because of the destabilising velocity ratio $\beta\simeq 1/6$ --- although this would have to be assessed against the effects of the high orbital eccentricity \citep{2011ApJ...726..105P}. We expect our results to hold for eccentric orbits if the system stays adiabatic. If the system moves from adiabatic to radiative along its orbit then thin shell instabilities develop with unknown consequences on the large scale structure.  

The spiral is stabilised again when the velocity ratio $\beta\ll 1$. Such a situation may occur in gamma-ray binaries composed of a young non-accreting pulsar and an early-type star \citep{2006A&A...456..801D}. In this case, the stellar wind interacts with the tenuous, relativistic pulsar wind. \citet{2011A&A...535A..20B} have argued that the KHI would destroy any large-scale structure. Assuming our results hold in the relativistic regime, we find that a stable spiral can form on large scales if the stellar wind dominates because $\beta \ll 1$ in this situation. The structure is unstable if the pulsar wind dominates, pointing to the intriguing possibility that the interaction may switch from one regime to another in gamma-ray binaries with Be companions such as \object{PSR B1259-63}. The highly eccentric orbit takes the pulsar close to the equatorial disc where the slow-moving stellar outflow dominates momentum balance \citep{1997ApJ...477..439T}, leading to a stable colliding wind region. However, at apastron, the pulsar wind may dominate over the radiatively-driven stellar wind and be unable to form a stable structure. Strong mixing of the two winds leads to rapid Coulomb or bremsstrahlung losses for the high energy particles, which has an impact on the gamma-ray emission \citep{2010MNRAS.403.1873Z}. Extended radio emission was detected around PSR B1259-63 near periastron \citep{2011ApJ...732L..10M}. Regular changes in radio morphology with orbital phase have been observed in other gamma-ray binaries that are compatible with non-thermal synchrotron emission from a stable collimated pulsar wind structure on scales $\leq v_w P_{\rm orb}$ \citep{2006smqw.confE..52D,2008A&A...481...17R,2011A&A...533L...7M}. The luminosity and frequency of the radiation are probably too low to be able to detect the spiral structure on larger scales. 

An example of colliding winds with $\beta \gg 1$ also involves pulsars, this time with a low-mass companion \citep{2011AIPC.1357..127R}. The weak stellar wind is overwhelmed by the relativistic wind of the recycled millisecond pulsar. No stable spiral is expected in this case. Another possible case is eruptive symbiotics like \object{AG Peg}, \object{HM Sge} or \object{V1016 Cyg}. These systems are composed of a red giant, with a very slow wind ($\simeq 20$ km s$^{-1}$) and a hot companion, a white dwarf in nova outburst at the origin of a fast outflow of several 1000 km s$^{-1}$ \citep{1987A&A...183..247G}. We expect the spiral structure to be destroyed if the hot companion dominates. The radio maps of AG Peg have been interpreted assuming a stable spiral structure and a reversal in $\eta$ with time \citep{2007ApJ...662.1231K}. In many cases there is probably no time to form a spiral, because of the long orbital period compared to the outburst timescale. The radio maps of HM Sge (possible 90 year orbit) show a more fragmented emission region than expected from colliding wind models \citep{2005ApJ...619..527K}, possibly because of thin shell instabilities triggered by radiative losses. 

Finally, we note that the KHI  also intervenes in the bow shock structure created by the adiabatic interaction of the wind of a fast-moving star with the interstellar medium (e.g. \object{Mira}). The controlling parameter is the ratio of wind speed to star velocity. Much like with spirals, fast growth of the KHI may strongly disturb the cometary structure at large distances, as seen in some hydrodynamical simulations \citep{2006MNRAS.372L..63W,2007ApJ...660L.129W}.

\subsection{Dust formation in pinwheel nebulae}

Dust formation in WR 104 and other pinwheel nebulae should be helped by the mixing with hydrogen-rich material from the early-type star that we observe in the 2D and 3D simulations. \citet{2009MNRAS.395.1749W} argued that stronger dust emission in the trailing arm would explain better the IR high-resolution images of WR 140, and attributed this to density variations. The winds have nearly identical velocities but the WR has a mass-loss rate 10 times higher than its O companion. The O wind is therefore bracketed by two high-density regions. Our simulations do not suggest very different densities. However, larger amplitude mixing is expected in the trailing arm because it propagates in the more tenuous O wind, possibly enhancing dust formation in this arm. The lower temperature in the trailing arm also helps. Hence a different dust-to-gas ratio between both arms could be an alternative explanation. High density eddies triggered by the KHI in the arms could also be responsible for the observation of IR obscuration events by dust clouds in other WR+O star systems \citep{1998A&A...329..199V}.

The offset of the peak IR emission in WR 104 is consistent with the distance at which we estimated the temperature to fall below the dust sublimation temperature. However, the densities in the colliding wind region are on the low side compared to what dust formation models require. Adiabatic shocks only enhance the density by  a factor 4 so radiative cooling is required. Close to the binary, the WR wind is likely to present some cooling, resulting in a thinner and denser shocked layer. \citet{2009MNRAS.396.1743P} show the post-shock density is about 100 times higher in their model cwb1, which has strong cooling, than in their adiabatic model cwb3. Radiative cooling also decreases the temperature, bringing the region where dust condensation is possible closer to the binary. Thin shell instabilities can develop when cooling is strong, enhancing mixing of the winds. Given the impact of the (weaker) KHI in adiabatic colliding winds, thin shell instabilities can also be expected to significantly influence the large scale structure. \citet{2009MNRAS.396.1743P} have shown that the differenciation of the arms remains when thin shell instabilities are present but the large scale outcome has not been studied yet.

Strong cooling is not necessarily present in all WR binaries. \citet{2011arXiv1111.5194W} present evidence from long-term IR observations of WR 48 for dust production throughout the orbit. The stellar winds in this system have similar characteristics than in WR 104 but the (tentative) orbital period is much longer, 32 years.  \citet{2011arXiv1111.5194W} estimate the system to be adiabatic with an average $\chi\simeq 11$. The value will be even higher at apastron in the eccentric orbit ($e=0.6$), yet dust formation is present. High densities will be much more difficult to reach than in WR 104, requiring dust formation at hitherto lower densities than have been considered possible.

\section{Conclusion}
We have studied the large scale impact of orbital motion and the Kelvin-Helmholtz instability on adiabatic shocks in colliding wind binaries. We used hydrodynamical simulations with adaptive mesh refinement to perform the first simulations of complete spiral steps. Orbital motion induces differentiation between both arms of the spiral. The arm propagating in the higher density wind gets compressed while the arm propagating in the lower density wind expands. We explain that this is due to a stronger growth of the KHI in the wider arm and discuss possible observational signatures in spectral lines. We confirm that the KHI arises even when both winds have identical speeds. We compute the step of the spiral and caution that there can be large differences with the standard estimates. We discover that the large-scale spiral structure is destroyed when the velocity gradient between the winds is important. Strong density gradients have a stabilizing effect. According to our simulations we can predict whether certain types of binaries present an extended spiral or not. Systems with stable spirals are those with near-equal velocity winds and those where the weaker wind is much faster. Performing high resolution simulations of pinwheel nebula WR 104, we show that in an adiabatic model significant mixing of the WR wind occurs with the hydrogen-rich wind of the companion. The temperature drop allows the formation of dust at roughly half a step of the spiral, consistent with the spatial offset in peak IR emission. However, the density in those regions falls short of the critical density for dust condensation. Including radiative cooling would lead to higher densities, and also to thin shell instabilities. The impact of these instabilities on the differentiation of the two arms and on the spiral structure is unknown: resolving the thin shock layer in a large scale simulation remains a very challenging numerical problem.

\begin{acknowledgements}
AL and GD are supported by the European Community via contract ERC-StG-200911. Calculations have been performed at CEA on the DAPHPC cluster and using HPC resources from GENCI- [CINES] (Grant 2011046391)
\end{acknowledgements}

\bibliographystyle{aa}
\bibliography{KHrot}

%\newpage
\appendix

\section{The Kelvin Helmholtz Instability in stratified flows}
\subsection{Linear theory}
\begin{figure}[h!]
 \centering
   \includegraphics[width=0.5\linewidth]{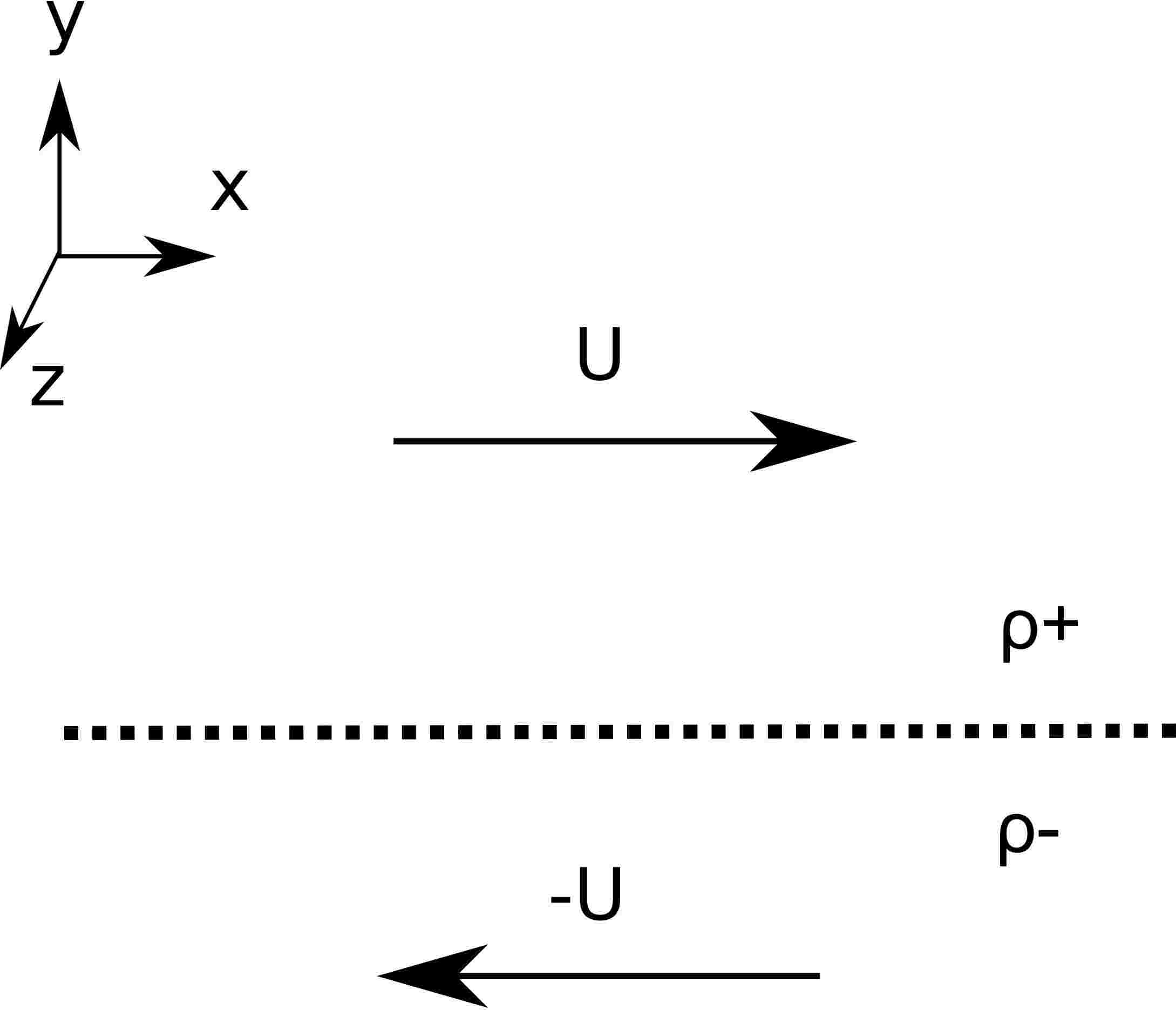}
   \caption{Configuration of the stratified flow}
   \label{config}
\end{figure}
We work in the incompressible approximation, assuming we have a system with a mean profile $\mathbf{U}=\pm U\mathbf{e_x}$. Above $y=0$, the flow has a density $\rho^+$ and $\rho^-$ for $y<0$ (see Fig.~\ref{config}). We neglect the Coriolis force since the local shear timescale $\tau_S=\Delta x/\Delta U\sim 10^{-6}\,\mathrm{yr}$ is much sorter than the orbital period $\tau_\Omega\sim 10^{-1}\,\mathrm{yr}$. In this approximation, the linearised equation of motion reads:
\begin{eqnarray}
\rho \frac{\partial\mathbf {v}}{ \partial t}+\rho U\frac{\partial \mathbf{v}}{\partial x} +\nabla P&=& 0 \\
\nabla \cdot \mathbf{v} &=& 0
\end{eqnarray}
In the following, each quantity is Fourier transformed in $x$ and $t$ thanks to homogeneity: $Q=Q\exp[i(\omega t - k x)$]. Rewriting the equation of motions and combining them leads to
\begin{equation}
\label{eqorr}\partial_y^2v_y-k^2v_y=0,
\end{equation}
which is solved with two decaying solutions
\begin{eqnarray}
v_y^+ = A^+\exp(-ky)&\mathrm{for}&y>0,\\
v_y^- = A^-\exp(ky)&\mathrm{for}&y<0,
\end{eqnarray}
$A^+$ and $A^-$ being two arbitrarily chosen constants which are adjusted by jump conditions at the interface $y=0$ : pressure should be continuous and  fluid particles should stick to the interface on both sides.
The pressure condition reads:
\begin{equation}
\label{cont1}
\rho^+\frac{\sigma^+}{k}\partial_y v_y^+=\rho^-\frac{\sigma^-}{k}\partial_y v_y^+,
\end{equation}
where $\sigma^\pm=\omega\pm U$. The second condition is obtained defining a displacement vector $\xi(x)$ which follows the interface. By definition, a fluid particle located at $(x,\xi(x)-\epsilon)$ satisfies
\begin{equation}
v_y=\frac{D\xi}{Dt}=\partial_t\xi+U\partial_x\xi=i\sigma\xi.
\end{equation}
Applying this to both side of interface ($\pm\epsilon$) leads to the jump condition
\begin{equation}
\label{cont2} \frac{v_y^+}{\sigma^+}=\frac{v_y^-}{\sigma^-}.
\end{equation}
Combining (\ref{cont1}) and (\ref{cont2}) and looking for non trivial solutions gives
\begin{equation}
\omega^2+2\alpha\omega kU+(kU)^2=0,
\label{omega}
\end{equation}
where $\alpha=(\rho^+-\rho^-)/(\rho^++\rho^-)$. An instability arises whenever
\begin{equation}
\Delta=(1-\alpha^2)(-k^2U^2)<0
\end{equation}
which is always true since $-1\leq\alpha\leq1$. The growth rate is $1/\tau_{\rm KHI}=\sqrt{-\Delta}=|kU|\sqrt{1-\alpha^2}$. Hence, a density contrast $|\alpha|$ close to 1 strongly dampens the growth rate of the KHI. 

In colliding wind binaries, the density and velocity of both winds are related through the momentum flux ratio $\eta$. Using Eq.~\ref{eq:eta} and mass conservation for both winds then, far enough from the binary so that $r_1\simeq r_2$, the density ratio is roughly
\begin{equation}
  \label{eq:mass_cons}
  \frac{\rho_2}{\rho_{1}}\simeq \eta \beta^{2}
\end{equation}

\begin{equation}
  \label{eq:growth_khi}
 \frac{\tau_{\rm adv}}{\tau_{\rm KHI}}=\frac{|\beta-1|\sqrt{1-\alpha^2}}{-\alpha(\beta-1)+(\beta+1)}=\frac{\eta^{1/2}\beta |\beta-1|}{1+\eta \beta^3}
\end{equation}

\subsection{Nonlinear evolution}
\subsubsection{2D Evolution}
In order to investigate the evolution of the KHI in the nonlinear regime, we have performed numerical simulations for increasing $\alpha$. The 2D setup is as follows: box size $(lx=8, ly=4)$, resolution $(1024\times 256)$, code: PLUTO~\citep{2007ApJS..170..228M}, adiabatic equation of state $P\propto \rho^{5/3}$, background pressure $P=1$ in the initial state (using units dimensioned to the box length, density, and velocity shear). Reflective boundary conditions are enforced in $y$ to confine the instability in the simulation box. We always have $\rho^+>\rho^-$ {\em i.e.}  the densest medium is found where $y>0$. 

\begin{figure}[h]
  \centering
   \includegraphics[width=0.85\linewidth]{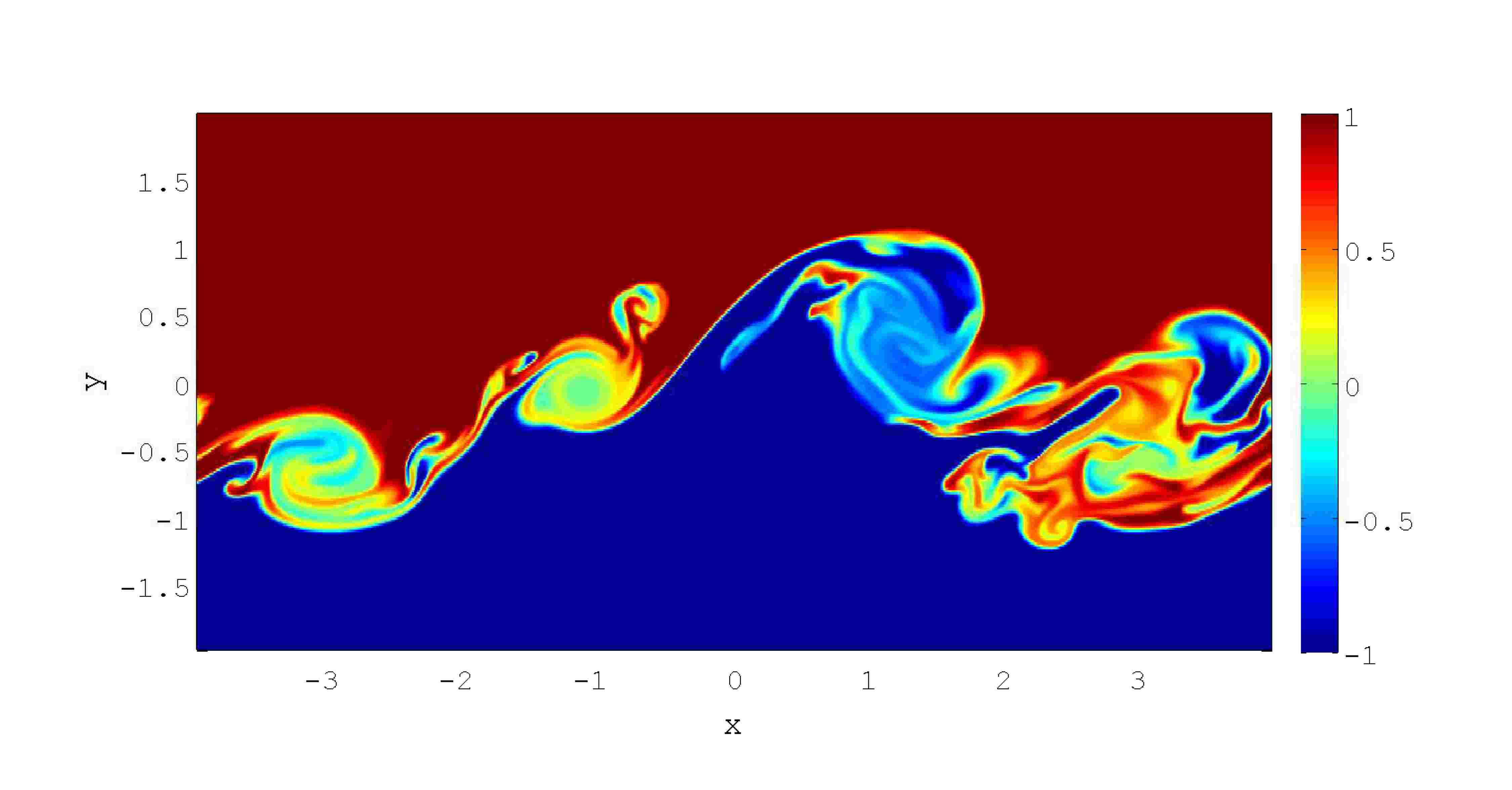}
    \caption{Snapshot of the density at t=21 (in dimensionless units) for $\alpha=0.5$.}
  \label{fig:snapshot_d}
\end{figure}

\begin{figure}[h!]
 \centering
   \includegraphics[width=0.45\linewidth]{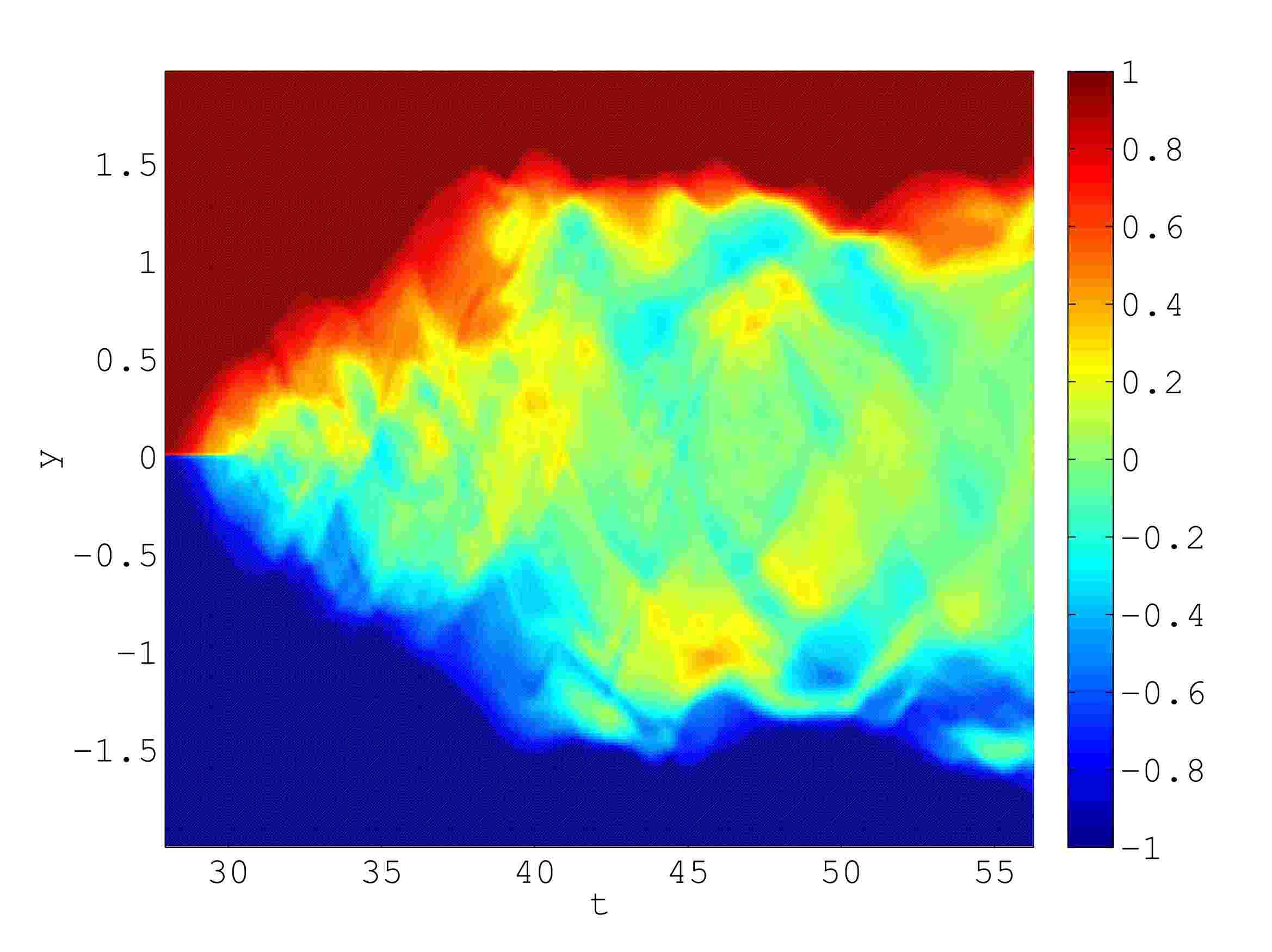}
   \includegraphics[width=0.45\linewidth]{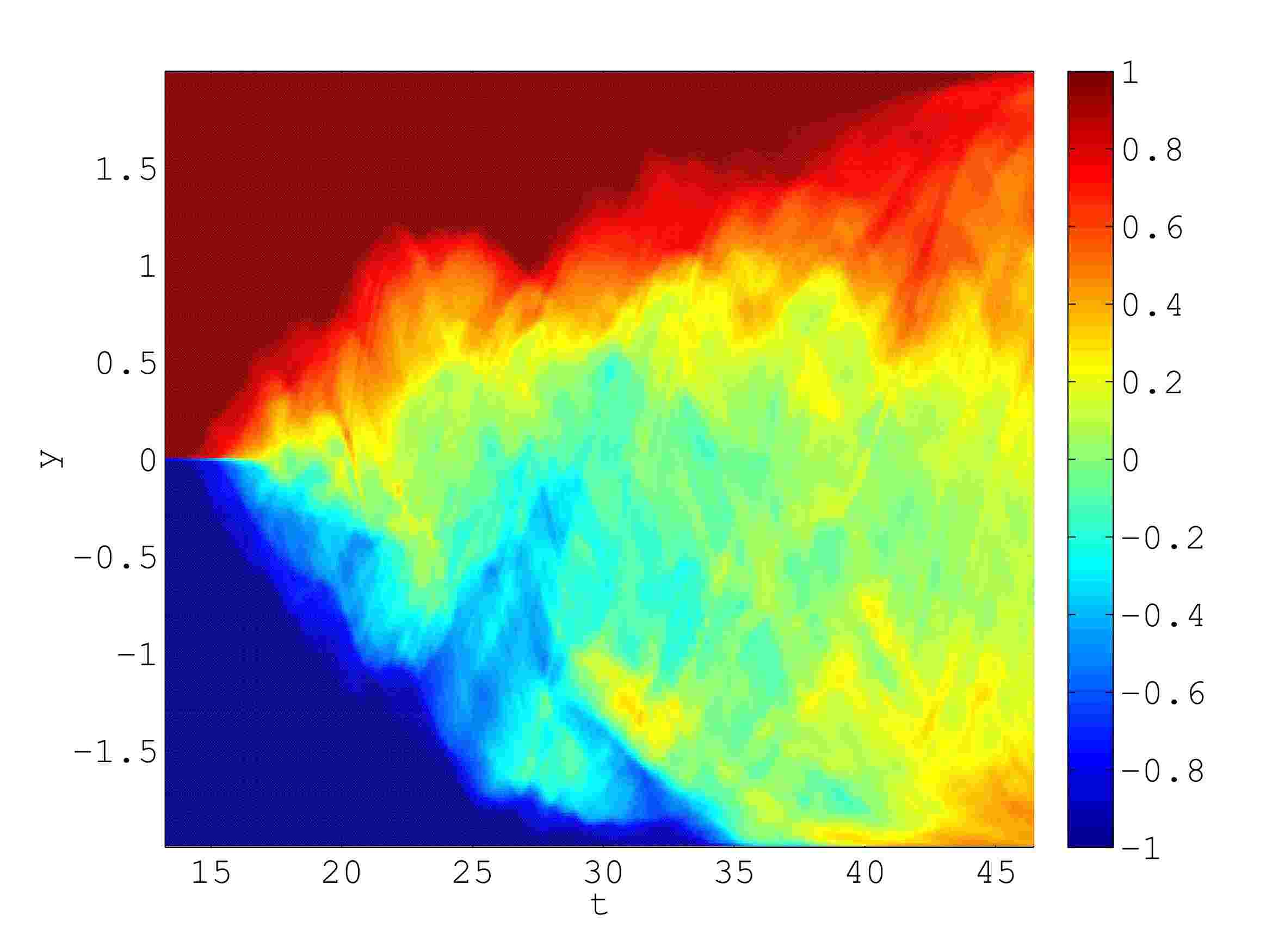} \\
   \includegraphics[width=0.45\linewidth]{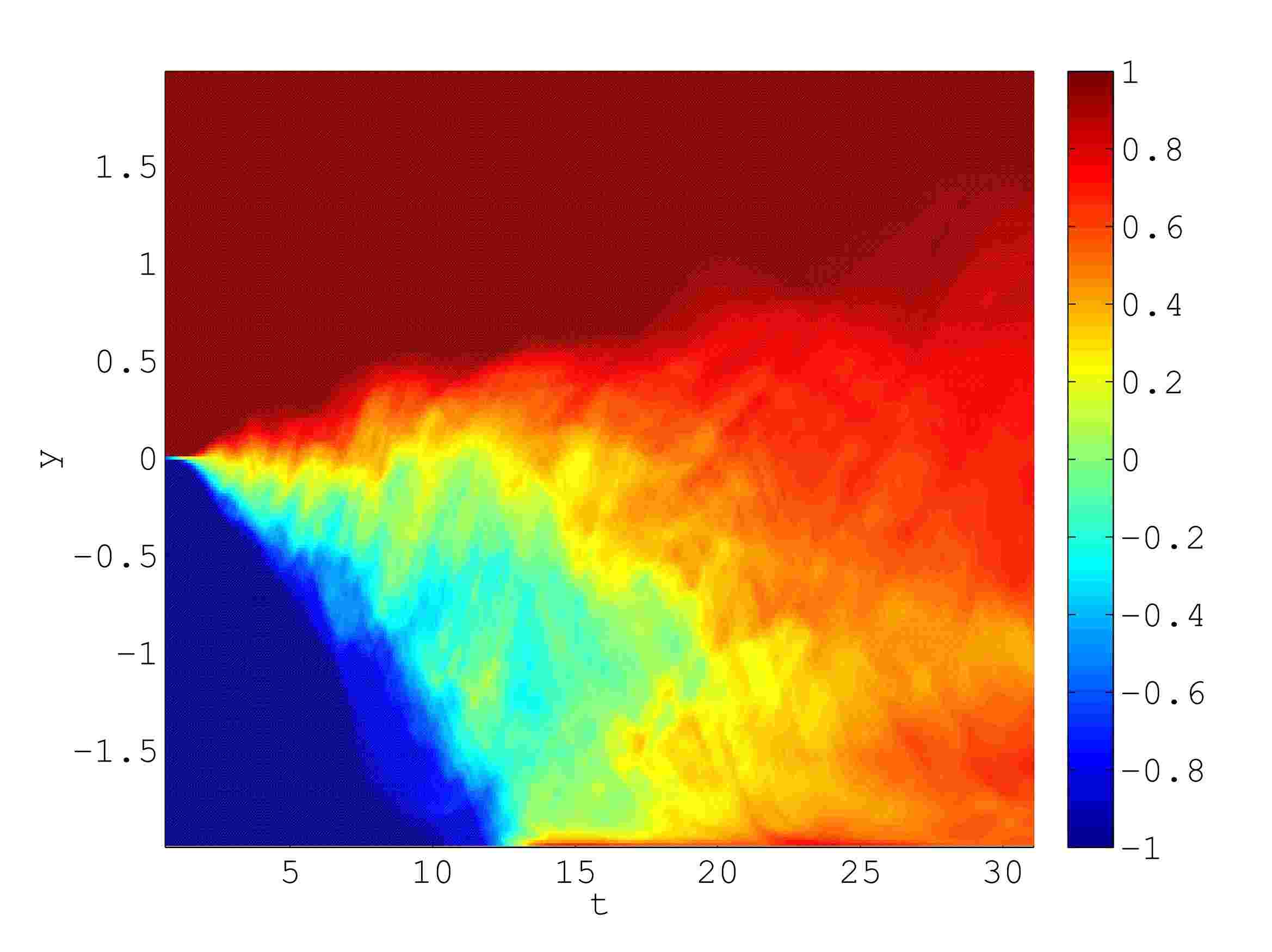}
   \includegraphics[width=0.45\linewidth]{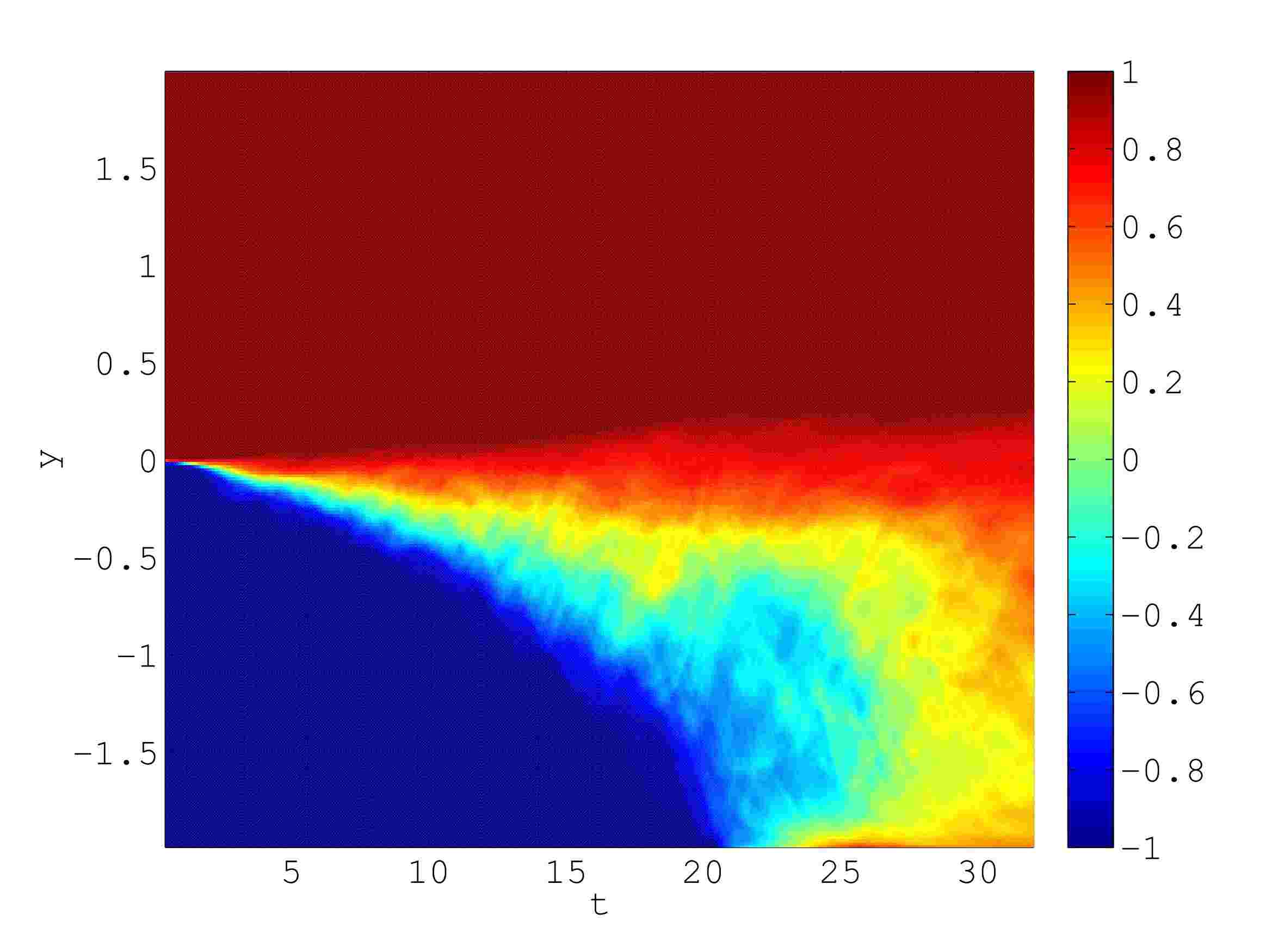}
   \caption{Diffusion of a passive scalar by the KHI. From left to right, top to bottom: $\alpha=0,0.5,0.9,0.99$.}
   \label{d2D}%
\end{figure}

In addition to that, we follow a passive scalar $s$, initialised with  $s=2\Theta(y)-1$ where $\Theta$ represents the Heaviside function. We performed simulations for $\{\alpha=0,0.5,0.9,0.99\}$. Kelvin-Helmholtz eddies are clearly present in the density snapshot shown in Fig.~\ref{fig:snapshot_d} for model $\alpha=0.5$. In order to show the diffusion of the passive scalar as a function of time, we plot the evolution of $\overline{s}(y,t)=\int s\,dx$ as a function of $y$ and $t$ in Fig.~\ref{d2D}. These results demonstrate that when $\alpha\ne 0$, the scalar diffusion propagates much less in the denser medium ($y>0$) and that diffusion looks less efficient when $|\alpha|$ increases, in the sense that the region with intermediate values of the scalar $s$ becomes smaller when $\alpha$ increases.

\begin{figure}[h!]
 \centering
   \includegraphics[width=0.45\linewidth]{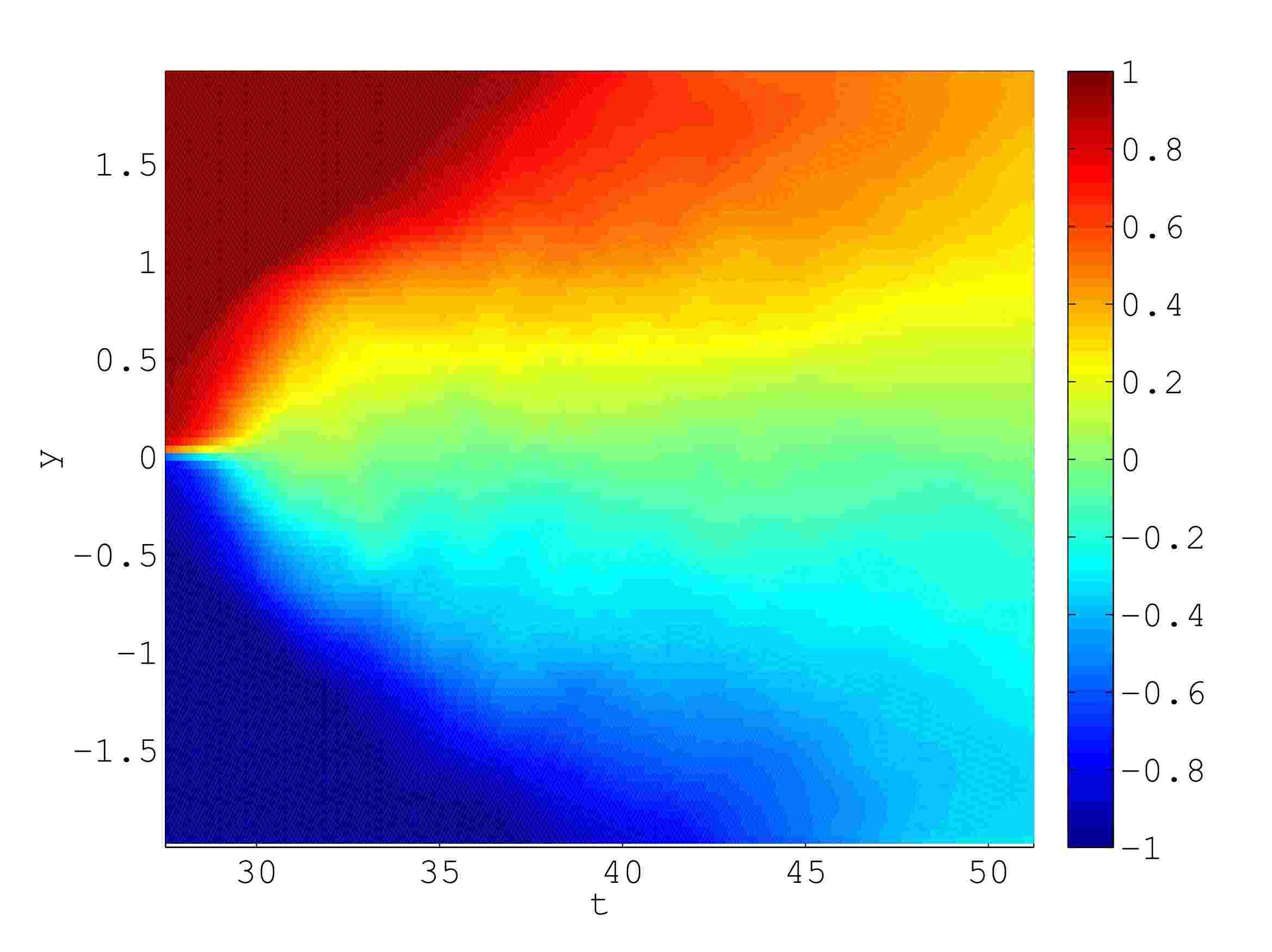}
   \includegraphics[width=0.45\linewidth]{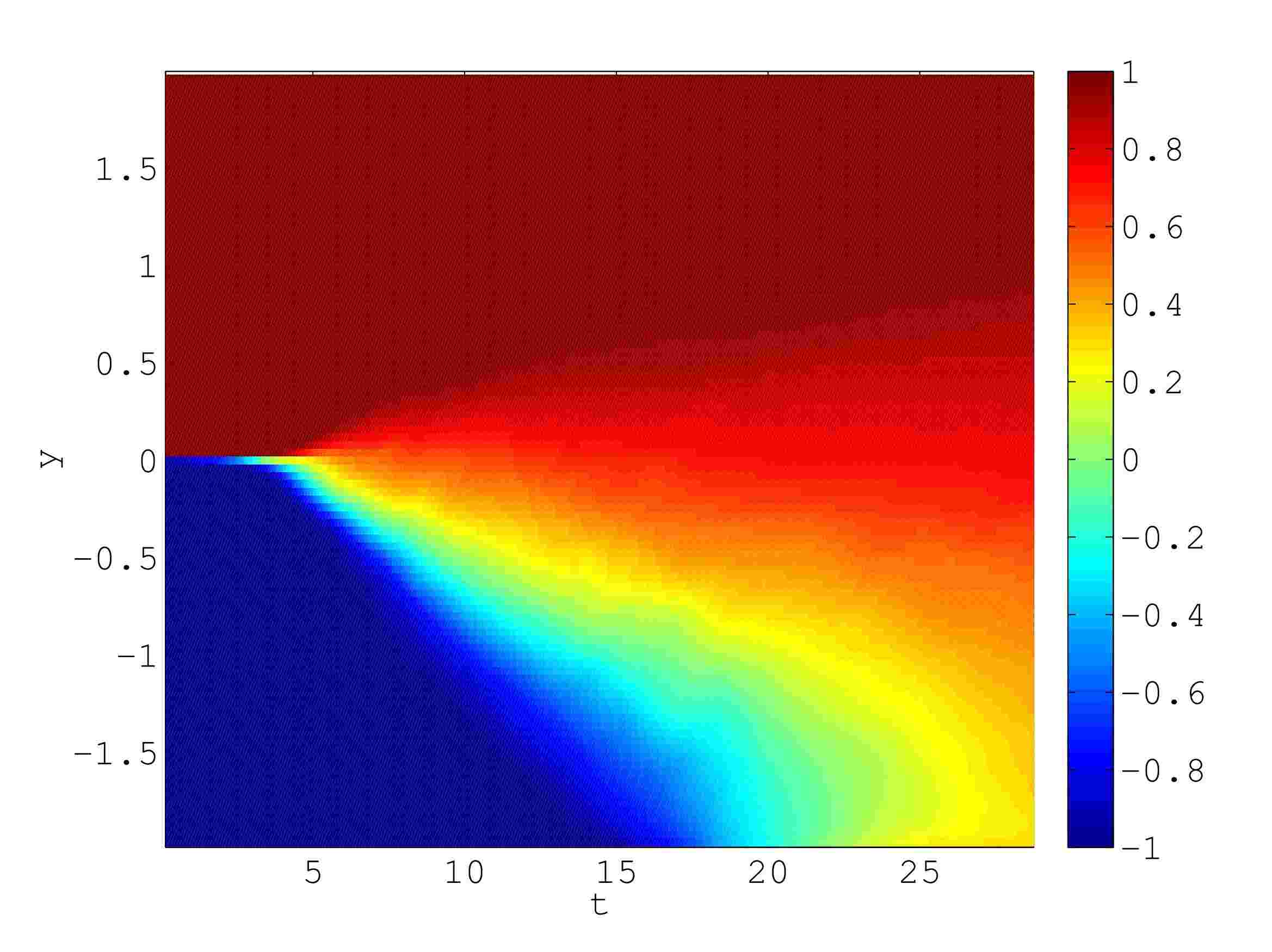} \\
   \caption{ Diffusion of the passive scalar in the 3D simulations with $\alpha=0$ (left) and $\alpha=0.9$ (right).}
   \label{d3D}%
\end{figure}

\subsection{3D evolution}
We have performed simulations for $\alpha=0$ and 0.9 in 3D to compare them to the 2D ones. They are very similar to the 2D configuration, except for the resolution which is reduced to $500\times 100 \times 100$ in order to reduce computational costs. We set $l_z=l_x=4.0$. $\overline{s}(y,t)$ is  shown on Fig.~\ref{d3D}. The direct comparison with the 2D cases indicates that faster diffusion into the more tenuous region is still verified in 3D; diffusion is also slightly less efficient in 3D.

\section{Parameters of the simulations}
 
\begin{table*}\label{tab:param_simu}
\begin{center}
\caption[]{Parameters of the simulations}
\begin{tabular}{c c c c c c c }
\hline
\noalign{\smallskip}
 $\{\eta,\beta\}$ &$v_1$ (km s$^{-1}$) &$v_2$ (km s$^{-1}$) &$\dot{M}_1$ ($10^{-7}M_{\odot}$ yr$^{-1}$) & $\dot{M}_2$  ($10^{-7}M_{\odot}$ yr$^{-1}$)& spiral?&  $S/S_1$ \\
\noalign{\smallskip}
\hline
 \noalign{\smallskip}
 $\{1,1\}$       &2000 &2000 & 1          & 1        &S        & 1 \\
$\{1,2\}$        &4000 &2000 & 0.5        & 1        &S        & 0.67\\
$\{1,4\}$        &2000 &500  & 0.25       & 1        &S        & 0.35\\
$\{1,8\}$        &4000 &500  & 0.25       & 2        &X        & \\
$\{1,20\}$       &40000&2000 & 0.05       & 1        &X        & \\
$\{1,200\}$      &8000 &40   & 0.05       & 10       &X        & \\
$\{0.5,0.01\}$   &40   &4000 & 100        & 0.5      &S        & 2.5\\
$\{0.5,0.05\}$   &200  &4000 & 40         & 1        &X        & \\
$\{0.5,0.1\}$    &400  &4000 & 20         & 1        &X        & \\
$\{0.5,0.5\}$    &1000 &2000 & 4          & 1        &S        & 1.3 \\
$\{0.5,1\}$      &2000 &2000 & 2          & 1        &S        & 1\\
$\{0.5,2\}$      &4000 &2000 & 1          & 1        &S        & 0.8\\
$\{0.5,8\}$      &4000 &500  & 1          & 4        &X        & \\
$\{0.5,20\}$     &8000 &400  & .5         & 5        &X        & \\
$\{0.5,200\}$     &8000 &40   & .05       & 5        &X        & \\
$\{0.0625,0.05\}$&100  &2000 & 320        & 1        &S        & 1.1\\
$\{0.0625,0.1\}$ &200  &2000 & 160        & 1        &X        & \\
$\{0.0625,0.5\}$ &1000 &2000 & 32         & 1        &S        & 1.04\\
$\{0.0625,1\}$   &2000 &2000 & 16         & 1        &S        & 1\\
$\{0.0625,2\}$   &4000 &2000 & 8          & 1        &S        & 0.9\\
$\{0.0625,4\}$   &4000 &1000 & 4          & 1        &S/X      & \\
$\{0.0625,8\}$   &4000 &500  & 4          & 2        &S        & 0.8\\
$\{0.0625,20\}$  &40000&2000 & .8         & 1        &S        & \\

\noalign{\smallskip}
\hline
\end{tabular}
\end{center}
\end{table*}

\end{document}